\documentclass{article}
\usepackage{amsbsy} 
\usepackage{amssymb}
\usepackage{amstext}
\usepackage{graphics}

\newcommand{\be}{\begin{equation}}
\newcommand{\ee}{\end{equation}}
\newcommand{\bea}{\begin{eqnarray}}
\newcommand{\eea}{\end{eqnarray}}
\newcommand{\ben}{\begin{eqnarray*}}
\newcommand{\een}{\end{eqnarray*}}

\newcommand{\dst}{\displaystyle}

\newcommand{\bs}{\boldsymbol}

\newcommand{\pa}{\partial}

\newcommand{\ve}{\varepsilon}

\newcommand{\F}{{\mathcal F}}
\newcommand{\Fo}{{\mathcal F}_o}
\newcommand{\Fs}{{\mathcal F}_{\text{sub}}}
\newcommand{\ds}{d_{\text{sub}}}

\newcommand{\vta}{V_{\text{total}(s)}^{\text{all}}}
\newcommand{\vtd}{V_{\text{total}(s)}^{\text{dir}}}
\newcommand{\vca}{V_{\text{cell}(s)}^{\text{all}}}
\newcommand{\vcd}{V_{\text{cell}(s)}^{\text{dir}}}
\newcommand{\ga}{\widetilde{\Gamma}_{(s)}^{\text{all}}}
\newcommand{\gd}{\widetilde{\Gamma}_{(s)}^{\text{dir}}}

\newcommand{\nca}{N_{\text{cells}(s)}^{\text{all}}}
\newcommand{\ncd}{N_{\text{cells}(s)}^{\text{dir}}}
\def\ncadef#1{N_{\text{cells}(#1)}^{\text{all}}}
\def\ncddef#1{N_{\text{cells}(#1)}^{\text{dir}}}

\newcommand{\gai}{\Gamma_{(s)}^{\text{all}}}
\newcommand{\gdi}{\Gamma_{(s)}^{\text{dir}}}
\newcommand{\gaiv}{\Gamma_{(4)}^{\text{all}}}
\newcommand{\gdiv}{\Gamma_{(4)}^{\text{dir}}}

\newcommand{\vtaf}{\overline{V}_{\text{total}(s)}^{\text{all}}}
\newcommand{\vtdf}{\overline{V}_{\text{total}(s)}^{\text{dir}}}
\newcommand{\vcaf}{\overline{V}_{\text{cell}(s)}^{\text{all}}}
\newcommand{\vcdf}{\overline{V}_{\text{cell}(s)}^{\text{dir}}}
\newcommand{\gaf}{\overline{\Gamma}_{(s)}^{\text{all}}}
\newcommand{\gdf}{\overline{\Gamma}_{(s)}^{\text{dir}}}

\newcommand{\nfa}{N_{\text{filters}(s)}^{\text{all}}}
\newcommand{\nfd}{N_{\text{filters}(s)}^{\text{dir}}}
\def\nfadef#1{N_{\text{filters}(#1)}^{\text{all}}}
\def\nfddef#1{N_{\text{filters}(#1)}^{\text{dir}}}

\newcommand{\fmin}{f_{\text{min}}}
\newcommand{\fmax}{f_{\text{max}}}
\newcommand{\tmin}{\tau_{\text{min}}}

\def\ctall#1{T_{\text{cross}(#1)}^{\text{all}}}
\def\ctdir#1{T_{\text{cross}(#1)}^{\text{dir}}}
\def\ctfall#1{\overline{T}_{\text{cross}(#1)}^{\text{all}}}
\def\ctfdir#1{\overline{T}_{\text{cross}(#1)}^{\text{dir}}}

\def\fkl#1{\stackrel{\scriptscriptstyle(#1)}{f_l}}
\def\fko#1{\stackrel{\scriptscriptstyle(#1)}{f_o}}

\def\tav#1{\left\langle#1\right\rangle}

\begin{document}

\title{
Data analysis of gravitational-wave signals\\
from spinning neutron stars.\\
III.\ Detection statistics and computational requirements}

\author{Piotr Jaranowski
\\{\it Institute of Physics, Bia{\l}ystok University}\\
  {\it Lipowa 41, 15-424 Bia{\l}ystok, Poland}
\and Andrzej Kr\'olak
\\{\it Institute of Mathematics,
       Polish Academy of Sciences}\\
  {\it \'{S}niadeckich 8, 00-950 Warsaw, Poland}}
\maketitle

\begin{abstract}

We develop the analytic and numerical tools for data analysis of the
gravitational-wave signals from spinning neutron stars for ground-based laser
interferometric detectors.  We study in detail the statistical properties of the
optimum functional that need to be calculated in order to detect the
gravitational-wave signal from a spinning neutron star and estimate its
parameters.  We derive formulae for false alarm and detection probabilities both
for the optimal and the suboptimal filters.  We assess the computational
requirements needed to do the signal search.  We compare a number of criteria to
build sufficiently accurate templates for our data analysis scheme.  We verify
the validity of our concepts and formulae by means of the Monte Carlo
simulations.  We present algorithms by which one can estimate the parameters of
the continuous signals accurately.

\vspace{2ex}\noindent PACS number(s): 95.55.Ym,04.80.Nn,95.75.Pq,97.60.Gb
\end{abstract}

\section{Introduction}

This paper is a continuation of the study of data analysis for one of the
primary sources of gravitational waves for long-arm ground-based laser
interferometers currently under construction \cite{GEO600,LIGO,VIRGO,TAMA300}:
spinning neutron stars.  In the first paper of this series \cite{P1} (hereafter
Paper I) we have introduced a two-component model of the gravitational-wave
signal from a spinning neutron star and we have derived the data processing
scheme, based on the principle of maximum likelihood, to detect the signal and
estimate its parameters.  In the second paper \cite{P2} (hereafter Paper II) we
have studied in detail accuracies of estimation of the parameters achievable
with the proposed data analysis method.

The main purpose of this paper which is Paper III of the series is to study the
statistical properties of the optimal functional that we need to calculate in
order to detect the signal.  We find that the two-component model of the signal
introduced in Paper I can be generalized in a straightforward way to the
$N$-component signal.  The main idea of this work is to approximate each
frequency component of the signal by a {\em linear} signal by which we mean a
signal with a constant amplitude and a phase linear in the parameters of the
signal.  We have demonstrated the validity of such an approximation in Paper II
by means of the Monte Carlo simulations which show that the rms errors
calculated using the linear model closely approximate those of the exact model.
The key observation is that for the linear model the detection statistics is a
homogeneous random field parametrized by the parameters of the signal.  For such
a field one can calculate a chracteristic correlation hyperellipsoid which
volume is independent of the values of the parameters.  The correlation
hyperellipsoid determines an elementary {\em cell} in the parameter space.  We
find that the number of cells covering the parameter space is a key concept that
allows the calculation of the false alarm probabilities that are needed to
obtain thresholds for the optimum statistics in order to search for significant
signals.  We use these ideas to calculate the number of filters needed to do the
search.  We show that the concept of an elementary cell is also useful in the
calculation of true rms errors of the estimators of the parameters that can be
achieved with matched filtering and explain their deviations from rms errors
calculated from the covariance matrix.  In this paper we develop a general
theory of suboptimal filters which is necessary as such filters usually occur in
practice.  Our concept of an elementary cell carries over to the case of
suboptimal filtering in a straightforward manner.  The analytic tools develop in
this work lead to independent criteria for construction of accurate templates to
do the signal search.  We demonstarte that those criteria give a consistent
picture of what a suitable template should be.  In an appendix to this paper we
indicate how to parametrize the templates in order that they realize an
approximately linear model so that the analytic formulae developed here can
directly be used.

The plan of the paper is as follows.  In Sec.\ 2 we introduce an $N$-component
model of the gravitational-wave signal from a spinning neutron star.  In Sec.\ 3
we study in detail the detection statistics for the $N$-component model.  We
show that the detection statistics constitutes a certain random field.  We
derive the probabilities of the false alarm and the probabilities of detection.
We present two approaches to the calculation of the probability of false alarm:
one is based on dividing the parameter space into elemetary cells determined by
the correlation function of the detection statistics and the other is based on
the geometry of random fields.  We compare the theoretical formulae with the
Monte-Carlo simulations.  In Sec.\ 4 we carry out detailed calculations of the
number of cells for the all-sky and directed searches.  In Sec.\ 5 we estimate
the number of filters needed to calculate the detection statistics and we obtain
the computational requirements needed to perform the searches so that the data
processing speed is comparable to data aquisition rate.  We compare our
calculations with the results of Brady {\it et al}.\ \cite{BCCS98} obtained
before by a different approach.  In Sec.\ 6 we present in detail the theory of
suboptimal filters and consider their use in the detection of continuous
signals.  In Sec.\ 7 we propose a detailed algorithm to estimate accurately the
parameters of the signal and we perform the Monte-Carlo simulations to determine
its performance.  In Appendix A we give analytic formulae for some coefficients
in the detection statistics.  In Appendix B we present analytic formulae for the
components of the Fisher matrix for the approximate, linear model of the
gravitational-wave signal from a spinning neutron star.  In Appendix C we give a
worked example of the application of our theory of suboptimal filtering derived
in Sec.\ 6.  In Apendix D we study the transformation of the paramaters of the
signal to a set of parameters such that the model is approximately linear.

\section{The $N$-component model of the gravitational-wave signal from a 
spinning neutron star}

In Paper I we have introduced a two-component model of the gravitational-wave
signal from a spinning neutron star.  The model describes the quadrupole
gravitational-wave emission from a freely precessing axisymmetric star.  Each of
the components of the model is a narrowband signal where frequency band of one
component is centered around a frequency $f_o$ which is the sum of the spin
frequency and the precession frequency and the frequency band of the second
component is centered around $2f_o$.  A special case of the above signal
consisting of one component only describes the quadrupole gravitational wave
from a triaxial ellipsoid rotating about one of its principal axes.  In this
case the narrowband signal is centered around twice the spin frequency of the
star.  However there are other physical mechanisms generating gravitational
waves and this can lead to signals consisting of many components.  Recently two
new mechanisms have been studied.  One is the $r-$mode instability of spinning
neutron stars \cite{A97,LOM98,OLCSVA98} that yield a spectrum of
gravitational-wave frequencies with the dominant one of $4/3$ of the star spin.
The other is a temperature asymmetry in the interior of the neutron star that is
misaligned from the spin axis \cite{B98}.  This can explain that most of the
rapidly accreting weakly magnetic neutron stars appear to be rotating at
approximately the same frequency due to the balance between the angular momentum
accreted by the star and lost to gravitational radiation.  Therefore in this
paper we shall introduce a signal consisting of $N$ narrowband components
centered around $N$ different frequencies.  More precisely we shall assume that
over the bandwidth of each component the spectral density of the detector's
noise is nearly constant and that the bandwidths of the components do not
overlap.

Analytic formulae in this paper will be given for the $N$-component signal.
However in numerical calculations and simulations we shall restrict ourselves to 
a one-component model.

We propose the following model of the $N$-component signal:
\be
\label{sig}
h(t) = \sum^{N}_{l=1}h_l(t),\quad
h_l(t) = \sum^{4}_{i=1} A_{li}\,h_{li}(t),\quad
l = 1,\ldots,N,
\ee
where $A_{li}$ are $4N$ nearly constant amplitudes.  The amplitudes are nearly
constant because they depend on the frequency of the gravitational wave which is
assumed to change little over the time of observation.  The amplitudes $A_{li}$
depend on the physical mechanism generating gravitational waves, as well as on
the polarization angle and the initial phase of the wave [cf.\ Eqs.\ (28)--(35)
of Paper I].  The above structure of the $N$-component signal is motivated by
the form of the two-component signal considered in Paper I [cf.\ Eq.\ (27) of
Paper I].  The time dependent functions $h_{li}$ have the form
\be
\begin{array}{rclrcl}
h_{l1}(t) &=& a(t) \cos\Phi_l(t), & h_{l2}(t) &=& b(t) \cos\Phi_l(t),\\[2ex]
h_{l3}(t) &=& a(t) \sin\Phi_l(t), & h_{l4}(t) &=& b(t) \sin\Phi_l(t),
\end{array}\quad l=1,\ldots,N,
\ee 
where the functions $a$ and $b$ are given by
\bea
\label{adef}
a(t) &=&
\frac{1}{16}\sin2\gamma(3 - \cos2\lambda)(3 - \cos2\delta)
\cos[2(\alpha-\phi_r-\Omega_r t)]
\nonumber\\&&
-\frac{1}{4}\cos2\gamma\sin\lambda(3 - \cos2\delta)
\sin[2(\alpha-\phi_r-\Omega_r t)]
\nonumber\\&&
+\frac{1}{4}\sin2\gamma\sin2\lambda\sin2\delta
\cos[\alpha-\phi_r-\Omega_r t]
\nonumber\\&&
-\frac{1}{2}\cos2\gamma\cos\lambda\sin2\delta
\sin[\alpha-\phi_r-\Omega_r t]
\nonumber \\&&
+\frac{3}{4}\sin2\gamma\cos^2\lambda\cos^2\delta,
\\
\label{bdef}
b(t) &=&
\cos2\gamma\sin\lambda\sin\delta\cos[2(\alpha-\phi_r-\Omega_r t)]
\nonumber \\&&
+\frac{1}{4}\sin2\gamma(3 - \cos2\lambda)\sin\delta
\sin[2(\alpha-\phi_r-\Omega_r t)]
\nonumber \\&&
+\cos2\gamma\cos\lambda\cos\delta\cos[\alpha-\phi_r-\Omega_r t]
\nonumber \\&&
+\frac{1}{2}\sin2\gamma\sin2\lambda\cos\delta
\sin[\alpha-\phi_r-\Omega_r t].
\eea

The functions $a$ and $b$ are the amplitude modulation functions.  They depend
on the position of the source in the sky (right ascension $\alpha$ and
declination $\delta$ of the source), the position of the detector on the Earth
(detector's latitude $\lambda$), the angle $\gamma$ describing orientation of
the detector's arms with respect to local geographical directions (see Sec.\ II
A of Paper I for the definition of $\gamma$), and the phase $\phi_r$ determined
by the position of the Earth in its diurnal motion at the beginning of
observation.  Thus the functions $a$ and $b$ are independent of the physical
mechanisms generating gravitational waves.  Formulae (\ref{adef}) and
(\ref{bdef}) are derived in Sec.\ II A of Paper I.

The phase $\Phi_l$ of the $l$th component is given by
\be
\label{phase1}
\Phi_l(t) = 2\pi \sum_{k=0}^{s_1}{\fkl k}\frac{t^{k+1}}{(k+1)!}
+ \frac{2\pi}{c} {\bf n}_0\cdot{\bf r}_{\rm ES}(t)
\sum_{k=0}^{s_2}{\fkl k}\frac{t^k}{k!}
+ \frac{2\pi}{c} {\bf n}_0\cdot{\bf r}_{\rm E}(t)
\sum_{k=0}^{s_3}{\fkl k}\frac{t^k}{k!},
\ee   
where ${\bf r}_{\rm ES}$ is the vector joining the solar system barycenter (SSB)
with the center of the Earth and ${\bf r}_{\rm E}$ joins the center of the Earth
with the detector, ${\bf n}_0$ is the constant unit vector in the direction from
the SSB to the neutron star.  We assume that the $l$th component is a narrowband
signal around some frequency $\fkl{0}$ which we define as instantaneous
frequency evaluated at the SSB at $t=0$, $\fkl{k}$ ($k=1,2,\ldots$) is the $k$th
time derivative of the instantaneous frequency of the $l$th component at the SSB
evaluated at $t=0$.  To obtain formula (\ref{phase1}) we model the frequency of
each component in the rest frame of the neutron star by a Taylor series.  For
the detailed derivation of the phase model see Sec.\ II~B and Appendix A of
Paper I.

\section{Optimal filtering for the $N$-component signal}

\subsection{Maximum liklihood detection}

Maximum likelihood detection and parameter estimation method applied in Paper I
to the two-component signal generalizes in a straightforward manner to the
$N$-component signal.

We assume that the noise $n$ in the detector is an additive, stationary, 
Gaussian, and zero-mean continuous random process. Then the data $x$ (if the 
signal $h$ is present) can be written as
\be
x(t) = n(t) + h(t).
\ee
The log likelihood function has the form
\be
\label{loglr1}
\ln\Lambda = (x|h) - \frac{1}{2}(h|h),
\ee
where the scalar product $(\,\cdot\,|\,\cdot\,)$ is defined by 
\be
\label{SP}
(h_1|h_2) := 4\Re \int^{\infty}_{0} 
\frac{\tilde{h}_1(f)\tilde{h}^*_2(f)}{S_h(f)} df.
\ee
In Eq.\ (\ref{SP}) $\tilde{}$ denotes the Fourier transform, ${}^*$ is complex
conjugation, and $S_h$ is the {\em one-sided} spectral density of the detector's
noise.  As by our assumption the bandwidths of the components of the signal are
disjoint we have $(h_l|h_{l'})\approx 0$ for $l\ne l'$, and the log likelihood
ratio (\ref{loglr1}) can be written as the sum of the log likelihood ratios for
each individual component:
\be
\label{loglr2}
\ln\Lambda\approx \sum^N_{l=1} \left[ (x|h_l)-\frac{1}{2}(h_l|h_l) \right].
\ee
Thus we can consider the $N$-component signal as $N$ independent signals. Since 
we assume that over the bandwidth of each component of the signal the spectral 
density $S_h(f)$ is nearly constant and equal to $S_h(f_l)$, where $f_l$ is the 
frequency of the signal $h_l$ measured at the SSB at $t=0$, the scalar products 
in Eq.\ (\ref{loglr2}) can be approximated by
\be
\label{approx1}
(x|h_l) \approx \frac{2}{S_h(f_l)} \int^{T_o/2}_{-T_o/2}x(t)h_l(t)\,dt, \quad
(h_l|h_l) \approx \frac{2}{S_h(f_l)} 
\int^{T_o/2}_{-T_o/2}\left[h_l(t)\right]^2\,dt,
\ee
where $T_o$ is the observation time, and the observation interval is 
$\left[-T_o/2,T_o/2\right]$.

It is useful to introduce the following notation
\be
\tav{x} := \frac{1}{T_o}\int^{T_o/2}_{-T_o/2}x(t)\,dt.
\ee
Using the above notation and Eq.\ (\ref{approx1}) the log likelihood ratio
from Eq.\ (\ref{loglr2}) can be written as
\be
\label{loglr3}
\ln\Lambda\approx \sum^N_{l=1} \frac{2T_o}{S_h(f_l)}
\left( \tav{xh_l}-\frac{1}{2}\tav{h_l^2} \right).
\ee

Proceeding along the line of argument of Paper I [cf.\ Sec.\ III~A of 
Paper I] we find the explicit analytic formulae for the maximum likelihood 
estimators $\widehat{A}_{li}$ of the amplitudes $A_{li}$:
\be
\label{amle}
\begin{array}{rcl}
\widehat{A}_{l1} &\approx& {\dst 2\frac{B\tav{x h_{l1}}-C\tav{x h_{l2}}}{D},} 
\\[2ex]
\widehat{A}_{l2} &\approx& {\dst 2\frac{A\tav{x h_{l2}}-C\tav{x h_{l1}}}{D},} 
\\[2ex]
\widehat{A}_{l3} &\approx& {\dst 2\frac{B\tav{x h_{l3}}-C\tav{x h_{l4}}}{D},} 
\\[2ex]
\widehat{A}_{l4} &\approx& {\dst 2\frac{A\tav{x h_{l4}}-C\tav{x h_{l3}}}{D},}
\end{array}\quad l=1,\ldots,N,
\ee
where we have defined
\be
\label{ABC}
A:=\tav{a^2},\quad B:=\tav{b^2},\quad C:=\tav{ab},\quad D:=AB-C^2.
\ee
To obtain Eqs.\ (\ref{amle}) we have used the following approximate relations:
\be
\label{ch1}
\begin{array}{ccc}
&\tav{h_{l1} h_{l3}} \approx \tav{h_{l1} h_{l4}} \approx
 \tav{h_{l2} h_{l3}} \approx \tav{h_{l2} h_{l4}} \approx 0,&\\[2ex]
&\tav{h_{l1}^2} \approx \tav{h_{l3}^2} \approx \frac{1}{2} A,\quad
 \tav{h_{l2}^2} \approx \tav{h_{l4}^2} \approx \frac{1}{2} B,\quad
 \tav{h_{l1} h_{l2}} \approx \tav{h_{l3} h_{l4}} \approx \frac{1}{2} C,&
\end{array}\quad l=1,\ldots,N.
\ee

One can show that when the observation time $T_o$ is an integer multiple of one
sidereal day the function $C$ vanishes.  To simplify the formulae from now on we
assume that $T_o$ is an integer multiple of one sidereal day (in Appendix A we
have given the explicit analytic expressions for the functions $A$ and $B$ in
this case).  In the real data analysis for long stretches of data of the order
of months such a choice of observation time is reasonable.  Then Eqs.\
(\ref{amle}) take the form
\be
\label{amle0}
\widehat{A}_{l1} \approx 2\frac{\tav{x h_{l1}}}{A},\quad
\widehat{A}_{l2} \approx 2\frac{\tav{x h_{l2}}}{B},\quad
\widehat{A}_{l3} \approx 2\frac{\tav{x h_{l3}}}{A},\quad
\widehat{A}_{l4} \approx 2\frac{\tav{x h_{l4}}}{B},\quad l=1,\ldots,N.
\ee

The reduced log likelihood function $\F$ is the log likelihood function where 
amplitude parameters $A_{li}$ were replaced by their estimators 
$\widehat{A}_{l1}$. By virtue of Eqs.\ (\ref{ch1}) and (\ref{amle0}) from Eq.\ 
(\ref{loglr3}) one gets
\be
\label{Fred}
\F \approx \sum^N_{l=1} \frac{2 T_o}{S_h(f_l)} \left[
\frac{\tav{x h_{l1}}^2+\tav{x h_{l3}}^2}{A}
+ \frac{\tav{x h_{l2}}^2+\tav{x h_{l4}}^2}{B} \right].
\ee

To obtain the maximum likelihood estimators of the parameters of the signal one
first finds the maximum of the functional $\F$ with respect to the frequency,
the spindown parameters and the angles $\alpha$ and $\delta$ and then one
calculates the estimators of the amplitudes $A_{li}$ from the analytic formulae
(\ref{amle}) with the correlations $\tav{xh_{li}}$ evaluated at the values of
the parameters obtained by the maximization of the functional $\F$.  Thus
filtering for the $N$-component narrowband gravitational-wave signal from a
neutron star requires $4N$ {\em linear} filters.  The amplitudes $A_{li}$ of the
signal depend on the physical mechanisms generating gravitational waves.  If we
know these mechanisms and consequently we know the dependence of $A_{li}$ on a
number of parameters we can estimate these parameters from the estimators of the
amplitudes by least-squares method.  We shall consider this problem in a future
paper.

Next we shall study the statistical properties of the functional $\F$.  The
probability density functions (pdfs) of $\F$ when the signal is absent or
present can be obtained in a similar manner as in Sec.\ III~B of Paper I for the
two-component signal.

Let us suppose that filters $h_{li}$ are known functions of time, i.e.\ the
phase parameters ${\fkl k}$, $\alpha$, $\delta$ are known, and let us define 
the following random variables:
\be
x_{li} := \tav{xh_{li}},\quad l=1,\ldots,N,\quad i=1,\ldots,4.
\ee
Since $x$ is a Gaussian random process the random variables $x_{li}$ being 
linear in $x$ are also Gaussian. Let ${\rm E}_0\{x_{li}\}$ and ${\rm 
E}_1\{x_{li}\}$ be respectively the means  of $x_{li}$ when the signal is 
absent and when the signal is present. One easily gets
\bea
\label{mean1a}
&{\rm E}_0\{x_{li}\} = 0,\quad i=1,\ldots,4,\quad l=1,\ldots,N,& \\[2ex]
\label{mean1b}
&{\rm E}_1\{x_{l1}\} = \frac{1}{2} A A_{l1},\quad
 {\rm E}_1\{x_{l2}\} = \frac{1}{2} B A_{l2},\quad
 {\rm E}_1\{x_{l3}\} = \frac{1}{2} A A_{l3},\quad
 {\rm E}_1\{x_{l4}\} = \frac{1}{2} B A_{l4},\quad l=1,\ldots,N.\quad&
\eea
Since here we assume that the observation time is an integer multiple of
one sidereal day it immediately follows from Eqs.\ (\ref{ch1}) that the 
Gaussian random variables $x_{li}$ are uncorrelated and their variances are
given by
\be
\begin{array}{c}
{\rm Var}\{x_{li}\} = {\dst \frac{S_h(f_l)A}{4T_o}},\quad i=1,3,\\[2ex]
{\rm Var}\{x_{li}\} = {\dst \frac{S_h(f_l)B}{4T_o}},\quad i=2,4,
\end{array}\quad l=1,\ldots,N.
\ee
The variances are the same irrespectively whether the signal is absent or 
present. We introduce new rescaled variables $z_{li}$:
\be
\begin{array}{c}
z_{li} = {\dst 2\sqrt{\frac{T_o}{S_h(f_l)A}} x_{li}},\quad i=1,3,\\[2ex] 
z_{li} = {\dst 2\sqrt{\frac{T_o}{S_h(f_l)B}} x_{li}},\quad i=2,4,
\end{array}\quad l=1,\ldots,N,
\ee
so that $z_{li}$ have a unit variance. By means of Eqs.\ (\ref{mean1a}) and 
(\ref{mean1b}) it is easy to show that
\be
{\rm E}_0\{z_{li}\} = 0,\quad i=1,\ldots,4,\quad l=1,\ldots,N,
\ee
and
\bea
\begin{array}{rcl}
m_{l1} &:=& {\rm E}_1\{z_{l1}\} = {\dst \sqrt{\frac{T_oA}{S_h(f_l)}}} A_{l1}, 
\\[2ex]
m_{l2} &:=& {\rm E}_1\{z_{l2}\} = {\dst \sqrt{\frac{T_oB}{S_h(f_l)}}} A_{l2}, 
\\[2ex]
m_{l3} &:=& {\rm E}_1\{z_{l3}\} = {\dst \sqrt{\frac{T_oA}{S_h(f_l)}}} A_{l3}, 
\\[2ex]
m_{l4} &:=& {\rm E}_1\{z_{l4}\} = {\dst \sqrt{\frac{T_oB}{S_h(f_l)}}} A_{l4},
\end{array}\quad l=1,\ldots,N.
\eea
The statistics $\F$ from Eq.\ (\ref{Fred}) can be expressed in terms of the 
variables $z_{li}$ as
\be
\label{F}
\F \approx \frac{1}{2} \sum^{N}_{l=1}\sum^{4}_{i=1} z_{li}^2.
\ee

The pdfs of $\F$ both when the signal is absent and present are known.  When the
signal is absent $2\F$ has a $\chi^2$ distribution with $4N$ degrees of freedom
and when the signal is present it has a noncentral $\chi^2$ distribution with
$4N$ degrees of freedom and noncentrality parameter
$\lambda=\sum^{N}_{l=1}\sum^4_{i=1}m_{li}^2$.  We find that the noncentrality
parameter is exactly equal to the {\em optimal signal-to-noise ratio} $d$
defined as
\be
\label{snr}
d := \sqrt{(h|h)}.
\ee
This is the maximum signal-to-noise ratio that can be achieved for a signal in 
additive noise with the {\em linear} filter \cite{Da}. This fact does not depend 
on the statistics of the noise. 

Consequently the pdfs $p_0$ and $p_1$ when respectively the signal is absent and 
present are given by
\bea
\label{p0}
p_0({\mathcal F})
&=&\frac{{\mathcal F}^{n/2-1}}{(n/2 -1)!}\exp(-{\mathcal F}),
\\
\label{p1}
p_1({d,\mathcal F})
&=&\frac{(2{\mathcal F})^{(n/2 -1)/2}}{d^{n/2-1}} 
I_{n/2-1}\left(d\sqrt{2 {\mathcal F}}\right)
\exp\left(-{\mathcal F}-\frac{1}{2}d^2\right),
\eea
where $n=4N$ is the number of degrees of freedom of $\chi^2$ distributions 
and $I_{n/2-1}$ is the modified Bessel function of the first kind and order 
$n/2-1$. The false alarm probability $P_F$ is the probability that $\F$
exceeds a certain threshold $\Fo$ when there is no signal. In our case we have
\be
\label{PF}
P_F(\Fo) := \int_{\Fo}^\infty p_0(\F)\,d\F
= \exp(-\Fo) \sum^{n/2-1}_{k=0}\frac{\Fo^k}{k!}.
\ee
The probability of detection $P_D$ is the probability that $\F$ exceeds the 
threshold $\Fo$ when the signal-to-noise ratio is equal to $d$:
\be
\label{PD}
P_D(d,\Fo) := \int^{\infty}_{\Fo} p_1(d,\F)\,d\F.
\ee
The integral in the above formula cannot be evaluated in terms of known special 
functions. We see that when the noise in the detector is Gaussian and the phase
parameters are known the probability of detection of the signal depends on a
single quantity: the optimal signal-to-noise ratio $d$.

Our signal detection problem is posed as the statistical hypothesis testing
problem.  The {\em null hypothesis} is that the signal is absent from the data
and the {\em alternative hypothesis} is that the signal is present.  The {\em
test statistics} is the functional $\F$.  We choose a certain {\em significance
level} $\alpha$ which in the theory of signal detection is the false alarm
probability defined above.  We then calculate the test statistics $\F$ and
compare it with the threshold $\Fo$ calculated from equation $\alpha=P_F(\Fo)$.
If $\F$ exceeds the threshold $\Fo$ we say that we reject the null hypothesis at
the significance level $\alpha$.  The quantity $1-\alpha$ is called the {\em
confidence level}.  Clearly because of the statistical nature of the problem the
null hypothesis can be rejected even if the signal is present.  In the theory of
hypothesis testing we call the false alarm probability the {\em error of type I}
and the $1-P_D$ which is the probability of false dismissal of the signal we
call the {\em error of type II}.  When the signal is known by Neyman-Pearson
lemma the likelihood ratio test is the most powerful test i.e.\ it maximizes
the probability of detection $P_D$ which in the theory of hypothesis testing is
called the {\em power of the test}.

\subsection{False alarm probability}

Our next step is to study the statistical properties of the functional $\F$ when
the parameters of the phase of the signal are unknown.  We shall first consider
the case when the signal is absent in the data stream.  Let $\bs{\xi}$ be the
vector consisting of all phase parameters.  Then the statistics $\F(\bs{\xi})$
given by Eq.\ (\ref{Fred}) is a certain generalized multiparameter random
process called the {\em random field}.  If the vector $\bs{\xi}$ is
one-dimensional the random field is simply a random process.  A comprehensive
study of the properties of the random fields can be found in the monograph
\cite{A81}.  For random fields we can define the mean $m$ and the autocovariance
function $C$ just in the same way as we define such functions for random
processes:
\bea
\label{meandef}
m(\bs{\xi}) &:=& \text{E}\left\{\F(\bs{\xi})\right\},\\[2ex]
\label{covdef}
C(\bs{\xi},\bs{\xi}') &:=& 
\text{E}\left\{[\F(\bs{\xi})-m(\bs{\xi})][\F(\bs{\xi}')-m(\bs{\xi}')]\right\}.
\eea
We say that the random field $\F$ is {\em homogeneous} if its mean $m$ is
constant and the autocovariance function $C$ depends only on the difference
$\bs{\xi}-\bs{\xi}'$.  The homogeneous random fields defined above are
also called {\em second order} or {\em wide-sense homogeneous} fields.

In a statistical signal search we need to calculate the false alarm probability
i.e.\ the probability that our statistics $\F$ crosses a given threshold if the
signal is absent in the data.  In Paper I for the case of a homogeneous field
$\F$ we proposed the following approach.  We divide the space of the phase
parameters $\bs{\xi}$ into {\em elementary cells} which size is determined by
the volume of the {\em characteristic correlation hypersurface} of the random 
field $\F$.
The correlation hypersurface is defined by the requirement that 
the correlation $C$ equals half of the maximum value of $C$.  Assuming that $C$
attains its maximum value when $\bs{\xi}-\bs{\xi}'=0$ the equation of the
the characteristic correlation hypersurface reads
\be
\label{gcov1}
C(\bs{\tau}) = \frac{1}{2}C(0),
\ee
where we have introduced $\bs{\tau}:=\bs{\xi}-\bs{\xi}'$. Let us expand the left 
hand side of Eq.\ (\ref{gcov1}) around $\bs{\tau}=0$ up to terms of second order 
in $\bs{\tau}$. We arrive at the equation
\be
\label{gcov2}
\sum_{i,j=1}^M G_{ij} \tau_i\tau_j = 1,
\ee
where $M$ is the dimension of the parameter space and the matrix $G$ is defined 
as follows
\be
\label{gmatrix}
G_{ij} := -\frac{1}{C(0)}
\frac{\partial^2{C(\bs{\tau})}}{\partial{\tau_i}\partial{\tau_j}}
\Bigg\vert_{\bs{\tau}=0}.
\ee
The above equation defines an $M$-dimensional hyperellipsoid which we
take as an approximation to the characteristic correlation hypersurface
of our random field and we call the {\em correlation hyperellipsoid}.
The $M$-dimensional Euclidean volume $V_{\text{cell}}$ of the hyperellipsoid 
defined by Eq.\ (\ref{gcov2}) equals
\be
\label{vc}
V_{\text{cell}} = \frac{\pi^{M/2}}{\Gamma(M/2+1)\sqrt{\det G}},
\ee
where $\Gamma$ denotes the Gamma function.

We estimate the number $N_c$ of elementary cells by dividing the total Euclidean 
volume $V_{\text{total}}$ of the parameter space by the volume $V_{\text{cell}}$ 
of the correlation hyperellipsoid, i.e.\ we have
\be
\label{NT}
N_c = \frac{V_{\text{total}}}{V_{\text{cell}}}.
\ee

We approximate the probability distribution of $\F(\bs{\xi})$ in each cell by
probability $p_0(\F)$ when the parameters are known [in our case by probability
given by Eq.\ (\ref{p0})].  The values of the statistics $\F$ in each cell can
be considered as independent random variables.  The probability that $\F$ does
not exceed the threshold $\Fo$ in a given cell is $1-P_F(\Fo)$, where $P_F(\Fo)$
is given by Eq.\ (\ref{PF}).  Consequently the probability that $\F$ does not
exceed the threshold $\Fo$ in {\em all} the $N_c$ cells is $[1-P_F(\Fo)]^{N_c}$.
The probability $P^T_F$ that $\F$ exceeds $\Fo$ in {\em one or more} cell is
thus given by
\be
\label{FP}
P^T_F(\Fo) = 1 - [1 - P_F(\Fo)]^{N_c}.
\ee
This is the false alarm probability when the phase parameters are unknown. 
The expected number of false alarms $N_F$ is given by
\be
\label{NF}
N_F = N_c P_F(\Fo).
\ee
By means of Eqs.\ (\ref{PF}) and (\ref{NT}), Eq.\ (\ref{NF}) can be written as
\be
\label{NFi}
N_F = \frac{V_{\text{total}}}{V_{\text{cell}}} \exp(-\Fo) 
\sum^{n/2-1}_{k=0}\frac{\Fo^k}{k!}.
\ee

Using Eq.\ (\ref{NF}) we can express the false alarm probability $P^T_F$ from 
Eq.\ (\ref{FP}) in terms of the expected number of false alarms. Using
$\lim_{n\rightarrow\infty}(1+\frac{x}{n})^n=\exp(x)$ we have that for large 
number of cells
\be
\label{FPap}
P^T_F(\Fo) \approx 1 - \exp(-N_F).
\ee
When the expected number of false alarms is small (much less than 1) we have
$P^T_F\approx N_F$.

Another approach to calculate the false alarm probability can be found in the
monograph \cite{H68}.  Namely one can use the theory of level crossing by random
processes.  A classic exposition of this theory for the case of a random
process, i.e.\ for a one-dimensional random field, can be found in Ref.\
\cite{CL67}.  The case of $M$-dimensional random fields is treated in \cite{A81}
and important recent contributions are contained in Ref.\ \cite{W94}.  For a
random process $n(t)$ it is clear how to define an {\em upcrossing} of the level
$u$.  We say that $n$ has an upcrossing of $u$ at $t_o$ if there exists
$\epsilon>0$ such that $n(t)\leq u$ in the interval $(t_o-\epsilon,t_o)$, and
$n(t)\geq u$ in $(t_o,t_o+\epsilon)$.  Then under suitable regularity conditions
of the random process involving differentiability of the process and the
existence of its appropriate moments one can calculate the mean number of
upcrossings per unit parameter interval (in the one-dimensional case the
parameter is usally the time $t$ and $n(t)$ is a time series).

For the case of an $M$-dimensional random field the situation is more
complicated.  We need to count somehow the number of times a random field
crosses a fixed hypersurface.  Let ${\F(\bs{\xi})}$ be $M$-dimensional
homogeneous real-valued random field where parameters
$\bs{\xi}=(\xi_1,\dots,\xi_M)$ belong to $M$-dimensional Euclidean space
${\mathbb R}^M$ and let $\bs{C}$ be a compact subset of ${\mathbb R}^M$.  We
define the {\em excursion set} of ${\F(\bs{\xi})}$ inside $\bs{C}$ above the
level $\Fo$ as
\be
A_{\F}(\Fo,\bs{C})
:= \left\{ \bs{\xi}\in\bs{C}: \F(\bs{\xi})\geq\Fo \right\}.
\ee
It was found \cite{A81} that when the excursion set does not intersect the
boundary of the set $\bs{C}$ then a suitable analogue of the mean number
of level crossings is the expectation value of the Euler characteristic $\chi$
of the set $A_{\mathcal F}$.  For simplicity we shall denote $\chi[A_{\mathcal
F}({\mathcal F}_o,\bs{C})]$ by $\chi_{{\mathcal F}_o}$.  It turns out
that using the Morse theory the expectation value of the Euler characteristic of
$A_{\mathcal F}$ can be given in terms of certain multidimensional integrals
(see Ref.\ \cite{A81}, Theorem 5.2.1). Closed form formulae were obtained
for homogeneous $M$-dimensional Gaussian fields and 2-dimensional $\chi^2$
fields (see \cite{A81}, Theorems 5.3.1 and 7.1.2).  Recently Worsley \cite{W94}
obtained explicit formulae for $M$-dimensional homogeneous $\chi^2$ field.  We
quote here the most general results and give a few special cases.

We say that $U(\bs{\xi})$, $\bs{\xi}\in{\mathbb R}^M$, is a
$\chi^2$ field if $U(\bs{\xi})=\sum^n_{l=1}X_l(\bs{\xi})^2$,
where $X_1(\bs{\xi}),\ldots,X_n(\bs{\xi})$ are independent,
identically distributed, homogeneous, real-valued Gaussian random fields with
zero mean and unit variance.  We say that $U(\bs{\xi})$ is a {\em
generalized} $\chi^2$ field if the Gaussian fields $X_l(\bs{\xi})$ are
not necessarily independent.

Let $2\F(\bs{\xi})$ be a $\chi^2$ field and let $X_l(\bs{\xi})$, $l=1,\dots,n$, 
be the component Gaussian fields then under suitable regularity conditions 
(differentiability of the random fields and the existence of appropriate moments 
of their distributions)
\be
\label{Wchi}
\text{E}[\chi_{\Fo}] = \frac{V\sqrt{\det\Lambda}}{\pi^{M/2}\Gamma(n/2)}
\Fo^{(n-M)/2} \exp(-\Fo) W_{M,n}(\Fo).
\ee
In Eq.\ (\ref{Wchi}) $V$ is the volume of the set $\bs{C}$ and matrix $\Lambda$ 
is defined by
\be
\Lambda_{ij} := -\frac{\partial^2{C(\bs{\xi})}}{\partial{\xi_i}\partial{\xi_j}}
\Bigg\vert_{\bs{\xi}=0},
\ee 
where $C$ is the correlation function of each Gaussian field $X_l(\bs{\xi})$.
$W_{M,n}(\Fo)$ is a polynomial of degree $M-1$ in $\Fo$ given by
\be
W_{M,n}(\Fo) = \frac{(M-1)!}{(-2)^{M-1}} 
\sum^{[(M-1)/2]}_{j=0}\sum^{M-1-2j}_{k=0} {n-1\choose M-1-2j-k}
2^k\frac{(-\Fo)^{j+k}}{j!k!},
\ee
where division by factorial of a negative integer is treated as multiplication
by zero and $[N]$ denotes the greatest integer $\leq N$. We have the following 
special cases:
\be
\begin{array}{rcl}
W_{1,n} &=& 1,\\[2ex]
W_{2,n} &=& \Fo - \frac{1}{2}(n-1),\\[2ex]
W_{3,n} &=& \Fo^2 - (n-\frac{1}{2})\Fo + \frac{1}{4}(n-1)(n-2),\\[2ex]
W_{4,n} &=& \Fo^3 + \frac{3}{4}(n-1)^2\Fo^2 - \frac{3}{2}n\Fo
- \frac{1}{8}(n-1)(n-2)(n-3).
\end{array}
\ee

It has rigorously been shown that for the homogeneous Gaussian random fields the
probability distribution of the Euler characteristic of the excursion set
asymptotically approaches a Poisson distribution (see Ref.\ \cite{A81}, Theorem
6.9.3).  It has been argued that the same holds for $\chi^2$ fields.  It has
also been shown for $M$-dimensional homogeneous $\chi^2$ fields that
asymptotically the level surfaces of the local maxima of the field are
$M$-dimesional ellipsoids.  Thus for large threshold the excursion set consists
of disjoint and simply connected (i.e.\ without holes) sets.  Remembering that
we assume that the excursion set does not intersect the boundary of the
parameter set the Euler characteristic of the excursion set is simply the number
of connected components of the excursion set.  
Thus we can expect that for a $\chi^2$ random field
the expected number of level crossings by the field i.e.\ in the language of
signal detection theory the expected number of false alarms has a Poisson
distribution.  Thus the probability that $\F_{\text{max}}$ does not cross a
threshold $\Fo$ is given by $\exp(-\text{E}[\chi_{\Fo}])$ and the probability 
that there is at least one level crossing (i.e.\ for our signal detection 
problem the false alarm probability $P_F^T$) is given by
\be
\label{FPf}
P_F^T(\Fo) = P(\F_{\text{max}} \geq \Fo)
\approx 1 - \exp(-\text{E}[\chi_{\Fo}]).
\ee

From Eqs.\ (\ref{FPap}) and (\ref{FPf}) we see that to compare the two
approaches presented above it is enough to compare the expected number of false
alarms $N_F$ with $\rm{E}[\chi_{\Fo}]$.  It is not difficult to see that for
$\chi^2$ fields $G=2\Lambda$.  Thus asymptotically (i.e.\ for large thresholds
$\Fo$) using Eqs.\ (\ref{vc}), (\ref{NFi}), and (\ref{Wchi}) we get
\be
\label{NFca}
\frac{N_F}{\rm{E}[\chi_{\Fo}]} \rightarrow 2^{M/2}\Gamma(M/2+1)\Fo^{-M/2}  
\quad \text{as} \quad \Fo\rightarrow\infty,
\ee
where we have used that $V$ from Eq.\ (\ref{Wchi}) coincides with 
$V_{\text{total}}$ from Eq.\ (\ref{NFi}).

Worsley (\cite{W94}, Corollary 3.6) also gives asymptotic 
(i.e.\ for threshold $\Fo$ tending to infinity) formula for the probability 
$P({\mathcal F}_{\text{max}} \geq {\mathcal F}_o)$ that the global maximum 
${\mathcal F}_{\text{max}}$ of  ${\mathcal F}$ crosses a threshold $\Fo$:
\be\
\label{FPa}
P({\mathcal F}_{\text{max}} \geq \Fo) \rightarrow
\frac{V \sqrt{\det\Lambda}}{\pi^{M/2}\Gamma(n/2)} \Fo^{(n+M)/2-1} \exp(-\Fo)
\quad\mbox{as}\quad \Fo \rightarrow \infty.
\ee

In the signal detection theory the above probability is simply the false
alarm probability and it should be compared with the probability given
by Eq.\ (\ref{FP}). It is not difficult to verify that asymptotically
the Eqs.\ (\ref{FP}) and (\ref{FPa}) are equivalent if we replace
expected number of false alarms $N_F$ by $\rm{E}[\chi_{\Fo}]$.
This reinforces the argument leading to Eq. \ (\ref{NFca}).

The above formulae were obtained for continuous stationary random fields.  In
practice we shall always deal with a discrete time series of finite duration.
Therefore to see how useful the above formulae are in the real data analysis of
discrete time series it is appropriate to perform the Monte Carlo simulations.

We have first tested Eqs.\ (\ref{p0}) and (\ref{PF}) for the probability density
of the false alarm and the false alarm probability in the simplest
case of $n=2$ and the known signal.  Using
a computer pseudo-random generator we
have obtained a signal $x$ consisting of $N=2^8$ independent random values drawn
from a Gaussian distribution with zero mean and unit variance.  The optimal
statistics $\F$ in this case is
\be
\label{Fk}
\F_k = \frac{|X_k|^2}{N}, \quad k=1,\ldots,N/2+1,
\ee
where $|X_k|$ is the modulus of the $k$th component of the discrete Fourier
transform of $x$.  In other words the optimal statistics is the periodogram
sampled at Fourier bins.  When $x$ consists of independent identically
distributed Gaussian random variables we know \cite{PW} that for
$k=2,\ldots,N/2$ the statistics $2\F_k$ has a $\chi^2$ distribution with 2
degrees of freedom whereas for $k=1$ (zero frequency bin) and $k= N/2+1$
(maximum, Nyquist frequency bin) $\F_k$ has a $\chi^2$ distribution with 1
degree of freedom.  In our Monte Carlo simulation we have generated the signal
$10^6$ times and we have made histograms of 127 bins of the statistics $\F_k$.
In the upper plot of Figure 1 we have shown the (appropriately normalized)
histogram for all the Fourier bins for $k=2,\ldots,N/2$ and in the middle plot
of Figure 1 we have presented the cumulative distribution.  Thus the probability
density generated in the upper plot is to be compared with Eq.\ (\ref{p0}) for
$n=2$ whereas the distribution obtained in the middle one is to be compared with
Eq.\ (\ref{PF}) for $n=2$.  Both theoretical distributions are exponential and
they are given by solid curves in the two plots.  The last bin in the upper plot
of Figure 1 deviates substantially from the exponential curve.  This is because
in this bin all the events above the maximum value of the histogram range are
accumulated.  We get $11$ events altogether in the last bin.  The expected value
of the events calculated from Eq.\ (\ref{NF}) is 8.25 (where we have put
$N_c=127$).  In the two lower plots of Figure 1 we have presented the cumulative
distributions for the first and the last bin.  The theoretical cumulative
distribution that follows from the $\chi^2$ distribution with 1 degree of
freedom is given by $1-\text{erf}(\sqrt{\Fo/2})$ (solid curve in the plots).  We
see that simulated and theoretical distributions agree very well.

We have next tested the formulae for the false alarm probabilities given by
Eqs.\ (\ref{FP}) and (\ref{FPf}) against the Monte Carlo simulations.  We have
considered again the case of $n=2$ and we have simulated the optimal statistics
for the case of a monochromatic signal ($M=1$) and the case of a signal with one
spindown included ($M=2$).  We have generated the random sequence of length
$N=2^8$ as in the first simulation described above.  We have however introduced
an extra parameter $P$---the zero padding.  Namely we add zeros to the random
sequence so that its total length is $(1+P)N$.  When we take Fourier transform
of the zero-padded signal we get additional points in the Fourier domain between
the Fourier bins.  Zero padding essentially amounts to interpolating the
periodogram between the Fourier bins.  Thus the larger the $P$ the closer the
discretelly sampled periodogram to a continuous function.  To generate the
statistics $\F$ for the signal with one spindown we have multiplied the
generated random sequence $x(k)$ by $T(k;l)=\exp[-2{\pi}il\sigma_1(k-1)^2]$,
where $k=1,\ldots,N$, $l=10,\ldots,29$, and $\sigma_1=(3/\pi)\sqrt{5/2}$.  The
function $T(k;l)$ is called a {\em filter} or a {\em template}.  The
multiplication operation is the {\em matched filtering} which in our case is
also called {\em dechirping}.  The quantity $\sigma_1$ is the accuracy of
estimation of the 1st spindown parameter for the optimal signal-to-noise ratio
$d = 1$ divided by $\sqrt{2}$ and it is the maximum extent of the ellipse
defined by Eq.\ (\ref{gcov2}) measured from the origin along the spindown axis
in the Cartesian (frequency, spindown)-plane.  The parameter $\sigma_1$ defines
the spacing of the templates that we choose in our simulations.  The zero
padding is always done after the dechirping operation.  Our optimal statistics
is the modulus of the discrete Fourier transform of the dechirped and zero
padded data divided by the number of points in the original data ($2^8$ in our
case).  We have made $10^5$ trials.

The results of the simulation are presented in Figure 2.  The three upper plots
are the results for the monochromatic signal search and the three lower ones are
the results for the one spindown signal search.  The false alarm curves are
given in the plots on the left.  We see that the false alarm probability
exhibits a threshold phenomenon, it drops very sharply within a narrow range of
the detection threshold.  We see from the plots on the left of Figure 2 that for
$P=0$ (no zero padding) the results of the simulation agree well with Eq.\
(\ref{FP}) (solid line) whereas for $P=3$ there is a reasonable agreement with
Eq.\ (\ref{FPf}) (dashed line).

In the plots on the right of Figure 2 we have divided the probability of the
false alarm obtained from the simulations by the probabilities obtained from the
theoretical formulae.  The upper plots give comparison with Eq.\ (\ref{FP})
based on dividing the parameter space into cells whereas the lower plots give
comparison with Eq.\ (\ref{FPf}) based on the expectation value of the Euler
characteristic of the excursion set.  We see that for the monochromatic signal
for thresholds up to 10 Eq.\ (\ref{FP}) gives a reasonable agreement for $P=0$
whereas Eq.\ (\ref{FPf}) gives a good agreement for $P=3$.  For the frequency
modulated signal for $P=0$ Eq.\ (\ref{FP}) underestimates the false alarm
probability whereas Eq.\ (\ref{FPf}) overestimates the false alarm probability.
For $P=3$ there is an underestimate of the false alarm probability by both
formulae.  For thresholds greater than 10 the curves become irregular what may
be attributed to a sparse number of events for such large thresholds.

\subsection{Detection probability}

When the signal is present a precise calculation of the pdf of ${\mathcal F}$ is
very difficult because the presence of the signal makes the data random process
$x(t)$ nonstationary.  As a first approximation we can estimate the probability
of detection of the signal when the parameters are unknown by the probability of
detection when the parameters of the signal are known [given by Eq.\
(\ref{PD})].  This approximation assumes that when the signal is present the
true values of the phase parameters fall within the cell where ${\mathcal F}$
has a maximum.  This approximation will be the better the higher the
signal-to-noise ratio $d$.  Parametric plot of probability of detection vs.\
probability of false alarm with optimal signal-to-noise ratio $d$ as a parameter
is called the {\em receiver operating characteristic} (ROC).

We have performed the numerical simulations to see how the ROC obtained from the
analytical formulae presented above compares with that obtained from the
discrete finite duration time series.  We have generated the noise as in the
simulations of the false alarm probability and we have added the signal.  We
have considered both the monochromatic and the linearly frequency modulated
signal.  The frequency of the signal was chosen not to coincide with one of the
Fourier frequencies.  However in the dechirping operation to detect the
frequency modulated signal we have chosen the spindown parameter in the filter
to coincide with the spindown parameter of the signal.  We have perfomed $10^4$
trials and we have examined the cumulative distributions of the two Fourier bins
between which the true value of the frequency of the signal had been chosen.
The results are presented in Figure 3.  The two upper plots are for the
monochromatic signal and the lower two are for the one spindown signal.  In the
plots on the left we compare the probability of detection calculated from Eq.\
(\ref{PD}) with the results of the simulations and in the plots on the right we
compare the theoretical and the simulated receiver operating characteristics.
For the false alarm probability we have used the formula (\ref{FP}).  In the
inserts we have zoomed the ROC for small values of the false alarm probability.
We see that the agreement between the theoretical and simulated ROC is quite
good.

\section{Number of cells for the one-component signal}

Let us return to the case of a gravitational-wave signal from a spinning 
neutron star. In Sec.\ 5 of Paper II we have shown that each component of the 
$N$-component signal can be approximated by the following one-component signal: 
\be
\label{cal1}
h(t;h_o,\Phi_0,\bs{\xi}) = h_o \sin\left[\Phi(t;\bs{\xi})+\Phi_0\right],
\ee
where the phase $\Phi$ of the signal is given by
\bea
\label{cal2}
\Phi(t;\bs{\xi}) &=&
\sum_{k=0}^{s} \omega_k \left(\frac{t}{T_o}\right)^{k+1}
+\frac{2\pi}{c}\left\{ \alpha_1
\left[R_{ES}\sin\left(\phi_o+\Omega_o t\right)
+R_{E}\cos\lambda\cos\ve\sin\left(\phi_r+\Omega_r t\right)\right]
\right.\nonumber\\&&\left.
+\alpha_2 \left[R_{ES}\cos\left(\phi_o+\Omega_o t\right)
+R_{E}\cos\lambda\cos\left(\phi_r+\Omega_r t\right)\right] \right\},
\eea
where $T_o$ denotes the observation time, $R_{ES}=$ 1 AU is the mean distance 
from the Earth's center to the SSB,
$R_E$ is the mean radius of the Earth, $\Omega_o$ is the mean orbital angular
velocity of the Earth, and $\phi_o$ is a deterministic phase which defines the
position of the Earth in its orbital motion at $t=0$, $\ve$ is the angle between
ecliptic and the Earth's equator.  The vector $\bs{\xi}$ collects all the phase
parameters, it equals $\bs{\xi}=(\alpha_1,\alpha_2,\omega_0,\ldots,\omega_s)$,
so the phase $\Phi$ depends on $s+3$ parameters.  The dimensionless parameters
$\omega_k$ are related to the spindown coefficients ${\fko k}$ introduced in
Eq.\ (\ref{phase1}) as follows:
\be
\omega_k := \frac{2\pi}{(k+1)!}{\fko k}T_o^{k+1},\quad k=0,\ldots,s.
\ee
The parameters $\alpha_1$ and $\alpha_2$ are defined by
\be
\label{a1a2}
\alpha_1 := f_o 
\left(\cos\ve\sin\alpha\cos\delta+\sin\ve\sin\delta\right),\quad
\alpha_2 := f_o \cos\alpha\cos\delta.
\ee
In Appendix D we show that the parameters $\alpha_1$, $\alpha_2$ can be used 
instead of the parameters $\alpha$, $\delta$ to label the templates needed to 
do the matched filtering.

The signal defined by Eqs.\ (\ref{cal1}) and (\ref{cal2}) has two important
properties:  it has a constant amplitude and its phase is a linear function of
the parameters $\bs{\xi}$.  In Paper II we have shown that for this signal's
model the rms errors calculated from the inverse of the Fisher information
matrix reproduce well the rms errors of the full model presented in Sec.\ 2.
We will use here the simpified signal (\ref{cal1})--(\ref{cal2}) to estimate 
the number of elementary cells in the parameter space.

For the signal given by Eqs.\ (\ref{cal1}) and (\ref{cal2}) the statistics 
$\F$ of Eq.\ (\ref{Fred}) can be written as
\be
\label{Sstat}
{\mathcal F}(\bs{\xi}) \approx \frac{1}{2}
\left\{ [x_c(\bs{\xi})]^2 + [x_s(\bs{\xi})]^2 \right\},
\ee
where
\be
\label{xcxs}
x_c(\bs{\xi}) := 2\sqrt{\frac{T_o}{S_h(f_o)}}
\tav{x \cos\Phi(t;\bs{\xi})},\quad
x_s(\bs{\xi}) := 2\sqrt{\frac{T_o}{S_h(f_o)}}
\tav{x \sin\Phi(t;\bs{\xi})}.
\ee

We calculate the autocovariance function $C$ [defined by Eq.\ (\ref{covdef})] 
of the random field (\ref{Sstat}) when the data $x$ consists only of the noise 
$n$. We recall that $n$ is a zero mean stationary Gaussian random process. 
Consequently we have the following useful formulae \cite{for}
\bea
\label{for1a}
\rm{E}\left\{(n|h_1)(n|h_2)\right\} &=& (h_1|h_2),\\[2ex]
\label{for1b}
\rm{E}\left\{(n|h_1)(n|h_2)(n|h_3)(n|h_4)\right\}
&=& (h_1|h_2)(h_3|h_4) + (h_1|h_3)(h_2|h_4) + (h_1|h_4)(h_2|h_3),
\eea
where $h_1$, $h_2$, $h_3$, and $h_4$ are deterministic functions. Let us also 
observe that
\be
\label{for2}
\tav{\sin^2\Phi(t;\bs{\xi})} \approx 
\tav{\cos^2\Phi(t;\bs{\xi})} \approx \frac{1}{2}.
\ee
Making use of Eqs.\ (\ref{for1a}), (\ref{for1b}), and  (\ref{for2}) one 
finds that
\bea
\label{auto1a}
\rm{E}_0\left\{{\mathcal F}(\bs{\xi})\right\} &\approx& 1,\\
\label{auto1b}
\rm{E}_0\left\{{\mathcal F}(\bs{\xi}){\mathcal F}(\bs{\xi}')\right\}
&\approx& 1 + 2 \left[
\tav{\cos\Phi(t;\bs{\xi})\cos\Phi(t;\bs{\xi}')}^2 +
\tav{\cos\Phi(t;\bs{\xi})\sin\Phi(t;\bs{\xi}')}^2
\right.\nonumber\\&&\left. +
\tav{\sin\Phi(t;\bs{\xi})\cos\Phi(t;\bs{\xi}')}^2 +
\tav{\sin\Phi(t;\bs{\xi})\sin\Phi(t;\bs{\xi}')}^2
\right],
\eea
where subscript $0$ means that there is no signal in the data. For our 
narrowband signal to a good approximation we have
\bea
\label{auto2a}
\tav{\cos\Phi(t;\bs{\xi})\cos\Phi(t;\bs{\xi}')} &\approx& 
\frac{1}{2} \tav{\cos[\Phi(t;\bs{\xi})-\Phi(t;\bs{\xi}')]},\\
\tav{\cos\Phi(t;\bs{\xi})\sin\Phi(t;\bs{\xi}')} &\approx& 
-\frac{1}{2} \tav{\sin[\Phi(t;\bs{\xi})-\Phi(t;\bs{\xi}')]},\\
\tav{\sin\Phi(t;\bs{\xi})\cos\Phi(t;\bs{\xi}')} &\approx& 
\frac{1}{2} \tav{\sin[\Phi(t;\bs{\xi})-\Phi(t;\bs{\xi}')]},\\
\label{auto2d}
\tav{\sin\Phi(t;\bs{\xi})\sin\Phi(t;\bs{\xi}')} &\approx& 
\frac{1}{2} \tav{\cos[\Phi(t;\bs{\xi})-\Phi(t;\bs{\xi}')]}.
\eea
Collecting Eqs.\ (\ref{auto1a})--(\ref{auto2d}) together one gets
\bea
\label{auto3}
C(\bs{\xi},\bs{\xi}') &=& \rm{E}_0\left\{\F(\bs{\xi})\F(\bs{\xi}')\right\}
- \rm{E}_0\left\{\F(\bs{\xi})\right\}E_0\left\{\F(\bs{\xi}')\right\}
\nonumber\\ &\approx& 
\tav{\cos[\Phi(t;\bs{\xi})-\Phi(t;\bs{\xi}')]}^2 +
\tav{\sin[\Phi(t;\bs{\xi})-\Phi(t;\bs{\xi}')]}^2.
\eea

The phase $\Phi$ given by Eq.\ (\ref{cal2}) is a linear function of the
parameters $\bs{\xi}$ hence the autocovariance function $C$ from Eq.\ 
(\ref{auto3}) depends only on the difference $\bs{\tau}=\bs{\xi}-\bs{\xi}'$ and 
it can be written as
\be
\label{auto5}
C(\bs{\tau}) \approx \tav{\cos[\Phi(t;\bs{\tau})]}^2 +
\tav{\sin[\Phi(t;\bs{\tau})]}^2.
\ee

To calculate the volume of the elementary cell by means of Eq.\ (\ref{vc}) we 
need to compute the matrix $G$ defined in Eq.\ (\ref{gmatrix}). Substituting 
(\ref{auto5}) into (\ref{gmatrix}) we obtain
\be
G = 2\widetilde{\Gamma},
\ee
where the matrix $\widetilde{\Gamma}$ has the components
\be
\label{redgamma}
\widetilde{\Gamma}_{ij} :=
\tav{\frac{\pa\Phi}{\pa\tau_i}\frac{\pa\Phi}{\pa\tau_j}}
- \tav{\frac{\pa\Phi}{\pa\tau_i}} \tav{\frac{\pa\Phi}{\pa\tau_j}}.
\ee
The matrix $\widetilde{\Gamma}$ is the reduced Fisher information matrix for 
our signal where the initial phase parameter $\Phi_0$ [cf.\ Eq.\ (\ref{cal1})] 
has been reduced, see Appendix B.

As the mean (\ref{auto1a}) of the random field $\F$ is constant and its
autocovariance (\ref{auto3}) depends only on the difference $\bs{\xi}-\bs{\xi}'$
the random field $\F$ is a homogeneous random field.  Let us observe that for
the fixed values of the parameters $\bs{\xi}$ the random variables $x_c$ and
$x_s$ are zero mean and unit variance Gaussian random variables.  However the
correlation between the Gaussian fields $x_c$ and $x_s$ does not vanish:
\be
\rm{E}\left\{x_c(\bs{\xi})x_s(\bs{\xi}')\right\} \approx
\tav{\sin[\Phi(t;\bs{\xi})-\Phi(t;\bs{\xi}')]},
\ee
and thus the Gaussian random fields $x_c$ and $x_s$ are not independent.
Therefore $\F$ is not a $\chi^2$ random field but only a generalized $\chi^2$
random field.  Our formula for the number of cells [Eq.\ (\ref{NT})] and the
formula for the false alarm probability [Eq.\ (\ref{FP})] apply to any
homogeneous random fields however formula (\ref{Wchi}) applies only to $\chi^2$
fields.  Nevertheless by examining the proof of formula (\ref{Wchi})
\cite{A81,W94} we find that it is very likely that the formula holds for
generalized $\chi^2$ random fields as well if we replace the determinant of the
matrix $\Lambda$ by the determinant of the reduced Fisher matrix
$\widetilde{\Gamma}$.

The total volume of the parameter space depends on the range of the individual 
parameters. Following Ref.\ \cite{BCCS98} we assume that 
\bea
\label{range1a}
&2\pi T_o f_{\text{min}} \le \omega_0 \le 2\pi T_o f_{\text{max}},&\\
\label{range1b}
&-\beta_k \omega_0 \le \omega_k \le \beta_k \omega_0,\quad \text{where}\quad
\beta_k := {\dst \frac{1}{k+1}\left(\frac{T_o}{\tau_{\text{min}}}\right)^k},
\quad k = 1,\ldots,s,& 
\eea
where $f_{\text{min}}$ and $f_{\text{max}}$ are respectively the minimum and 
the maximum value of the gravitational-wave frequency, $\tau_{\text{min}}$ is 
the minimum spindown age of the neutron star. The parameters $\alpha_1$ and 
$\alpha_2$ defined in Eq.\ (\ref{a1a2}) fill, for the fixed value of the 
frequency parameter $\omega_0$, 2-dimensional ball concentrated around the 
origin 
in the $(\alpha_1,\alpha_2)$-plane, with radius equal to $\omega_0/(2\pi T_o)$:
\be
\label{range2}
(\alpha_1,\alpha_2) \in B_2(0,\omega_0/(2\pi T_o)).
\ee

Taking Eqs.\ (\ref{range1a})--(\ref{range2}) into account the total volume 
$\vta$ of the parameter space for all-sky searches with $s$ spindowns included 
can be calculated as follows
\bea
\vta &=&
\int\limits_{2\pi T_o f_{\text{min}}}^{2\pi T_o f_{\text{max}}} d\,\omega_0
{\int\!\!\int}_{B_2(0,\omega_0/(2\pi T_o))} d\,\alpha_1\,d\,\alpha_2
\int\limits_{-\beta_1\omega_0}^{\beta_1\omega_0} d\,\omega_1 \ldots
\int\limits_{-\beta_s\omega_0}^{\beta_s\omega_0} d\,\omega_s
\nonumber\\&=&
\frac{2^{2s+1} \pi^{s+2}}{(s+3)(s+1)!}
T_o^{s+1} \left(\frac{T_o}{\tau_{\text{min}}}\right)^{s(s+1)/2}
\left(f_{\text{max}}^{s+3}-f_{\text{min}}^{s+3}\right).
\eea

The volume $\vca$ of one cell we calculate from Eq.\ (\ref{vc}) for $M=s+3$ and 
$G=2\ga$, where $\ga$ is the reduced Fisher matrix (\ref{redgamma}) for the 
phase $\Phi$ given by Eq.\ (\ref{cal2}) with $s$ spindowns included:
\be
\label{vca}
\vca = \frac{(\pi/2)^{(s+3)/2}}{\Gamma((s+5)/2)\sqrt{\det\ga}}. 
\ee
In Appendix B we have given formulae needed to calculate matrices $\ga$ for 
$s=0,\ldots,4$ analytically. In Figure \ref{fignca0} we have plotted the volume 
$\vca$ of one cell as a function of the observation time for signals with 
various numbers $s$ of spindown parameters included.

The number $\nca$ of cells for all-sky searches is given by 
\be
\label{nca}
\nca = \frac{\vta}{\vca} = \frac{2^{(3s+1)/2}\pi^{s/2+1}}{\Gamma(s/2+1)}
\sqrt{\det\ga} \left(\frac{T_o}{\tmin}\right)^{s(s+1)/2} T_o^{s+1}
\left(\fmax^{s+3}-\fmin^{s+3}\right).
\ee
In Figure \ref{fignca1} we have plotted the number $\nca$ of cells as a function
of the observation time $T_o$ for various models of the signal depending on the
minimum spindown age $\tmin$ and the maximum gravitational-wave frequency
$\fmax$, and for various numbers $s$ of spindowns included (the minimum
gravitational-wave frequency $\fmin=0$).  We see that for a given $\tmin$ and
$\fmax$ curves corresponding to different numbers $s$ intersect.  This effect
was observed and explained by Brady {\it et al}.\ \cite{BCCS98}.  To obtain the
number of cells for a given observation time $T_o$ we always take the number of
cells given by the uppermost curve.  We have calculated the observation times
$\ctall{k}$ for which the numbers of cells with $k$ and $k+1$ spindowns included
coincide:
\be
\ncadef{k+1}(T_o=\ctall{k}) = \ncadef{k}(T_o=\ctall{k}),\quad k=0,\ldots,s-1.
\ee
In Table \ref{tab:ctcells} we have given the values of $\ctall{k}$ for all the 
signal models considered.

\begin{table}[!ht]
\begin{center}
\begin{tabular}{|c|c|c|c|c|c|c|c|c|c|}\hline
\raisebox{-1.5ex}[0cm][0cm]{$\tmin$ (years)} &
\raisebox{-1.5ex}[0cm][0cm]{$\fmax$ (Hz)} & 
\multicolumn{4}{c|}{$\ctall{k}$ (days)} &
\multicolumn{4}{c|}{$\ctdir{k}$ (days)} \\ \cline{3-10}
&& $k=0$ & $k=1$ & $k=2$ & $k=3$ & $k=0$ & $k=1$ & $k=2$ & $k=3$ \\ \hline\hline
40 & $10^3$ &
0.21 & 3.11 & 116   & 311 &
0.03 & 3.53 &  40.5 & 175 \\ \hline
40 & 200 &
0.31 & 5.19 & 158   & 389 &
0.06 & 6.04 &  60.5 & 242 \\ \hline
$10^3$ & $10^3$ &
0.46 & 114   & 575 & 2210 &
0.14 &  30.2 & 452 & 2300 \\ \hline
$10^3$ & 200 &
0.69 & 157   & 725 & 3040 &
0.32 &  51.7 & 676 & 3180 \\ \hline
\end{tabular}
\end{center}
\caption{\label{tab:ctcells}The observation times for which the numbers of cells 
with $k$ and $k+1$ spindowns included coincide for various models of the signal 
depending on the minimum spindown age $\tmin$ and the maximum gravitational-wave 
frequency $\fmax$. The minimum gravitational-wave frequency $\fmin=0$. In the 
case of all-sky searches we have used the latitude $\lambda=46.45^\circ$ of the 
LIGO Hanford detector and we have put $\phi_o=0.123$ and $\phi_r=1.456$.}
\end{table}

The Fisher matrix $\ga$ depends on the phases $\phi_r$, $\phi_o$, and the
latitude $\lambda$ of the detector (see Appendix B).  We know from Paper II (see
Appendix C of Paper II) that the Fisher matrix also depends on the choice of the
instant of time at which the instantanenous frequency and spindown parameters
are defined (in the present paper this moment is chosen to coincide with the
middle of the observational interval).  We find that the determinant $\det\ga$
and consequently the number of cells does not depend on this choice.  The
dependence on the remaining parameters is studied in Figure \ref{fignca2}.  The
dependence on the phases $\phi_r$ and $\phi_o$ is quite weak.  The dependence on
$\lambda$ is quite strong however for the detectors under construction for which
$\lambda$ varies from $35.68^\circ$ (TAMA300) to $52.25^\circ$ (GEO600) the
number of cells changes by a factor of 2 for 7 days of observation time and by
around 10\% for 120 days of observation time.

In Sec.\ 5 of Paper II we have shown that for directed searches the 
constant amplitude signal given by Eqs.\ (\ref{cal1}) and (\ref{cal2}) can be 
further simplified by discarding in the phase (\ref{cal2}) terms due to the 
motion of the detector w.r.t.\ the SSB and the rms errors calculated form the 
inverse of the Fisher matrix do not change substancially. Such a signal reads
\be
\label{poly}
h(t;h_o,\Phi_0,\bs{\xi})
= h_o \sin\left[\Phi(t;\bs{\xi})+\Phi_0\right],\quad
\Phi(t;\bs{\xi}) = \sum_{k=0}^{s} \omega_k \left(\frac{t}{T_o}\right)^{k+1}.
\ee
The vector $\bs{\xi}$ has now $s+1$ components: 
$\bs{\xi}=(\omega_0,\ldots,\omega_s)$.

Using Eqs.\ (\ref{range1a}) and (\ref{range1b}) the total volume $\vtd$ of the 
parameter space with $s$ spindowns included for directed searches is calculated 
as follows
\bea
\vtd &=&
\int\limits_{2\pi T_o f_{\text{min}}}^{2\pi T_o f_{\text{max}}} d\,\omega_0
\int\limits_{-\beta_1\omega_0}^{\beta_1\omega_0} d\,\omega_1 \ldots
\int\limits_{-\beta_s\omega_0}^{\beta_s\omega_0} d\,\omega_s
\nonumber\\&=&
\frac{2^{2s+1} \pi^{s+1}}{(s+1)(s+1)!}
T_o^{s+1} \left(\frac{T_o}{\tau_{\text{min}}}\right)^{s(s+1)/2}
\left(f_{\text{max}}^{s+1}-f_{\text{min}}^{s+1}\right).
\eea

The volume $\vtd$ of one cell we calculate from Eq.\ (\ref{vc}) for $M=s+1$ and 
$G=2\gd$, where $\gd$ is the reduced Fisher matrix (\ref{redgamma}) for the 
polynomial phase (\ref{poly}) with $s$ spindowns included:
\be
\vcd = \frac{(\pi/2)^{(s+1)/2}}{\Gamma((s+3)/2)\sqrt{\det\gd}}.
\ee
The matrix $\gd$ for $s=0,\ldots,4$ can be calculated analytically by means of 
formulae given in Appendix B.

The number $\ncd$ of independent cells is given by
\be
\label{ncd}
\ncd = \frac{\vtd}{\vcd} = \frac{2^{(3s+1)/2}\pi^{s/2+1}}{(s+1)\Gamma(s/2+1)} 
\sqrt{\det\gd} \left(\frac{T_o}{\tau_{\text{min}}}\right)^{s(s+1)/2} T_o^{s+1}
\left(\fmax^{s+1}-\fmin^{s+1}\right).
\ee
In Figure \ref{figncd} we have plotted the number of cells $\ncd$ as a function
of the observation time $T_o$ for various models of the signal depending on the
minimum spindown age $\tmin$, the maximum gravitational-wave frequency $\fmax$,
and the number $s$ of spindowns included (the minimum gravitational-wave
frequency $\fmin=0$).  We see that like for all-sky searches for a given $\tmin$
and $\fmax$ curves corresponding to different numbers $s$ intersect.  We have
calculated analytically the observation times $\ctdir{k}$ for which the numbers
of cells with $k$ and $k+1$ spindowns included coincide:
\be
\ncddef{k+1}\left(T_o=\ctdir{k}\right) = \ncddef{k}\left(T_o=\ctdir{k}\right), 
\quad k=0,\ldots,s-1.
\ee
Using Eq.\ (\ref{ncd}) one obtains
\be
\ctdir{k} = \left[
\frac{(k+2)\Gamma((k+3)/2)}{2\sqrt{2\pi}(k+1)\Gamma((k+2)/2)}
\sqrt{ \frac{\det\widetilde{\Gamma}_{(k)}^{\text{dir}}}
{\det\widetilde{\Gamma}_{(k+1)}^{\text{dir}}} }
\frac{f_{\text{max}}^{k+1}-f_{\text{min}}^{k+1}}
{f_{\text{max}}^{k+2}-f_{\text{min}}^{k+2}} \tau_{\text{min}}^{k+1}
\right]^{1/(k+2)},\quad k=0,\ldots,s-1.
\ee
In Table \ref{tab:ctcells} we have given the values of $\ctdir{k}$ for all the 
signal models considered.

In Table 2 we have given the number of cells both for all-sky and directed
searches for various models of the signal depending on the minimum spindown age
$\tmin$ and the maximum gravitational-wave frequency $\fmax$, and for the
observation time $T_o$ of 7 and 120 days (the minimum gravitational-wave
frequency $\fmin=0$).  The number of cells is calculated from Eq.\ (\ref{nca})
for all-sky searches and from Eq.\ (\ref{ncd}) in the case of directed searches.
For a given observation time $T_o$ the number $s$ of spindowns one should
include in the signal's model is obtained as such number $s$ chosen out of
$s=0,\ldots,4$ for which $\nca$ (or $\ncd$) is the greatest.

We have also calculated the threshold $\Fo$ for the $1\%$ false alarm
probability (or equivalently for $99\%$ detection confidence).  By means of
Eqs.\ (\ref{PF}) and (\ref{FP}) for $n=2$ (what corresponds to a one-component
signal) the relation between the threshold $\Fo$ and the false alarm probability
$\alpha$ reads
\be
\label{thre1}
\Fo = - \ln\left[1-(1-\alpha)^{1/N_c}\right], \quad\alpha=0.01,
\ee 
where $N_c$ is the number of cells. Following the relation between the 
expectation value of the optimum statistics when the signal is present and the 
signal-to-noise ratio which is given by
\be 
\label{thre2}
\rm{E}_1\{\F\} = 1 + \frac{1}{2}d^2,
\ee  
we have calculated the "threshold" signal-to-noise ratio
\be
\label{thre3}
d_o:=\sqrt{2(\Fo-1)},
\ee
where $\Fo$ is given by Eq.\ (\ref{thre1}). The values of $d_o$ for various 
models of the signal and observation times of 7 and 120 days are given in Table 
2.  If the signal-to-noise ratio is $d_o$ then there is roughly a $50\%$ 
probability that the optimum statistic will cross the threshold $\Fo$.

\begin{table}[!ht]
\begin{center}
\begin{tabular}{|c|c|c|c|c|c|c|c|c|}\hline
\raisebox{-1.5ex}[0cm][0cm]{$T_o$ (days)} &
\raisebox{-1.5ex}[0cm][0cm]{$\tau_{\text{min}}$ (years)} &
\raisebox{-1.5ex}[0cm][0cm]{$f_{\text{max}}$ (Hz)} & 
\multicolumn{3}{c|}{all-sky} & \multicolumn{3}{c|}{directed} \\ \cline{4-9}
&&& $s$ & $\nca$ & $d_o$ & $s$ & $\ncd$ & $d_o$ \\ \hline\hline
7 & 40 & $10^3$ &
2 & $4.2\times10^{18}$ & 9.6 &
2 & $3.7\times10^{14}$ & 8.6 \\ \hline
7 & 40 & 200 &
2 & $1.3\times10^{15}$ & 8.8 &
2 & $2.9\times10^{12}$ & 8.0 \\ \hline
7 & $10^3$ & $10^3$ &
1 & $1.5\times10^{16}$ & 9.0 &
1 & $1.9\times10^{12}$ & 8.0 \\ \hline
7 & $10^3$ & 200 &
1 & $2.4\times10^{13}$ & 8.3 &
1 & $7.6\times10^{10}$ & 7.6 \\ \hline\hline
120 & 40 & $10^3$ &
3 & $3.8\times10^{29}$ & 12 &
3 & $7.2\times10^{23}$ & 11 \\ \hline
120 & 40 & 200 &
2 & $1.1\times10^{26}$ & 11 &
3 & $1.2\times10^{21}$ & 10 \\ \hline
120 & $10^3$ & $10^3$ &
2 & $2.2\times10^{25}$ & 11 &
2 & $6.0\times10^{17}$ & 9.4 \\ \hline
120 & $10^3$ & 200 &
1 & $2.7\times10^{22}$ & 11& 
2 & $4.8\times10^{15}$ & 8.9 \\ \hline
\end{tabular}
\end{center}
\caption{Number of cells for all-sky and directed searches for various models of 
the signal depending on the minimum spindown age $\tmin$ and the maximum 
gravitational-wave frequency $\fmax$, and for the observation time $T_o$ of 7 
and 120 days. The minimum gravitational-wave frequency $\fmin=0$. To calculate 
the Fisher matrix $\ga$ we have used the latitude $\lambda=46.45^\circ$ of the 
LIGO Hanford detector and we have put $\phi_o=0.123$ and $\phi_r=1.456$. For 
each case we also give the $99\%$ confidence threshold signal-to-noise ratio 
$d_o$ calculated by virtue of Eq.\ (\ref{thre3}).}
\end{table}

\section{Number of filters for the one-component signal}

To calculate the number of FFTs to do the search we first need to
calculate the volume of the elementary cell in the subspace of the
parameter space defined by $\omega_0 =$ const.  
This subspace of the parameter space is called the {\em filter
space}.

Let us expand the autocovariance function $C$ of Eq.\ (\ref{auto5}) around 
$\bs{\tau}=0$ up to terms of second order in $\bs{\tau}$:
\be
\label{nf01}
C\left(\bs{\tau}\right)
= 1 - \sum_{i,j=1}^M \widetilde{\Gamma}_{ij}\tau_i\tau_j,
\ee
where $\widetilde{\Gamma}_{ij}$ are defined in Eq.\ (\ref{redgamma}) and $M$ is 
the number of phase parameters. In Eq.\  (\ref{nf01}) we have used the property 
that $C$ attains its maximum value of 1 for $\bs{\tau}=0$. Let us assume that 
$\tau_1$ corresponds to frequency parameter and let us maximize $C$ given by 
Eq.\ 
(\ref{nf01}) with respect to $\tau_1$. It is easy to show that $C$ attains its 
maximum value, keeping $\tau_2$, \ldots, $\tau_M$ fixed, for
\be
\label{nf02}
{\overline \tau}_1 = -\frac{1}{\Gamma_{11}} \sum_{i=2}^M \Gamma_{1i}\tau_i.
\ee  
Let us define
\be
\label{nf03}
{\overline C}\left(\tau_2,\ldots,\tau_M\right)
:= C\left({\overline \tau}_1,\tau_2,\ldots,\tau_M\right).
\ee
Substituting Eqs.\ (\ref{nf01}) and (\ref{nf02}) into Eq.\ (\ref{nf03}) we 
obtain
\be
\label{nf04}
{\overline C}\left(\tau_2,\ldots,\tau_M\right)
= 1 - \sum_{i,j=2}^M\overline{\Gamma}_{ij}\tau_i\tau_j,
\ee
where
\be
\label{nf05}
\overline{\Gamma}_{ij} := \widetilde{\Gamma}_{ij} - 
\frac{\widetilde{\Gamma}_{1i}\widetilde{\Gamma}_{1j}}{\widetilde{\Gamma}_{11}}.
\ee

We define an elementary cell in the filter space by the requirement that at the 
boundary of the cell the correlation $\overline{C}$ equals 1/2:
\be
\label{nf06}
{\overline C}\left(\tau_2,\ldots,\tau_M\right) = \frac{1}{2}.
\ee
Substituting (\ref{nf04}) into (\ref{nf06}) we arrive at the equation 
describing the surface of the elementary hyperellipsoid in the filter space:
\be
\label{nf07}
\sum_{i,j=2}^M \overline{\Gamma}_{ij}\tau_i\tau_j = \frac{1}{2}.
\ee
The volume of the elementary cell is thus equal to [cf.\ Eq.\ (\ref{vc})]
\be
\label{nf08}
\overline{V}_{\text{cell}}
= \frac{(\pi/2)^{(M-1)/2}}{\Gamma((M+1)/2)\sqrt{\det\overline{\Gamma}}}.
\ee
The volume $\overline{V}_{\text{cell}}$ of the elementary cell in the filter 
space is independent on the value of the frequency parameter.

Taking Eqs.\ (\ref{range1a})--(\ref{range2}) into account the total volume 
$\vtaf$ of the filter space for all-sky searches with $s$ spindowns included  
can be calculated as follows
\bea
\label{vtafaux}
\vtaf &=& \left\{
{\int\!\!\int}_{B_2(0,\omega_0/(2\pi T_o))} d\,\alpha_1\,d\,\alpha_2
\int\limits_{-\beta_1\omega_0}^{\beta_1\omega_0} d\,\omega_1 \ldots
\int\limits_{-\beta_s\omega_0}^{\beta_s\omega_0} d\,\omega_s
\right\}_{\omega_0=2\pi T_o f_{\text{max}}}
\\ \label{vtaf}
&=& \frac{2^{2s} \pi^{s+1}}{(s+1)!}
T_o^{s} \left(\frac{T_o}{\tau_{\text{min}}}\right)^{s(s+1)/2}
f_{\text{max}}^{s+2}.
\eea
Putting in Eq.\ (\ref{vtafaux}) $\omega_0=2\pi T_o f_{\text{max}}$ we have 
defined $\vtaf$ as that slice $\omega_0=\textit{const}$ of the parameter space 
which has maximum volume.

The volume $\vcaf$ of one cell in the filter space we calculate from Eq.\ 
(\ref{nf08}) for $M=s+3$:
\be
\vcaf = \frac{(\pi/2)^{(s+2)/2}}{\Gamma((s+4)/2)\sqrt{\det\gaf}},
\ee
where the matrix $\gaf$ is calulated from Eq.\ (\ref{nf05}) for 
$\widetilde{\Gamma}=\ga$.

The number $\nfa$ of filters for all-sky searches is given by 
\be
\label{nfa}
\nfa = \frac{\vtaf}{\vcaf}
= \frac{2^{3s/2}\pi^{(s+1)/2}(s+2)}{(s+1)\Gamma((s+1)/2)} \sqrt{\det\gaf} 
\left(\frac{T_o}{\tmin}\right)^{s(s+1)/2} T_o^{s} \fmax^{s+2}.
\ee
In Figure \ref{fignfa} we have plotted the number $\nfa$ of filters as a
function of the observation time $T_o$ for various models of the signal
depending on the minimum spindown age $\tmin$ and the maximum gravitational-wave
frequency $\fmax$, and for various numbers $s$ of spindowns included.  We see
that for a given $\tmin$ and $\fmax$ curves corresponding to different numbers
$s$ intersect.  This effect was observed and explained in Ref.\
\cite{BCCS98}: in the regime where adding an extra parameter reduces the
number of filters the parameter space in the extra dimension extends
less than the width of the elementary cell in this dimension. 
To obtain the number of filters for a given observation time $T_o$ we
always take the number of filters given by the uppermost curve.  We have
also calculated the 
observation times $\ctfall{k}$ for which the numbers of filters with $k$ and 
$k+1$ spindowns included coincide:
\be
\nfadef{k+1}(T_o=\ctfall{k}) = \nfadef{k}(T_o=\ctfall{k}),\quad k=0,\ldots,s-1.
\ee
In Table \ref{tab:ctfilters} we have given the values of $\ctfall{k}$ for all 
the signal models considered.

\begin{table}[!ht]
\begin{center}
\begin{tabular}{|c|c|c|c|c|c|c|c|c|}\hline
\raisebox{-1.5ex}[0cm][0cm]{$\tmin$ (years)} &
\raisebox{-1.5ex}[0cm][0cm]{$\fmax$ (Hz)} & 
\multicolumn{4}{c|}{$\ctfall{k}$ (days)} &
\multicolumn{3}{c|}{$\ctfdir{k}$ (days)} \\ \cline{3-9}
&& $k=0$ & $k=1$ & $k=2$ & $k=3$ & $k=1$ & $k=2$ & $k=3$ \\ \hline\hline
40 & $10^3$ &
0.19 & 3.01 & 113   & 307 &
       3.26 &  38.8 & 171 \\ \hline
40 & 200 &
0.30 & 5.07 & 156   & 384 &
       5.58 &  58.0 & 236 \\ \hline
$10^3$ & $10^3$ &
0.45 & 111   & 566 & 2170 &
        27.9 & 434 & 2250 \\ \hline
$10^3$ & 200 &
0.66 & 153   & 715 & 2990 &
        47.7 & 649 & 3100 \\ \hline
\end{tabular}
\end{center}
\caption{\label{tab:ctfilters}The observation times for which the numbers of 
filters with $k$ and $k+1$ spindowns included coincide for various models of the 
signal depending on the minimum spindown age $\tmin$ and the maximum 
gravitational-wave frequency $\fmax$. In the case of all-sky searches we have 
used the latitude $\lambda=46.45^\circ$ of the LIGO Hanford detector and we have 
put $\phi_o=0.123$ and $\phi_r=1.456$.}
\end{table}

For directed searches the total volume $\vtdf$ of the filter space with $s$ 
spindowns included we calculate using Eqs.\ (\ref{range1a}) and (\ref{range1b}):
\be
\vtdf = \left\{
\int\limits_{-\beta_1\omega_0}^{\beta_1\omega_0} d\,\omega_1 \ldots
\int\limits_{-\beta_s\omega_0}^{\beta_s\omega_0} d\,\omega_s
\right\}_{\omega_0=2\pi T_o f_{\text{max}}}
= \frac{2^{2s} \pi^{s}}{(s+1)!}
\left(\frac{T_o}{\tau_{\text{min}}}\right)^{s(s+1)/2} 
\left(f_{\text{max}}T_o\right)^{s}.
\ee

The volume $\vcdf$ of one cell in the filter space for directed searches with 
$s$ spindowns included we calculate from Eq.\ (\ref{nf08}) for $M=s+1$:
\be
\vcdf = \frac{(\pi/2)^{s/2}}{\Gamma((s+2)/2)\sqrt{\det\gdf}},
\ee
where the matrix $\gdf$ is calulated from Eq.\ (\ref{nf05}) for 
$\widetilde{\Gamma}=\gd$.

The number $\nfa$ of filters in the case of directed searches is thus given by: 
\be
\label{nfd}
\nfd = \frac{\vtdf}{\vcdf} = \frac{2^{(3s-2)/2}\pi^{(s+1)/2}}{\Gamma((s+3)/2)}  
\sqrt{\det\gdf} \left(\frac{T_o}{\tmin}\right)^{s(s+1)/2}
\left(\fmax T_o\right)^{s}.
\ee
In Figure \ref{fignfd} we have plotted the number of filters for various models
of the signal depending on the minimum spindown age $\tmin$ and the maximum
gravitational-wave frequency $\fmax$, and for various numbers $s$ of spindowns
included. We have also calculated analytically the
observation times $\ctfdir{k}$ for which the numbers of filters with $k$ and
$k+1$ spindowns included coincide:
\be
\nfddef{k+1}\left(T_o=\ctfdir{k}\right) = \nfddef{k}\left(T_o=\ctfdir{k}\right), 
\quad k=1,\ldots,s-1.
\ee
Using Eq.\ (\ref{nfd}) one obtains
\be
\ctfdir{k} = \left[
\frac{\Gamma((k+4)/2)}{2\sqrt{2\pi}\Gamma((k+3)/2)}
\sqrt{\frac{\det\overline{\Gamma}^{\text{dir}}_{(k)}}
{\det\overline{\Gamma}^{\text{dir}}_{(k+1)}}}
\frac{\tau_{\text{min}}^{k+1}}{f_{\text{max}}}
\right]^{1/(k+2)},\quad k=1,\ldots,s-1.
\ee
In Table \ref{tab:ctfilters} we have given the values of $\ctfdir{k}$ 
for all the signal models considered.

In Table 4 we have given the number of filters both for all-sky and directed
searches for various models of the signal depending on the minimum spindown age
$\tmin$ and the maximum gravitational-wave frequency $\fmax$, and for the
observation time $T_o$ of 7 and 120 days.  The number of filters is calculated
from Eq.\ (\ref{nfa}) for all-sky searches and from Eq.\ (\ref{nfd}) in the case
of directed searches.  For a given observation time $T_o$ the number $s$ of
spindowns one should include in the signal's model is obtained as such number
$s$ chosen out of $s=0,\ldots,4$ for which $\nfa$ (or $\nfd$) is the greatest.

\begin{table}[!ht]
\begin{center}
\begin{tabular}{|c|c|c|c|c|c|c|c|c|}\hline
\raisebox{-1.5ex}[0cm][0cm]{$T_o$ (days)} &
\raisebox{-1.5ex}[0cm][0cm]{$\tau_{\text{min}}$ (years)} &
\raisebox{-1.5ex}[0cm][0cm]{$f_{\text{max}}$ (Hz)} &
\multicolumn{3}{c|}{all-sky} & \multicolumn{3}{c|}{directed} \\ \cline{4-9}
&&& $s$ & $\nfa$ & ${\mathcal P}$ (Tf) & $s$ & $\nfd$ & ${\mathcal P}$ (Tf)
\\ \hline\hline
7 & 40 & $10^3$ &
2 & $1.4\times10^{10}$ & $2.6\times10^3$ &
2 & $9.5\times10^{5}$  & $1.7\times10^{-1}$ \\ \hline
7 & 40 & 200 &
2 & $2.3\times10^{7}$ & $7.8\times10^{-1}$ &
2 & $3.8\times10^{4}$ & $1.3\times10^{-3}$ \\ \hline
7 & $10^3$ & $10^3$ &
1 & $4.6\times10^{7}$ & 8.5 &
1 & $3.8\times10^{3}$ & $7.0\times10^{-4}$ \\ \hline
7 & $10^3$ & 200 &
1 & $3.7\times10^{5}$ & $1.3\times10^{-2}$ &
1 & $7.7\times10^{2}$ & $2.6\times10^{-5}$ \\ \hline\hline
120 & 40 & $10^3$ &
3 & $8.4\times10^{19}$ & $1.7\times10^{13}$ &
3 & $1.3\times10^{14}$ & $2.7\times10^7$ \\ \hline
120 & 40 & 200 &
2 & $1.1\times10^{17}$ & $4.3\times10^9$ &
3 & $1.0\times10^{12}$ & $3.9\times10^4$ \\ \hline
120 & $10^3$ & $10^3$ &
2 & $4.4\times10^{15}$ & $9.2\times10^8$ &
2 & $9.0\times10^{7}$  & $1.8\times10$ \\ \hline
120 & $10^3$ & 200 &
1 & $2.4\times10^{13}$ & $9.3\times10^5$ &
2 & $3.6\times10^{6}$  & $1.4\times10^{-1}$ \\ \hline
\end{tabular}
\caption{Number of filters for all-sky and directed searches for various models 
of the signal depending on the minimum spindown age $\tmin$ and the 
maximum gravitational-wave frequency $\fmax$, and for the observation 
time $T_o$ of 7 and 120 days.  To calculate the Fisher matrix $\gaf$ we have 
used the latitude $\lambda=46.45^\circ$ of the LIGO Hanford detector and we have 
put $\phi_o=0.123$ and $\phi_r=1.456$. For each case we also give the number 
${\mathcal P}$ of floating point operations per second (flops) needed to do the 
search; ${\mathcal P}$ is calculated by means of Eq.\ (\ref{copower}).}
\end{center}
\end{table}

We shall next compare the number of filters obtained above with the number of
filters calculated by Brady {\em et al.}  \cite{BCCS98}.  In their calculations
they have assumed a constant amplitude of the signal however they have used a
full model of the phase.  To calculate the number of templates they have used so
called metric approach of Owen \cite{O96}.  They have assumed a certain geometry
of spacing of the templates:  combination of a hexagonal and a hypercubic
spacing and they have introduced an additional parameter---a mismatch $\mu$,
which was the measure of the correlation of the two neighbouring templates.  
Also in their calculation they have assumed that the data processing
method involves resampling of the time series so that the resampled
signal is monochromatic. We shall compare the number of filters in Table
4 of our paper with the corresponding number of filters given in Table 1
of \cite{BCCS98}.  Our calculations correspond to mismatch $\mu = 0.5$.
This means
that to compare our numbers of filters with the corresponding numbers of Brady
{\it et al}.\ our numbers have to be multiplied by 2.4, 5.8, 15, and 40 for the
signal with 0, 1, 2, and 3 spindowns respectively for all-sky searches and by
1.3, 1.7, 2.2 for 1, 2, and 3 spindowns respectively for directed searches.  The
difference in the volume of our hyperellipsoidal cells and their volumes of
elementary patches means [see Ref.\ \cite{BCCS98}, Eq.\ (5.18) for all-sky
searches and the paragraph above Eq.\ (7.2) for directed searches] that our
numbers aditionally have to be multiplied by 1.7, 2.2, 2.8, and 3.6 for all-sky
searches and by 1.0, 1.4, 1.3 for directed searches for comparison.  After
introducing the corrections for the mismatch parameter and the size of an
elementary cell we find that our corrected number of templates is greater than
the number of templates given in Table 1 of \cite{BCCS98} by (going from top to
bottom of Table 1) $2.8\times10^4$, 14, 2.7, and 1.5 for all-sky searches and by
2.2, 1.7, 0.31, and 0.25 for directed searches.  We thus conclude that
considering the differences in the way the calculations were done there is a
reasonable agreement between the number of filters obtained by the two
approaches except for one case:  all-sky searches with the maximum frequency
$f_{\text{max}}=200$ Hz and the minimum spindown age $\tau_{\text{min}}=1000$
years where the difference is 4 orders of magnitude.

We would also like to point out to the uncertainties in the calculation of the
number of filters.  Our model of the intrinsic spin frequency evolution of the
neutron star is extremely simple:  we approximate the frequency evolution by a
Taylor series.  In reality the frequency evolution will be determined by complex
physical processes.  The size of the parameter space is likewise uncertain.  The
range for the spindown parameters [see Eqs.\ (\ref{range1a})--(\ref{range1b})]
was chosen so that the total size of our parameter space is the same as in
\cite{BCCS98}.  The approximation of the time derivative of the frequency as
$\fmax/\tmin$ that is used to estimate the maximum value of the spindowns is
probably an order of magnitude estimate.  This implies that the size of the
parameter space and consequently the number of filters is accurate within
$s(s+1)/2$ orders of magnitude, where $s$ is the number of spindowns in the
phase of the signal.  Even this large uncertainty does not change the conclusion
that all-sky searches for 120 days of observation time are computationally too
prohibitive.

To estimate the computational requirement to do the signal search we adopt a
simple formula [see Eq.\ (6.11) of \cite{BCCS98}] for the number ${\mathcal P}$
of floating point operations per second (flops) required assuming that the data
processing rate should be comparable to the data aquisition rate (it is assumed
that fast Fourier transform (FFT) algorithm is used):
\be
\label{copower}
{\mathcal P} = 6 \fmax N_f [\log_2(2 \fmax T_o) + 1/2],
\ee 
where $N_f$ is the number of filters.
The above formula assumes that we calculate only one modulus of the Fourier
transform.  Calculation of the optimal statistics $\F$ for the amplitude
modulated signal requires two such muduli for each component of the signal [see
Eq.\ (99) of Paper I, we assume that the observation time is an integer multiple
of the sidereal day so that $C = 0$] and several multiplications.  Moreover if
dechirping operations are used instead of resampling, the data processing would
involve complex FFTs.  All these operation will not increase the complexity of
the analysis i.e.\ the number of floating point operations will still go as
$O(N\log_2(N))$, where $N$ is the number of points to be processed.

In Table 4 we have given the computer power ${\mathcal P}$ (in Teraflops, Tf)
required for all the cases considered.  We see that for 120 days of observation
time all-sky searches are computationally too prohibitive whereas for directed
searches only one case ($\tmin=1000$ years, $\fmax=200$ Hz) is within reach of a
1 Teraflops computer.  For 7 days of observation time all cases except for the
most demanding all-sky search with $\tmin=40$ years and $\fmax=1$ kHz are within
a reach of a 1 Teraflops computer.

Finally we would like to point to a technique that can distribute the data
processing into several smaller computers.  We shall call this technique {\em
signal splitting}.  We can divide the available bandwidth of the detector
$(f_{\text{min}},f_{\text{max}})$ into $M$ adjacent intervals of length $B$.  We
then apply a standard technique of heterodyning.  For each of the chosen bands
we multiply our data time series by $\exp(-2\pi i f_I)$, where
$f_I=f_{\text{min}}+IB$ ($I=0,\ldots,M-1$).  Such an operation moves the
spectrum of the data towards zero by frequency $f_I$.  We then apply a low pass
filter with a cutoff frequency $B$ and we resample the resulting sequence with
the frequency $2B$.  The result is $M$ time series sampled at frequency $2 B$
instead of one sampled at $2f_{\text{max}}$.  The resampled sequencies are
shorter than the original ones by a factor of $M$ and each can be processed by a
separate computer.  We only need to perform the signal splitting operation once
before the signal search.  The splitting operation can also be performed
continuously when the data are collected so that there is no delay in the
analysis.  The signal splitting does not lead to a substancial saving in the
total computational power but yields shorter data sequencies for the analysis.
For example for the case of 7 days of observation time and sampling rate of 1
kHz the data itself would occupy around 10 GB of memory (assuming double
precision) which is available on expensive supercomputers whereas if we split
the data into a bandwidth of 50 Hz so that sampling frequency is only 100 Hz
each sequence will occupy 0.5 GB memory which is available on inexpensive
personal computers.

In the case of a narrowband detector, e.g.\ for the GEO600 detector tuned to a
certain frequency $f_o$ around a bandwidth $B$, it is natural to apply the above
data reduction technique so that the resulting sampling frequency is $2B$.  Such
a technique is applied in data preprocessing of bar detectors \cite{exp}.  From
the formulae given in the present section one can show that to perform the
all-sky search with the integration time of 7 days for pulsars in the bandwidth
of 50 Hz around the frequency of 300 Hz and minimum spindown age of 40 years so
that the processing proceeds at the rate of data aquisition requires a 1
Teraflops computer.  Since the data sequence occupies only 0.5 GB of memory the
data processing task can be distributed over several smaller computers.  If we
also relax the requirement of data processing to be done in real time the signal
search can be performed by a 20 Gigaflops workstation in a year.

\section{Suboptimal filtering}

It will very often be the case that the filter we use to extract the signal from
the noise is not optimal.  This may be the case when we do not know the exact
form of the signal (this is almost always the case in practice) or we choose a
suboptimal filter to reduce the computational cost and simplify the analysis.
We shall consider here an important special case of a suboptimal filter that
may be usful in the analysis of gravitational-wave signals from a spinning 
neutron star.

\subsection{General theory}

We shall assume a constant amplitude one-component model of the signal.  Then
the optimal (maximum likelihood) statistics is given by Eq.\ (\ref{Sstat}).  Let
us suppose that we do not model the phase accurately and instead of the two
optimal filters $\cos\left[\Phi(t;\bs{\xi})\right]$ and
$\sin\left[\Phi(t;\bs{\xi})\right]$ we use filters with a phase
$\Phi'(t;\bs{\xi}')$, where function $\Phi'$ is different form $\Phi$ and the
set of filter parameters $\bs{\xi}'$ is in general different from $\bs{\xi}$,
i.e.\ $\Fs$ has the form [cf.\ Eqs.\ (\ref{Sstat}) and (\ref{xcxs})]
\be
\Fs = \frac{2T_o}{S_h(f_o)}
\left[ \tav{x\cos\Phi'(t;\bs{\xi}')}^2 + \tav{x\sin\Phi'(t;\bs{\xi}')}^2 
\right],
\ee
where we have assumed that the suboptimal filters are narrowband at some 
"carrier" frequency $f_o$ as in the case of optimal filters.

Let us first establish the probability density functions of $\Fs$ when the phase
parameters $\bs{\xi}'$ are known.  Since the dependence on the data random
process is the same as in the optimal case the false alarm and detection
probability densities will be the same as for the optimal case i.e.\ $2\Fs$ has
a central or a noncentral $\chi^2$ distribution with 2 degrees of freedom
depending on whether the signal is absent or present.  From the narrowband
property of the suboptimal filter we get the following expressions for the
expectation values and the variances of $\Fs$ ($0$ means that signal is absent
and $1$ means that signal is present):
\bea
\text{E}_0\left\{\Fs\right\} = 1,\quad
\text{E}_1\left\{\Fs\right\} = 1 + \frac{1}{2}\ds^2, \\  
\text{Var}_0\left\{\Fs\right\} = 1,\quad
\text{Var}_1\left\{\Fs\right\} = 1 + \ds^2,
\eea
where
\be 
\label{SNRsub}
\ds := d \left\{ 
\tav{\cos[\Phi(t;\bs{\xi})-\Phi'(t;\bs{\xi}')]}^2 +
\tav{\sin[\Phi(t;\bs{\xi})-\Phi'(t;\bs{\xi}')]}^2 \right\}^{1/2},
\ee
here $d$ is the optimal signal-to-noise ratio.

We see that for the suboptimal filter introduced above the false alarm
probability has exactly the same $\chi^2$ distribution as in the optimal case
whereas the probability of detection has noncentral $\chi^2$ distribution but
with a different noncentrality parameter $\ds^2$.  We shall call $\ds$ (the
square root of the noncentrality parameter) the {\em suboptimal signal-to-noise
ratio}.  It is clear that when the phases of the signal and the suboptimal
filter are different the suboptimal signal-to-noise ratio is strictly less and
the probability of detection is less than for the optimal filter.

When the parameters $\bs{\xi}'$ are unknown the functional $\Fs$ is a random
field and we can obtain the false alarm probabilities as in the case of an
optimal filter.  Here we only quote the formula based on the number of
independent cells of the random field.  One thing we must remember is that the
number of cells for the suboptimal and the optimal filters will in general be
different because they may have a different functional dependence and a
different number of parameters.  Thus we have [cf.\ Eqs.\ (\ref{PF}) and 
(\ref{FP}) for $n=2$]
\be
P^T_{sF}(\Fo) = 1 - [1 - \exp(-\Fo)]^{N_{sc}},
\ee
where $N_{sc}$ is the number of cells for the suboptimal filter.

The detection probability for the suboptimal filter is given by [cf.\ Eqs.\ 
(\ref{p1}) and (\ref{PD}) for $n=2$]
\be
\label{PDs}
P_{sD}(\ds,\Fo) := \int^{\infty}_{\Fo}p_{s1}(\ds,\F)\,d\F,
\ee
where
\be
\label{p1s}
p_{s1}({\ds,\F}) = I_0\left(\ds\sqrt{2\F}\right) 
\exp\left(-\F-\frac{1}{2}\ds^2\right).
\ee
Probability of detection for the suboptimal filter is obtained from
the probability of detection for the optimal one by replacing the optimal
signal-to-noise ratio $d$ by the suboptimal one $\ds$.

When we design a suboptimal filtering scheme we would like to know what is the
expected number of false alarms with such a scheme and what is the expected
number of detections.  As in the optimal case the expected number $N_{sF}$ of
false alarms with suboptimal filter is given by [cf.\ Eqs.\ (\ref{PF}) and 
(\ref{NF}) for $n=2$]
\be
\label{NFs}
N_{sF} = N_{sc}\exp(-\Fo).
\ee

To obtain the expected number of detections we assume that the signal-to-noise
ratio $d$ varies inversely proportionally to the distance from the source and
that the sources are uniformly distributed in space.  We also assume that the
space is Euclidean.  Let us denote by $d_1$ the signal-to-noise ratio for which
the number of events is one.  Then the number of events corresponding to the
signal-to-noise ratio $d$ is $(d_1/d)^3$.  The expected number of the detected 
events is given by
\be
\label{ND}
N_D(d_1,\Fo) = 3 \int^{\infty}_0 x^2 P_D\left(\frac{d_1}{x},\Fo\right) \,dx
\ee
in the case of the optimal filter, and by
\be
\label{NsD}
N_{sD}(d_{1\text{sub}},\Fo) = 3 \int^{\infty}_0 x^2
P_{sD}\left(\frac{d_{1\text{sub}}}{x},\Fo\right) \,dx
\ee
for the suboptimal filter. Let us note that [cf.\ Eq.\ (\ref{SNRsub})]
\be
d_{1\text{sub}} = d_1 \left\{ 
\tav{\cos[\Phi(t;\bs{\xi})-\Phi'(t;\bs{\xi}')]}^2 +
\tav{\sin[\Phi(t;\bs{\xi})-\Phi'(t;\bs{\xi}')]}^2 \right\}^{1/2}.
\ee

Because of the statistical nature of the detection any signal can only be
detected with a certain probability less than 1.  In the case of Gaussian noise
for signals with the signal-to-noise ratio around the threshold this probability
is roughly 1/2 and it increases exponentially with increasing signal-to-noise
ratio.  In Appendix C we give a worked example of the application of the
statistical formulae for the suboptimal filtering derived above.

\subsection{Fitting factor}

To study the quality of suboptimal filters (or search templates as they are
sometimes called) one of the present authors \cite{K94,KKS} introduced an
$l-$factor defined as the square root of the correlation between the signal and
the suboptimal filter.  It turned out that a more general and more natural
quantity is the {\em fitting factor} introduced by Apostolatos \cite{A1}.
The fitting factor FF between a signal $h(t;\bs{\theta})$ and a filter
$h'(t;\bs{\theta}')$ ($\bs{\theta}$ and $\bs{\theta}'$ are the parameters of the
signal and the filter, respectively) is defined as
\be
\label{ff1}
\text{FF} := \max_{\bs{\theta}'}
\frac{\left(h(t;\bs{\theta})\vert h'(t;\bs{\theta}')\right)}
{\sqrt{\left(h(t;\bs{\theta})\vert h(t;\bs{\theta})\right)}
\sqrt{\left(h'(t;\bs{\theta}')\vert 
h'(t;\bs{\theta}')\right)}}.
\ee
If both the signal $h$ and the filter $h'$ are narrowband around the same 
frequency $f_o$ the scalar products $(\cdot\vert\cdot)$ from Eq.\ (\ref{ff1}) 
can be computed from the formula
\be
\label{sca}
(h_1\vert h_2) \approx \frac{2}{S_h(f_o)}\int_{-T_o/2}^{T_o/2}h_1(t)h_2(t)\,dt,
\ee
where $S_h$ is the one-sided noise spectral density and $T_o$ is the 
observation time.

Let us assume that the signal and the filter can be written as
\be
\label{ff2}
h(t;\bs{\theta}) = h_o \sin\Psi(t;\bs{\zeta}), \quad
h'(t;\bs{\theta}') = h'_o \sin\Psi'(t;\bs{\zeta}'),
\ee
where $h_o$ and $h'_o$ are constant amplitudes, $\bs{\zeta}$ and 
$\bs{\zeta}'$ denote the parameters entering the phases $\Psi$ and 
$\Psi'$ of the signal and the filter, respectively. We substitute Eqs.\ 
(\ref{ff2}) into Eq.\ (\ref{ff1}). Using Eq.\ (\ref{sca}) we obtain
\be
\label{ff3}
\text{FF} \approx \max_{\bs{\zeta}'} \frac{1}{T_o} 
\int_{-T_o/2}^{T_o/2} 
\cos\left[\Psi(t;\bs{\zeta})-\Psi'(t;\bs{\zeta}')\right]\,dt.
\ee
It is easy to maximize the FF (\ref{ff3}) with respect to the initial phase of 
the filter. Let us denote the initial phases of the functions $\Psi$ and $\Psi'$ 
by $\Phi_0$ and $\Phi'_0$, respectively. Then
\be
\label{ff4}
\Psi(t;\bs{\zeta}) = \Phi(t;\bs{\xi}) + \Phi_0,\quad
\Psi'(t;\bs{\zeta}') = \Phi'(t;\bs{\xi}') + \Phi'_0,
\ee
where $\bs{\xi}$ and $\bs{\xi}'$ denote the remaining parameters of the signal 
and the filter, respectivley. After substitution Eqs.\ (\ref{ff4}) into Eq.\ 
(\ref{ff3}) we easily get
\bea
\label{ff5}
\text{FF} &\approx& \max_{\Phi'_0,\bs{\xi}'} 
\left\langle \cos\left[\Phi(t;\bs{\xi})-\Phi'(t;\bs{\xi}')
+ \left(\Phi_0-\Phi'_0\right) \right] \right\rangle
\nonumber\\
&=& \max_{\Phi'_0,\bs{\xi}'} \left\{
\cos\left(\Phi_0-\Phi'_0\right) \left\langle
\cos\left[\Phi(t;\bs{\xi})-\Phi'(t;\bs{\xi}')\right]
\right\rangle
- \sin\left(\Phi_0-\Phi'_0\right) \left\langle
\sin\left[\Phi(t;\bs{\xi})-\Phi'(t;\bs{\xi}')\right]
\right\rangle \right\}
\nonumber\\
&=& \max_{\bs{\xi}'} \left\{
\tav{\cos\left[\Phi(t;\bs{\xi})-\Phi'(t;\bs{\xi}')\right]}^2
+ \tav{\sin\left[\Phi(t;\bs{\xi})-\Phi'(t;\bs{\xi}')\right]}^2 \right\}^{1/2}.
\eea
Thus we obtain that the FF is nothing else but the ratio of the maximized value
of the suboptimal signal-to-noise ratio $\ds$ and the optimal signal-to-noise
ratio $d$ [cf.\ Eq.\ (\ref{SNRsub})].  We stress however that the value of the
fitting factor by itself is not adequate for determining the quality of a
particular search template---one also needs the underlying probability
distributions (both the false alarm and the detection) derived in the previous
subsection.  This is clearly shown by an example in Appendix C.

In the remaining part of this subsection we shall propose a way of approximate 
computation of the fitting factor. Let us now assume that the filter and the 
signal coincide, i.e.\ $\Phi'=\Phi$, and the filter parameters $\bs{\xi}'$ 
differ from the parameters $\bs{\xi}$ of the signal by small quantities 
$\Delta\bs{\xi}$: $\bs{\xi}'=\bs{\xi}+\Delta\bs{\xi}$. The Eq.\ (\ref{ff5}) can 
be rewritten as
\be
\label{ff6}
\text{FF} \approx \max_{\Delta\bs{\xi}} \left\{
\tav{\cos\left[\Phi(t;\bs{\xi})-\Phi(t;\bs{\xi}+\Delta\bs{\xi})\right]}^2
+ \tav{\sin\left[\Phi(t;\bs{\xi})-\Phi(t;\bs{\xi}+\Delta\bs{\xi})\right]}^2 
\right\}^{1/2}.
\ee
Obviously the FF (\ref{ff6}) attains its maximum value of 1 when 
$\Delta\bs{\xi}=0$. Let us expand the expression in curly brackets on the 
right-hand side of Eq.\ (\ref{ff6}) w.r.t.\ $\Delta\bs{\xi}$ around 
$\Delta\bs{\xi}=0$ up to terms of second order in $\Delta\bs{\xi}$. The result 
is
\be
\label{ff7}
\text{FF} \approx \left\{ 1 - \min_{\Delta\bs{\xi}}
\left( \sum_{i,j}\Gamma_{ij}\Delta\xi_i\Delta\xi_j \right) \right\}^{1/2},
\ee
where
\be
\Gamma_{ij} := \tav{\frac{\pa\Phi}{\pa\xi_i}\frac{\pa\Phi}{\pa\xi_j}}
- \tav{\frac{\pa\Phi}{\pa\xi_i}} \tav{\frac{\pa\Phi}{\pa\xi_j}}.
\ee

One can employ the formula (\ref{ff7}) to estimate the FF in the case when the 
filter $\Phi'$ is obtained from the signal $\Phi$ by replacing some of the 
signal parameters by zeros, provided the signal $\Phi$ depends weakly on these 
discarded parameters. Let the signal $\Phi$ depend on $n$ parameters 
$\xi_1,\ldots,\xi_n$, and the filter $\Phi'$ is defined by
\be
\label{ff8}
\Phi'(t;\xi'_1,\ldots,\xi'_k) :=
\Phi(t;\xi'_1,\ldots,\xi'_k,\underbrace{0,\ldots,0}_{n-k}), 
\ee
where $k<n$, so the filter $\Phi'$ depends on $k$ parameters 
$\xi'_1,\ldots,\xi'_k$. One can write
\be
\label{ff9}
\Phi(t;\bs{\xi})-\Phi'(t;\bs{\xi}') = 
\Phi(t;\xi_1,\ldots,\xi_n)
- \Phi(t;\xi'_1,\ldots,\xi'_k,\underbrace{0,\ldots,0}_{n-k}) =
\Phi(t;\bs{\xi}) - \Phi(t;\bs{\xi}+\Delta\bs{\xi})
\ee
with
\be
\label{ff10}
\Delta\xi_i = \left\{
\begin{array}{ll}
\xi'_i - \xi_i, & i=1,\ldots,k, \\
-\xi_i, & i=k+1,\ldots,n.
\end{array} \right.
\ee

We want to approximate the differnce
$\Phi(t;\bs{\xi})-\Phi(t;\bs{\xi}+\Delta\bs{\xi})$ with $\Delta\bs{\xi}$ given
by Eq.\ (\ref{ff10}) by its Taylor expansion around $\Delta\bs{\xi}=0$.  It is
reasonable provided the two following conditions are satisfied.  Firstly, the
filter parameters differ slightly from the respective parameters of the signal,
i.e.\ the quantitites $\Delta\xi_i$ are small compared to $\xi_i$ for
$i=1,\ldots,k$.  Secondly, the function $\Phi$ depends on the parameters
$\xi_{k+1},\ldots,\xi_n$ (discarded from the filter) weakly enough to make a
reasonable approximation by Taylor expansion up to $\Delta\xi_i=-\xi_i$ for
$i=k+1,\ldots,n$.  If the above holds, one can use the formula (\ref{ff7}) to
approximate the FF.  Taking Eqs.\ (\ref{ff9}) and (\ref{ff10}) into account,
from Eq.\ (\ref{ff7}) one gets
\be
\label{ff11}
\text{FF} \approx \left\{ 1 - \min_{\Delta\xi_1,\ldots,\Delta\xi_k}
\left( \sum_{i,j=1}^n \Gamma_{ij}\Delta\xi_i\Delta\xi_j
\Bigg\vert_{\Delta\xi_{k+1}=-\xi_{k+1},\ldots,\Delta\xi_n=-\xi_n} \right)
\right\}^{1/2}.
\ee

\subsection{Fitting factor vs.\ 1/4 of a cycle criterion}

Let us consider the phase of the gravitational-wave signal of the form [cf.\ 
Eq.\ (\ref{phase1})]
\be
\label{phase2}
\Phi(t) = 2\pi \sum_{k=0}^{s_1}{\fko k}\frac{t^{k+1}}{(k+1)!}
+ \frac{2\pi}{c} {\bf n}_0\cdot{\bf r}_{\rm ES}(t)
\sum_{k=0}^{s_2}{\fko k}\frac{t^k}{k!}
+ \frac{2\pi}{c} {\bf n}_0\cdot{\bf r}_{\rm E}(t)
\sum_{k=0}^{s_3}{\fko k}\frac{t^k}{k!}.
\ee
In Paper I we have introduced the following criterion:  {\em we exclude an
effect from the model of the signal in the case when it contributes less than
1/4 of a cycle to the phase of the signal during the observation time}.  In
Paper II we have shown that if we restrict to observation times $T_o\le120$
days, frequencies $f_o\le1000$ Hz, and spindown ages $\tau\ge40$ years, the
phase model (\ref{phase2}) meets the criterion for an appropriate choice of the
numbers $s_1$, $s_2$, and $s_3$.  We have also shown that the effect of
the star proper motion in
the phase is negligible if we assume that the star moves w.r.t.\ the SSB not
faster than $10^3$ km/s and its distance to the Earth $r_o\ge1$ kpc.  In Table
\ref{tab:models}, which is Table 1 of Paper II, one can find the numbers $s_1$,
$s_2$, and $s_3$ needed to meet 1/4 of a cycle criterion for different
observation times $T_o$, maximum values $\fmax$ of the gravitational-wave
frequency, and minimum values $\tmin$ of the neutron star spindown age.

\begin{table}[!ht]
\begin{center}
\begin{tabular}{|c|c|c|c|c|c|}\hline
$T_o$ (days) & $\tmin$ (years) & $\fmax$ (Hz) & 
$s_1$  & $s_2$  & $s_3$ \\ \hline\hline
120 & 40     & $10^3$ & 4 & 3 & 0 \\ \hline
120 & 40     & 200    & 4 & 2 & 0 \\ \hline
120 & $10^3$ & $10^3$ & 2 & 1 & 0 \\ \hline
120 & $10^3$ & 200    & 2 & 1 & 0 \\ \hline\hline
7   & 40     & $10^3$ & 2 & 1 & 0 \\ \hline
7   & 40     & 200    & 2 & 1 & 0 \\ \hline
7   & $10^3$ & $10^3$ & 1 & 1 & 0 \\ \hline
7   & $10^3$ & 200    & 1 & 1 & 0 \\ \hline
\end{tabular}
\end{center}
\caption{\label{tab:models}The number of spindown terms needed in various 
contributions to the phase of the signal depending on the type of population of 
neutron stars searched for [cf.\ Eq.\ (\ref{phase2})].  The number $s_1$ refers 
to the dominant polynomial in time term in Eq.\ (\ref{phase2}), $s_2$ refers to 
the Earth orbital motion contribution, and $s_3$ refers to the Earth diurnal 
motion contribution.}
\end{table}

In Appendix A of Paper I we have indicated that the 1/4 of a cycle criterion is
only a sufficient condition to exclude a parameter from the phase of the signal
but not necessary.  In this subsection we study the effect of neglecting certain
parameters in the template by calculating FFs.  We employ the approximate
formula (\ref{ff11}) developed in the previous subsection to calculate FF
between the one-component constant amplitude signals with the phases given by
Eq.\ (\ref{phase2}) for numbers $s_1$, $s_2$, and $s_3$ taken from Table
\ref{tab:models} and the same signals with a smaller number (as compared to that
given in Table \ref{tab:models}) of spindowns included.  We have found that for
the first two models of Table \ref{tab:models} if in the template one neglects
the fourth spindown, FF is greater than 0.99, both for all-sky and directed
searches.  For other cases in Table \ref{tab:models} we have found that
neglecting any spindown parameter can result in the FF appreciably less than 
one.

In Paper II we have considered the effect of the proper motion of the
neutron star on the phase of the signal assuming that it moves uniformly with
respect to the SSB reference frame.  We have found that for the observation time
$T_o=$ 120 days and the extreme case of a neutron star at a distance $r_o=$ 40
pc moving with the transverse velocity $|{\bf v}_{{\rm ns}\perp}|=10^3$ km/s
(where ${\bf v}_{{\rm ns}\perp}$ is the component of the star's velocity ${\bf
v}_{\rm ns}$ perpendicular to the vector ${\bf n}_{0}$), gravitational-wave
frequency $f_o=1$ kHz, and spindown age $\tau=$ 40 years, proper motion
contributes only $\sim$4 cycles to the phase of the signal.  We have shown in
Paper II that in this extreme case the phase model consistent with the 1/4 of a
cycle criterion reads [cf.\ Eq.\ (33) in Paper II]
\be
\label{pmphase}
\Phi(t) = 2\pi \sum_{k=0}^{4}{\fko k}\frac{t^{k+1}}{(k+1)!}
+ \frac{2\pi}{c} {\bf n}_0\cdot{\bf r}_{\rm ES}(t)
\sum_{k=0}^{3}{\fko k}\frac{t^k}{k!}
+ \frac{2\pi}{c} \left( {\bf n}_0\cdot{\bf r}_{\rm E}(t) +
\frac{{\bf v}_{{\rm ns}\perp}}{r_o}\cdot{\bf r}_{\rm ES}(t)\,t \right)f_o.
\ee
The ratio ${\bf v}_{{\rm ns}\perp}/r_o$ determines the proper motion of the star
and can be expressed in terms of the proper motions $\mu_\alpha$ and
$\mu_\delta$ in right ascension $\alpha$ and declination $\delta$, respectively
(see Sec.\ 4 of Paper II).

For the extreme case described above we have applied formula (\ref{ff11}) to
calculate the FF between the one-component constant amplitude signal with the
phase given by Eq.\ (\ref{pmphase}) and the same signal with a simplified phase.
We have found that when both proper motion parameteres $\mu_\alpha$,
$\mu_\delta$ and the fourth spindown parameter ${\fko 4}$ are neglected, 
the FF is greater than 0.99 for both all-sky and directed
searches.  Thus we conclude that neglecting the fourth spindown and the proper
motion does not reduce appreciably the probability of detection of the signal.

It is also interesting to compare the results obtained from the calculation of
the fitting factor with the results summarized in Table 1 for the observation
times when the number of cells for models with $k$ and $k+1$ spindowns
coincides.  The observation times given in Table 1 can be interpreted as
observation times at which one should include the $k+1$ parameter in the
template.  We see that for the first two models in Table \ref{tab:models} the
Table 1 says that only 3 spindowns are needed as indicated by the calculation of
the FF.  The remaining cases also agree except for the cases of 120 days of
observation time and 200 Hz frequency where Table 1 indicates one less spindown
than Table \ref{tab:models}.  Finally we note that the crossover observation
times in Table 1 agree within a few percent with those for the number of filters
given in Table 3.

\section{Monte Carlo simulations and the Cram\'er-Rao bound}

As signal-to-noise ratio goes to infinity the maximum-liklihood estimators
become unbiased and their rms errors tend to the errors calculated from the
covariance matrix.  The rms errors calculated from the covariance matrix are the
smallest error achievable for unbiased estimators and they give what is called
the Cram\'er-Rao bound.

In this section we shall study some practical aspects of detecting phase
modulated and multiparameter signals in noise and estimating their parameters.
For simplicity we consider the polynomial phase signal with a constant
amplitude.  Our aim is to estimate the parameters of the signal accurately.  We
compare the results of the Monte Carlo simulations with the Cram\'er-Rao bound.

We consider a monochromatic signal and signals with 1, 2, and 3 spindown
parameters.  In our simulations we add white noise to the signals and we repeat
our simulations for several values of the optimal signal-to-noise ratio $d$.  To
detect the signal and estimate its parameters we calculate the optimal
statistics $\F$ derived in Sec.\ 3.  The maximum likelihood detection involves
finding the global maximum of $\F$.  Our algorithm consists of two parts:  a
{\em coarse} search and a {\em fine} search.  The coarse search involves
calculation of $\F$ on an appropriate grid in parameter space and finding the
maximum value of $\F$ on the grid and the values of the parameters of the signal
that give the maximum.  This gives coarse estimators of the parameters.  Fine
search involves finding the maximum of $\F$ using optimization routines with the
starting value determined from the coarse estimates of the parameters.  The grid
for the coarse search is determined by the region of convergence of the
optimization routine used in the fine search.  We have determined the regions of
convergence of our optimization routines in the noise free case.  For the case
of a monochromatic signal when $\F$ depends only on one parameter (frequency)
our optimization algorithm is based on golden section search and parabolic
interpolation.  For a signal with some spindowns included $\F$ depends on $s+1$
parameters ($s$ is number of spindowns) and we use Nelder-Mead simplex
algorithm.

To perform our simulations we have used MATLAB software where the above
optimization algorithms are implemented in {\em fmin} (1-parameter case) and
{\em fmins} ($n$-parameter case) routines.  Both algorithms involve only
calculation of the function to be maximized at certain points but not its
derivatives.  For the multiparameter case the regions of convergence are
approximately parallelepipeds.  We have summarized our results in the Table
\ref{tab:radc} below.  We have given the values of the intersection of the
parallelepipeds with the coordinate axes in the parameter space.  We have
expressed these values in the units of square roots of diagonal values of the
inverse of the matrix $G$ given by Eq.\ (\ref{gmatrix}).

\begin{table}[!ht]
\begin{center}
\begin{tabular}{|c|c|c|c|c|}\hline
$s$ & $r_0$ & $r_1$ & $r_2$ & $r_3$ \\ \hline\hline
0 & 10 & -- &  -- & -- \\ \hline
1 & 0.7 & 0.5 &  -- & -- \\ \hline
2 & 0.2 & 0.08 & 0.1 & -- \\ \hline
3 & $\sim$0.08 & $\sim$0.02 & $\sim$0.01 & $\sim$0.03  \\ \hline
\end{tabular}
\end{center}
\caption{\label{tab:radc} Coordinates of the regions of convergence for the 
polynomial phase signals with $s$ spindowns included in the units of the square 
roots of diagonal elements of the inverse of the matrix $G$ given by Eq.\ 
(\ref{gmatrix}). The region of convergence for the $k$th ($k=0,\ldots,3$) 
spindown is the interval $[-r_k,r_k]$.}
\end{table}

In the case of the signal with 3 spindowns our estimation of the radius of
convergence is very crude because the computational burden to do such a
calculation is very heavy.  The above results hold for the statistics $\F$
calculated when data is only signal and no noise.

In the coarse search we have chosen a rectangular grid in the spindown parameter
space with the nodes separated by twice the values given in Table \ref{tab:radc}
and we have chosen the spindown parameter ranges to be from $-3$ to 3 times the
square roots of the corresponding diagonal elements of matrix $G$ given by Eq.\
(\ref{gmatrix}).  We have made $10^4$ simulations in the case of a monochromatic
signal, 1-spindown, and 2-spindown signals and for each signal-to-noise ratio.
The case of 3 spindowns turned out to be computationally too prohibitive.  
In each case we have taken the length of the signal to be $2^5$ points.

In our simulations we observe that above a certain signal-to-noise ratio that we
shall call the threshold signal-to-noise ratio the results of the Monte Carlo
simulations agree very well with the calculations of the rms errors from the
covarince matrix however below the threshold signal-to-noise ratio they differ
by a large factor.  This threshold effect is well-known in signal processing
\cite{VT69} and has also been observed in numerical simulations for the case of
a coalescing binary chirp signal \cite{KKT94,BSD96}.  There exist more refined
theoretical bounds on the rms errors that explain this effect and they were also
studied in the context of the gravitational-wave signal from a coalescing binary
\cite{NV97}.  Here we present a simple model that explains the deviations from
the covariance matrix and reproduces well the results of the Monte Carlo
simulations.  The model makes use of the concept of the elementary cell of the
parameter space that we introduced in Sec.\ 3.  The calculation given below is a
generalization of the calculation of the rms error for the case of a
monochromatic signal given by Rife and Boorstyn \cite{RB74}.

When the values of
parameters of the template that correspond to the maximum of the functional $\F$
fall within the cell in the parameter space where the signal is present the rms
error is satisfactorily approximated by the covariance matrix.  However
sometimes as a result of noise the global maximum is in the cell where there is
no signal.  We then say that an {\em outlier} has occurred.  
In the simplest case we can assume that the probability
density of the values of the outliers is uniform
over the search interval of a parameter and then the rms error is given by
\be
\sigma_{out}^2 = \frac{\Delta^2}{12},
\ee
where $\Delta$ is the length of the search interval for a given parameter. The 
probability that an outlier occurs will be the higher the lower the 
signal-to-noise ratio. Let $q$ be the probability that an outlier occurs. Then 
the total variance $\sigma^2$ of the estimator of a parameter is the wighted sum 
of the two errors
\be
\label{err}
\sigma^2 = \sigma_{out}^2 q + \sigma_{CR}^2 (1 - q),
\ee
where $\sigma_{CR}$ is the rms errors calculated form the covariance
matrix for a given parameter.

Let us now calculate the probability $q$.  Let $\F_s$ be the value of $\F$ in
the cell where the signal is present and let $\F_o$ be its value in the cells
where signal is absent.  We have
\be
1 - q = P\{\mbox{all:}\,\F_o<\F_s\}
= \int_0^{\infty} P\{\mbox{all:}\,\F_o<\F_s|\F_s=\F\} P\{\F_s = \F\}\,d\F,
\ee
where $P$ stands for probability.
Since the values of the output of the filter in each cell are independent
and they have the same probability density function we have
\be
P\{\mbox{all:}\,\F_o<\F_s|\F_s=\F\} =
\left[P\{\F_o<\F_s|\F_s =\F\}\right]^{N_c-1},
\ee
where $N_c$ is the number of cells of the parameter space. Thus
\be
\label{pout}
1 - q = \int^{\infty}_0 p_1(d,\F) 
\left[\int^\F_o p_0(y)\,dy\right ]^{N_c - 1}\, d\F,
\ee
where $p_0$ and $p_1$ are probability density functions of respectively
false alarm and detection given by Eqs.\ (\ref{p0}) and (\ref{p1}).

In Figures \ref{fig1acr}, \ref{fig1bcr}, and \ref{fig2cr} we have presented the
results of our simulations and we have compared them with the rms errors
calculated from the covariance matrix.  We have also calculated the errors from
our simple model presented above using Eqs.\ (\ref{err}) and (\ref{pout}).  In
the case of frequency, spindowns, and phase to calculate $\sigma_{out}$ we have
assumed uniform probability density.  The estimator of the amplitude is
proportional to the modulus $|\widetilde{X}|$ of the Fourier transform of the
data and in the case of the amplitude we have calculated $\sigma_{out}$ for the
probability density of $|\widetilde{X}|$ assuming that there is no signal in the
data.  We see that the agreement between the simulated and calculated errors is
very good.  This confirms that our simple model is correct.  We also give biases
of the estimators in our simulations.  We see from Figures
\ref{fig1acr}--\ref{fig2cr} that as signal-to-noise ratio increases the
simulated biases tend to zero and the standard deviations tend to rms errors
calculated from the covariance matrices.

\section*{Acknowledgments}

We would like to thank the Albert Einstein Institute, Max Planck Institute for
Gravitational Physics where most of the work presented above has been done for
hospitality.  We would also like to thank Bernard F.\ Schutz for many useful
discussions.  This work was supported in part by Polish Science Committee grant
KBN 2 P303D 021 11.

\appendix

\section{Functions $A$, $B$, and $C$}

The functions  $A$, $B$, and $C$ in Eqs.\ (\ref{ABC}) for the observation 
time chosen as an integer number of sidereal days take the form (here $n$ is a 
positive integer)
\bea 
A\bigg\vert_{T_o=n\,2\pi/\Omega_r}
&=& \frac{1}{16} \sin^22\gamma \left[
9\cos^4\lambda\cos^4\delta + \frac{1}{2}\sin^22\lambda\sin^22\delta
+ \frac{1}{32}\left(3-\cos2\lambda\right)^2\left(3-\cos2\delta\right)^2 \right]
\nonumber\\&&
+ \frac{1}{32} \cos^22\gamma \left[
4\cos^2\lambda\sin^22\delta + \sin^2\lambda\left(3-\cos2\delta\right)^2 
\right],
\\
B\bigg\vert_{T_o=n\,2\pi/\Omega_r}
&=& \frac{1}{32} \sin^22\gamma \left[
\left(3-\cos2\lambda\right)^2\sin^2\delta
+ 4\sin^22\lambda \cos^2\delta \right]
\nonumber\\&&
+ \frac{1}{4} \cos^22\gamma \left(1 + \cos2\lambda\cos2\delta\right), \\
C\bigg\vert_{T_o=n\,2\pi/\Omega_r} &=& 0.
\eea

We see that the functions $A$, $B$, and $C$ depend only on one unknown parameter
of the signal---the declination $\delta$ of the gravitational-wave source.  They
also depend on the latitude $\lambda$ of the detector's location and the
orientation $\gamma$ of the detector's arms with respect to local geographical
directions.

\section{The Fisher matrix}

In this appendix we give the explicit analytic formula for the Fisher matrix for
the simplified model of the gravitational-wave signal from a spinning neutron
star.  The model is defined by Eqs.\ (\ref{cal1}) and (\ref{cal2}) in Sec.\ 4.
It has a constant amplitude and its phase is linear in the parameters.  In Paper
II we have shown that this model reproduces well the accuracy of the estimators
of the parameters calculated from the full model which has amplitude modulation
and nonlinear phase.  In this paper in Sec.\ 5 we show that the number of
templates needed to perform all-sky searches calculated from the linear model
reproduces well the number of templates calculated from the nonlinear phase
model in Ref.\ \cite{BCCS98}.  Thus we see that the Fisher matrix presented
below can be used in the theoretical studies of data analysis of
gravitational-wave signals from spinning neutron stars instead of a very complex
Fisher matrix for the full model.

In Paper II we have found that the Fisher matrix depends on the choice of the
initial time within the observational interval (initial time is that instant of
time at which the instantaneous frequency and the spindown parameters are
defined, see Appendix C of Paper II).  However one finds that the determinant of
the transformation between the two Fisher matrices with different values of the
initial time chosen is 1.  Consequently the number of cells and the number of
filters do not depend on the choice of initial time.  We present our analytic
formula for the initial time chosen to coincide with the middle of the
observation interval.  This simplifies the analytic expressions considerably.

The Fisher matrix $\gai$ for all-sky searches with $s$ spindowns included is 
defined by 
\be
\label{ab1}
\left(\gai\right)_{ij} := \frac{1}{T_o} \int_{-T_o/2}^{T_o/2} 
\frac{\pa\Psi\left(t;\bs{\zeta}\right)}{\pa\zeta_i} 
\frac{\pa\Psi\left(t;\bs{\zeta}\right)}{\pa\zeta_j} dt,
\ee
where $\bs{\zeta}=(\Phi_0,\bs{\xi})$, 
$\bs{\xi}=(\alpha_1,\alpha_2,\omega_0,\ldots,\omega_s)$, and the phase 
$\Psi$ is equal to
\be
\label{ab2}
\Psi\left(t;\bs{\zeta}\right) = \Phi_0 + \Phi\left(t;\bs{\xi}\right);
\ee
the function $\Phi$ is given by Eq.\ (\ref{cal2}). We have calculated the Fisher 
matrix $\gai$ for $s=4$. The result is
\be
\label{ab3}
\gaiv = \left(\begin{array}{cccccccc}
1 & \Gamma_{\Phi_0\alpha_1}& \Gamma_{\Phi_0\alpha_2} &
0 & \frac{1}{12} & 0 & \frac{1}{80} & 0 \\
& \Gamma_{\alpha_1\alpha_1} & \Gamma_{\alpha_1\alpha_2} & 
\Gamma_{\alpha_1\omega_0} & \Gamma_{\alpha_1\omega_1} & 
\Gamma_{\alpha_1\omega_2} & \Gamma_{\alpha_1\omega_3} & 
\Gamma_{\alpha_1\omega_4} \\
&& \Gamma_{\alpha_2\alpha_2} & \Gamma_{\alpha_2\omega_0} & 
\Gamma_{\alpha_2\omega_1} & \Gamma_{\alpha_2\omega_2} & 
\Gamma_{\alpha_2\omega_3} & \Gamma_{\alpha_2\omega_4} \\
&&& \frac{1}{12} & 0 & \frac{1}{80} & 0 & \frac{1}{448} \\
&&&& \frac{1}{80} & 0 & \frac{1}{448} & 0 \\
&&&&& \frac{1}{448} & 0 & \frac{1}{2304} \\
&&&&&& \frac{1}{2304} & 0 \\
&&&&&&& \frac{1}{11264}
\end{array}\right),
\ee
where (here $\Xi_o:=\Omega_oT_o$ and $\Xi_r:=\Omega_rT_o$)
\ben
\Gamma_{\Phi_0\alpha_1} &=&
\frac{4\pi r_{ES}}{c\Xi_o} \sin\phi_o \sin\frac{\Xi_o}{2}
+ \frac{4\pi r_{E}}{c\Xi_r} \cos\ve\,\cos\lambda\,\sin\phi_r 
\sin\frac{\Xi_r}{2},
\\ \Gamma_{\Phi_0\alpha_2} &=&
\frac{4\pi r_{ES}}{c\Xi_o} \cos\phi_o \sin\frac{\Xi_o}{2}
+ \frac{4\pi r_{E}}{c\Xi_r} \cos\lambda\,\cos\phi_r \sin\frac{\Xi_r}{2},
\\
\Gamma_{\alpha_1\alpha_1} &=&
\frac{2\pi^2 r_{ES}^2}{c^2}
\left(1 - \cos2\phi_o\frac{\sin\Xi_o}{\Xi_o}\right)
+ \frac{2\pi^2 r_{E}^2}{c^2} \cos^2\ve\,\cos^2\lambda
\left(1 - \cos2\phi_r\frac{\sin\Xi_r}{\Xi_r}\right)
\nonumber\\&&
+ \frac{8\pi^2r_{E}r_{ES}}{c^2}  \cos\ve\,\cos\lambda \left[
\cos(\phi_o-\phi_r) \frac{\sin\frac{1}{2}(\Xi_r-\Xi_o)}{\Xi_r-\Xi_o}
- \cos(\phi_o+\phi_r) \frac{\sin\frac{1}{2}(\Xi_r+\Xi_o)}{\Xi_r+\Xi_o} \right],
\\
\Gamma_{\alpha_1\alpha_2} &=&
\frac{2\pi^2 r_{ES}^2}{c^2} \sin2\phi_o\frac{\sin\Xi_o}{\Xi_o}
+ \frac{2\pi^2 r_{E}^2}{c^2} \cos\ve\,\cos^2\lambda\,\sin2\phi_r  
\frac{\sin\Xi_r}{\Xi_r}
+ \frac{8\pi^2r_{E}r_{ES}}{c^2\left(\Xi_r^2-\Xi_o^2\right)} \cos\lambda
\nonumber\\&&
\times \left\{
\left[ \cos\phi_r\sin\phi_o \left(\Xi_r\cos\ve-\Xi_o\right)
+ \cos\phi_o\sin\phi_r \left(\Xi_r-\Xi_o\cos\ve\right) \right]
\cos\frac{\Xi_r}{2}\sin\frac{\Xi_o}{2}
\nonumber\right.\\&&\left.
+ \left[ \cos\phi_r\,\sin\phi_o \left(\Xi_r-\Xi_o\cos\ve\right)
+ \cos\phi_o\,\sin\phi_r \left(\Xi_r\cos\ve-\Xi_o\right) \right]
\sin\frac{\Xi_r}{2}\cos\frac{\Xi_o}{2} \right\},
\\
\Gamma_{\alpha_1\omega_0} &=&
\frac{2\pi r_{ES}}{c\Xi_o^2} \cos\phi_o
\left(2\sin\frac{\Xi_o}{2}-\Xi_o\cos\frac{\Xi_o}{2}\right)
\nonumber\\&&
+ \frac{2\pi r_{E}}{c\Xi_r^2} \cos\ve\,\cos\lambda\,\cos\phi_r
\left(2\sin\frac{\Xi_r}{2}-\Xi_r\cos\frac{\Xi_r}{2}\right),
\\
\Gamma_{\alpha_1\omega_1} &=&
\frac{\pi r_{ES}}{c\Xi_o^3} \sin\phi_o \left[
4\Xi_o \cos\frac{\Xi_o}{2} - \left(8-\Xi_o^2\right) \sin\frac{\Xi_o}{2} \right]
\nonumber\\&&
+ \frac{\pi r_{E}}{c\Xi_r^3} \cos\ve\,\cos\lambda\,\sin\phi_r \left[
4\Xi_r \cos\frac{\Xi_r}{2} - \left(8-\Xi_r^2\right) \sin\frac{\Xi_r}{2} 
\right],
\\
\Gamma_{\alpha_1\omega_2} &=&
\frac{\pi r_{ES}}{2c\Xi_o^4} \cos\phi_o \left[
\Xi_o \left(24-\Xi_o^2\right) \cos\frac{\Xi_o}{2}
- 6\left(8-\Xi_o^2\right) \sin\frac{\Xi_o}{2} \right]
\nonumber\\&&
+ \frac{\pi r_{E}}{2c\Xi_r^4} \cos\ve\,\cos\lambda\,\cos\phi_r \left[
\Xi_r \left(24-\Xi_r^2\right) \cos\frac{\Xi_r}{2}
- 6\left(8-\Xi_r^2\right) \sin\frac{\Xi_r}{2} \right],
\\
\Gamma_{\alpha_1\omega_3} &=&
\frac{\pi r_{ES}}{4c\Xi_o^5} \sin\phi_o \left[
8\Xi_o\left(-24+\Xi_o^2\right) \cos\frac{\Xi_o}{2}
+ \left(384-48\Xi_o^2+\Xi_o^4\right) \sin\frac{\Xi_o}{2} \right]
\nonumber\\&&
+ \frac{\pi r_{E}}{4c\Xi_r^5} \cos\ve\,\cos\lambda\,\sin\phi_r \left[
8\Xi_r \left(-24+\Xi_r^2\right) \cos\frac{\Xi_r}{2}
+ \left(384-48\Xi_r^2+\Xi_r^4\right) \sin\frac{\Xi_r}{2} \right],
\\
\Gamma_{\alpha_1\omega_4} &=&
-\frac{\pi}{8c} \left\{
\frac{r_{ES}}{\Xi_o^6} \cos\phi_o \left[
\Xi_o \left(1920-80\Xi_o^2+\Xi_o^4\right) \cos\frac{\Xi_o}{2}
- 10 \left(384-48\Xi_o^2+\Xi_o^4\right) \sin\frac{\Xi_o}{2} \right]
\nonumber\right.\\&&\left.
+ \frac{r_{E}}{\Xi_r^6} \cos\ve\,\cos\lambda\,\cos\phi_r \left[
\Xi_r \left(1920-80\Xi_r^2+\Xi_r^4\right) \cos\frac{\Xi_r}{2}
- 10 \left(384-48\Xi_r^2+\Xi_r^4\right) \sin\frac{\Xi_r}{2} \right] \right\},
\\ \Gamma_{\alpha_2\alpha_2} &=&
\frac{2\pi^2 r_{ES}^2}{c^2}
\left(1 + \cos2\phi_o\frac{\sin\Xi_o}{\Xi_o}\right)
+ \frac{2\pi^2 r_{E}^2}{c^2} \cos^2\lambda
\left(1 + \cos2\phi_r\frac{\sin\Xi_r}{\Xi_r}\right)
\nonumber\\&&
+ \frac{8\pi^2r_{E}r_{ES}}{c^2} \cos\lambda \left[
\cos(\phi_o-\phi_r) \frac{\sin\frac{1}{2}(\Xi_r-\Xi_o)}{\Xi_r-\Xi_o}
+ \cos(\phi_o+\phi_r) \frac{\sin\frac{1}{2}(\Xi_r+\Xi_o)}{\Xi_r+\Xi_o} \right],
\\
\Gamma_{\alpha_2\omega_0} &=&
\frac{2\pi r_{ES}}{c\Xi_o^2} \sin\phi_o
\left(\Xi_o\cos\frac{\Xi_o}{2}-2\sin\frac{\Xi_o}{2}\right)
+ \frac{2\pi r_{E}}{c\Xi_r^2} \cos\lambda\,\sin\phi_r
\left(\Xi_r\cos\frac{\Xi_r}{2}-2\sin\frac{\Xi_r}{2}\right),
\\
\Gamma_{\alpha_2\omega_1} &=&
\frac{\pi r_{ES}}{c\Xi_o^3} \cos\phi_o \left[
4\Xi_o \cos\frac{\Xi_o}{2} - \left(8-\Xi_o^2\right) \sin\frac{\Xi_o}{2} \right]
\nonumber\\&&
+ \frac{\pi r_{E}}{c\Xi_r^3} \cos\lambda\,\cos\phi_r \left[
4\Xi_r \cos\frac{\Xi_r}{2} - \left(8-\Xi_r^2\right) \sin\frac{\Xi_r}{2} 
\right],
\\
\Gamma_{\alpha_2\omega_2} &=&
\frac{\pi r_{ES}}{2c\Xi_o^4} \sin\phi_o \left[
-\Xi_o \left(24-\Xi_o^2\right) \cos\frac{\Xi_o}{2}
+ 6\left(8-\Xi_o^2\right) \sin\frac{\Xi_o}{2} \right]
\nonumber\\&&
+ \frac{\pi r_{E}}{2c\Xi_r^4} \cos\lambda\,\sin\phi_r \left[
-\Xi_r \left(24-\Xi_r^2\right) \cos\frac{\Xi_r}{2}
+ 6\left(8-\Xi_r^2\right) \sin\frac{\Xi_r}{2} \right],
\\
\Gamma_{\alpha_2\omega_3} &=&
\frac{\pi r_{ES}}{4c\Xi_o^5} \cos\phi_o \left[
8\Xi_o\left(-24+\Xi_o^2\right) \cos\frac{\Xi_o}{2}
+ \left(384-48\Xi_o^2+\Xi_o^4\right) \sin\frac{\Xi_o}{2} \right]
\nonumber\\&&
+ \frac{\pi r_{E}}{4c\Xi_r^5} \cos\lambda\,\cos\phi_r \left[
8\Xi_r \left(-24+\Xi_r^2\right) \cos\frac{\Xi_r}{2}
+ \left(384-48\Xi_r^2+\Xi_r^4\right) \sin\frac{\Xi_r}{2} \right],
\\
\Gamma_{\alpha_2\omega_4} &=&
\frac{\pi}{8c} \left\{
\frac{r_{ES}}{\Xi_o^6} \sin\phi_o \left[
\Xi_o \left(1920-80\Xi_o^2+\Xi_o^4\right) \cos\frac{\Xi_o}{2}
- 10 \left(384-48\Xi_o^2+\Xi_o^4\right) \sin\frac{\Xi_o}{2} \right]
\nonumber\right.\\&&\left.
+ \frac{r_{E}}{\Xi_r^6} \cos\lambda\,\sin\phi_r \left[
\Xi_r \left(1920-80\Xi_r^2+\Xi_r^4\right) \cos\frac{\Xi_r}{2}
- 10 \left(384-48\Xi_r^2+\Xi_r^4\right) \sin\frac{\Xi_r}{2} \right] \right\}.
\een

The above formulae could further be simplified if we assume that the observation
time is an integer multiple of one sidereal day.  We also note that if we have
data corresponding to a full year we can start our observation at a time
corresponding to any position of the detector in its motion around the Sun.
This means that in such a case we can choose the phases $\phi_r$ and $\phi_o$
arbitrarily.

The Fisher matrix $\gai$ for $s=0,\ldots,3$ equals to the submatrix of $\gaiv$
consisting of the first $s+3$ columns and the first $s+3$ rows of $\gaiv$.  The
reduced matrix $\ga$ defined in Eq.\ (\ref{redgamma}) can also be obtained from
the matrix $\gai$ by means of the following procedure:  take the inverse of
$\gai$, remove the first column and the first row of the inverse, take again the
inverse of such a submatrix---it equals $\ga$.

In the case of directed searches the Fisher matrix $\gdi$ is also
defined by Eq.\ (\ref{ab1}), but now $\bs{\zeta}=(\Phi_0,\bs{\xi})$,
$\bs{\xi}=(\omega_0,\ldots,\omega_s)$, and the phase $\Phi$ is given by
Eq.\ (\ref{poly}).  The Fisher matrix $\gdiv$ with $s=4$ spindowns
included reads
\be
\gdiv = \left(\begin{array}{cccccccc}
1 & 0 & \frac{1}{12} & 0 & \frac{1}{80} & 0 \\
& \frac{1}{12} & 0 & \frac{1}{80} & 0 & \frac{1}{448} \\
&& \frac{1}{80} & 0 & \frac{1}{448} & 0 \\
&&& \frac{1}{448} & 0 & \frac{1}{2304} \\
&&&& \frac{1}{2304} & 0 \\
&&&&& \frac{1}{11264}
\end{array}\right).
\ee

The Fisher matrix $\gdi$ for $s=0,\ldots,3$ equals to the submatrix of $\gdiv$
consisting of $s+1$ first columns and $s+1$ first rows of $\gdiv$.  The reduced
matrix $\gd$ defined by Eq.\ (\ref{redgamma}) can be obtained from the matrix
$\gdi$ by means of the same procedure as described above for the case of all-sky
searches.

\section{Suboptimal filtering}

Very often suboptimal filter (or a search template) is proposed in hierarchical
signal searches.  In such a search one passes the data through a suboptimal
filter that requires much less computational cost than the optimal filter and
one registers the condidate events.  Then one passes the data through optimal
filters however only for the values (or around the values) of the parameters of
the candidate events to assess the significance of the candidate events.  In
such a search one would like to ensure that there is no loss of events.  A way
to achieve this when using a suboptimal filter is to lower the threshold with
respect to the threshold chosen for the optimal filter so that the number of
expected significant events is the same as with the optimum filter.  The
probability densities derived in Sec.\ 6.1 can be used to calculate what the
lowered threshold should be.

To illustrate the general theory developed in Sec.\ 6 we have considered the
following example.  We have assumed the observation time $T_o$ to be 3 days and
we have restricted ourselves to directed searches.  For such a case the model of
the phase consistent with the 1/4 of a cycle criterion has $s_1=2$ spindowns in
the dominant term, $s_2=1$ spindown in the contribution due to the Earth orbital
motion and no contribution due to the Earth diurnal motion ($s_3=0$), cf.\ Eq.\
(\ref{phase2}).  We have correlated this signal with a template that has
$s_1=1$, $s_2=1$, and $s_3=0$.  Assuming the gravitational-wave frequency
$f_o=1$ kHz and the maximum values of the spindowns for the spindown age
$\tau=40$ yr the fitting factor is 0.91, the number of cells $N_c$ for the
optimal random field is $2.3\times10^{12}$ and the number of cells $N_{sc}$ for
the suboptimal random field is $3.7\times10^{12}$.  We have found that the
fitting factor is practically independent on the right ascension and the
declination of the gravitational-wave source.

In our computations we assume that we lower the threshold according to the law
\be
\label{LT}
\F_{oL} = (\Fo - 1) \text{FF}^2 + 1. 
\ee
The above rule is motivated by the relation between the expectation value of the
statistics $\F$ and the optimal signal-to-noise ratio given by Eq.\
(\ref{thre2}).

The numerical results obtained using formulae derived in Sec.\ 6.1 are presented
in Figure \ref{fig05sim}.  We have assumed the false alarm probability to be 1\%
for the optimal filter.  There is one more input parameter that we need in order
to calculate the numbers of expected events:  the signal-to-noise ratio $d_1$
for which the number of events is 1.  In the upper left plot in Figure
\ref{fig05sim} we have shown the ratio of the expected number of the detected
events for the suboptimal filtering [calculated from Eq.\ (\ref{NsD})] and the
optimal one [calculated from Eq.\ (\ref{ND})] as a function of $d_1$.  We have
assumed that in the suboptimal filter we lower the threshold according to Eq.\
(\ref{LT}).  We have also put $\text{FF}=0.91$.

To assess the number of events that one loses using a search template
Apostolatos \cite{A1} assumed that the number of detected events decreases as
FF$^3$.  In the right upper plot of Figure \ref{fig05sim} we have compared the
number of detected events calculated from Eq.\ (\ref{NsD}) and the ones
calculated using FF$^3$ law.  We see that in general FF$^3$ law underestimates
the event loss.  However for the fitting factors close to one the difference is
small.

We have calculated the numbers of expected detections and false alarms for the
optimal and suboptimal filter both with original and lowered thresholds.  The
results are presented in the two lower plots in Figure \ref{fig05sim}.  In the
plot on the left diamonds mark the ratio of the number of the detected events
for the suboptimal filter with lowered threshold [calculated from Eqs.\
(\ref{NsD}) and (\ref{LT})] and the number of events detected with the optimum
filter [calculated from Eq.\ (\ref{ND})]; squares denote the ratio of the number
of events detected by suboptimal filtering without lowering the threshold and
the number of events detected with the optimum filter; the solid line gives the
fraction of the detected events calculated from FF$^3$ law; all dependencies are
shown as functions of the fitting factor.  The lower plot on the right gives the
ratio of the expected number of false alarms with the suboptimal filter and
lowered threshold and the expected number of false alarms for the optimal
filter.

From our example we see that when using a suboptimal filter by appropriate
lowering of the threshold we can detect all that events that can be detected
with an optimal filter.  There is however a limitation to threshold lowering
arising from the fact that below a certain threshold the false alarm rate can
increase to an unmanageable level.  In the real data analysis there may be other
limitations.  For example below a certain threshold a forest of non-Gaussian
events may appear completely obscuring the real signals.

\section{The use of parameters $\alpha_1$ and $\alpha_2$ to label the 
filters}

If one knows the values of the parameters $\alpha_1$, $\alpha_2$, and $f_o$ it 
is possible to solve Eqs.\ (\ref{a1a2}) with respect to the angles $\alpha$ and 
$\delta$. One can show that each triple $(\alpha_1,\alpha_2,f_o)$ gives two 
such solutions which can be written as follows (note that because 
$\delta\in[-\frac{\pi}{2},\frac{\pi}{2}]$ to determine $\delta$ uniquely it is 
enough to know $\sin\delta$):
\bea
\label{ad1a}
\sin\delta &=& \beta_1\sin\ve \pm \sqrt{1-\beta_1^2-\beta_2^2},
\\[2ex]
\label{ad1b}
\cos\delta &=& \sqrt{1-\sin^2\delta},
\\[2ex]
\label{ad1c}
\sin\alpha &=& \frac{\beta_1-\sin\ve\sin\delta}{\cos\ve\cos\delta},
\\[2ex]
\label{ad1d}
\cos\alpha &=& \frac{\beta_2}{\cos\delta},
\eea
where
\be
\beta_1 := \frac{\alpha_1}{f_o}, \quad \beta_2 := \frac{\alpha_2}{f_o}.
\ee

The correspondence between the parameters $\alpha_1$, $\alpha_2$, $f_o$ and
$\alpha$, $\delta$ given by Eqs.\ (\ref{ad1a})--(\ref{ad1d}) implies that one
can use $\alpha_1$, $\alpha_2$ instead of $\alpha$, $\delta$ to label the
templates needed for matched filtering.  To do this the family of templates
labelled by $\alpha$, $\delta$ (and the other parameters) must be replaced by
{\em two} template families labelled by $\alpha_1$, $\alpha_2$ (and the other
parameters).  The first family arises when in the original family one replaces
$\sin\delta$, $\cos\delta$, $\sin\alpha$, and $\cos\alpha$ by the left-hand
sides of Eqs.\ (\ref{ad1a})--(\ref{ad1d}) with $+$ sign chosen in the front of
the square root in Eq.\ (\ref{ad1a}).  In the second family the replacements are
made with $-$ sign chosen.  The filters labelled by parameters $\alpha_1$ and
$\alpha_2$ will to a good appproximation be linear and the theory of data
processing developed in this paper applies to such a filtering scheme.

When as a result of filtering of the data one gets a significant event one
obtains at the same time the maximum likelihood estimators of the parameters
$\alpha_1$, $\alpha_2$, $f_o$ (and the others).  One can obtain the maximum
likelihood estimators of the position $(\alpha,\delta)$ of the
gravitational-wave source in the sky by means of Eqs.\
(\ref{ad1a})--(\ref{ad1d}).  Note that one should expect to get the maximum
correlation for a template belonging to one out of two families described above,
what means that after filtering one would also know which sign on the left-hand
side of Eq.\ (\ref{ad1a}) should be chosen.

The covariance matrix for the parameters $\alpha$, $\delta$, and $f_o$ can be 
obtained from the covariance matrix for the parameters $\alpha_1$, $\alpha_2$, 
and $f_o$ by means of the law of propagation of errors. Let us introduce
\be
\label{ad2}
{\mathbf x} := (\alpha_1,\alpha_2,f_o), \quad
{\mathbf y} := (\alpha,\delta,f_o).
\ee
Let $C_{\mathbf x}$ be the covariance matrix for the parameters ${\mathbf x}$, 
then the covariance matrix $C_{\mathbf y}$ for the parameters ${\mathbf y}$ can 
be calculated as follows:
\be
\label{ad3}
C_{\mathbf y} = J C_{\mathbf x} J^T,
\ee
where $T$ denotes matrix transposition and the Jacobi matrix $J$ has 
components: 
\be
\label{ad4}
J = \left(\begin{array}{ccc}
{\dst \frac{\pa\alpha}{\pa\alpha_1}} & {\dst \frac{\pa\alpha}{\pa\alpha_2}} &
{\dst \frac{\pa\alpha}{\pa f_o}} \\[2ex]
{\dst \frac{\pa\delta}{\pa\alpha_1}} & {\dst \frac{\pa\delta}{\pa\alpha_2}} &
{\dst \frac{\pa\delta}{\pa f_o}} \\[2ex] 0 & 0 & 1
\end{array}\right).
\ee
All derivatives entering Eq.\ (\ref{ad4}) can be calulated using Eqs.\ 
(\ref{ad1a})--(\ref{ad1d}).

\begin{figure}[!ht]
\begin{center}
\scalebox{0.90}{\includegraphics{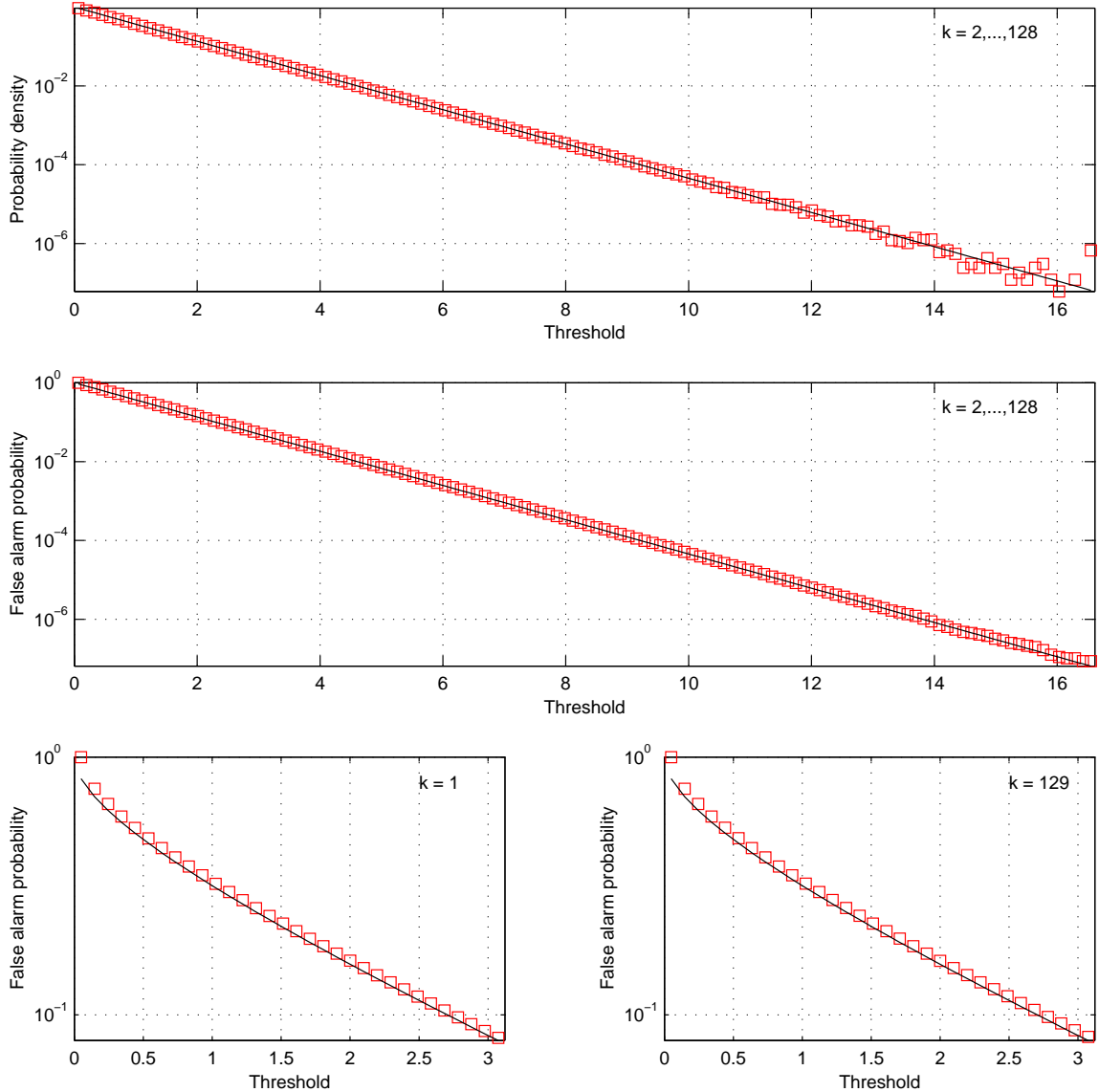}}
\caption{\label{fig01sim}
Probability density of the false alarm (upper plot) and the false alarm
probability (middle plot) of 127 bins of the statistics $\F_k$ (for 
$k=2,\ldots,128$) given by Eq.\ (\ref{Fk}). The false alarm probability of the 
first ($k=1$), zero frequency bin and the last ($k=129$), Nyquist frequency bin
is given in the left lower and the right lower plot, respectively. The 
continuous lines in the upper and the middle plots are theoretical distributions 
given by Eqs.\ (\ref{p0}) and (\ref{PF}) for $n=2$ and the continuous lines in 
the lower plots are theoretical distributions that follow from the cumulative 
$\chi^2$ distribution with 1 degree of freedom. In the simulation one million of 
sequencies of 256 random independent samples drawn from zero mean and unit 
variance normal distribution were generated and modulus of their discrete 
Fourier transforms evaluated. The results of the simulations are marked by the 
squares.}
\end{center}
\end{figure}

\begin{figure}[!ht]
\begin{center}
\scalebox{0.90}{\includegraphics{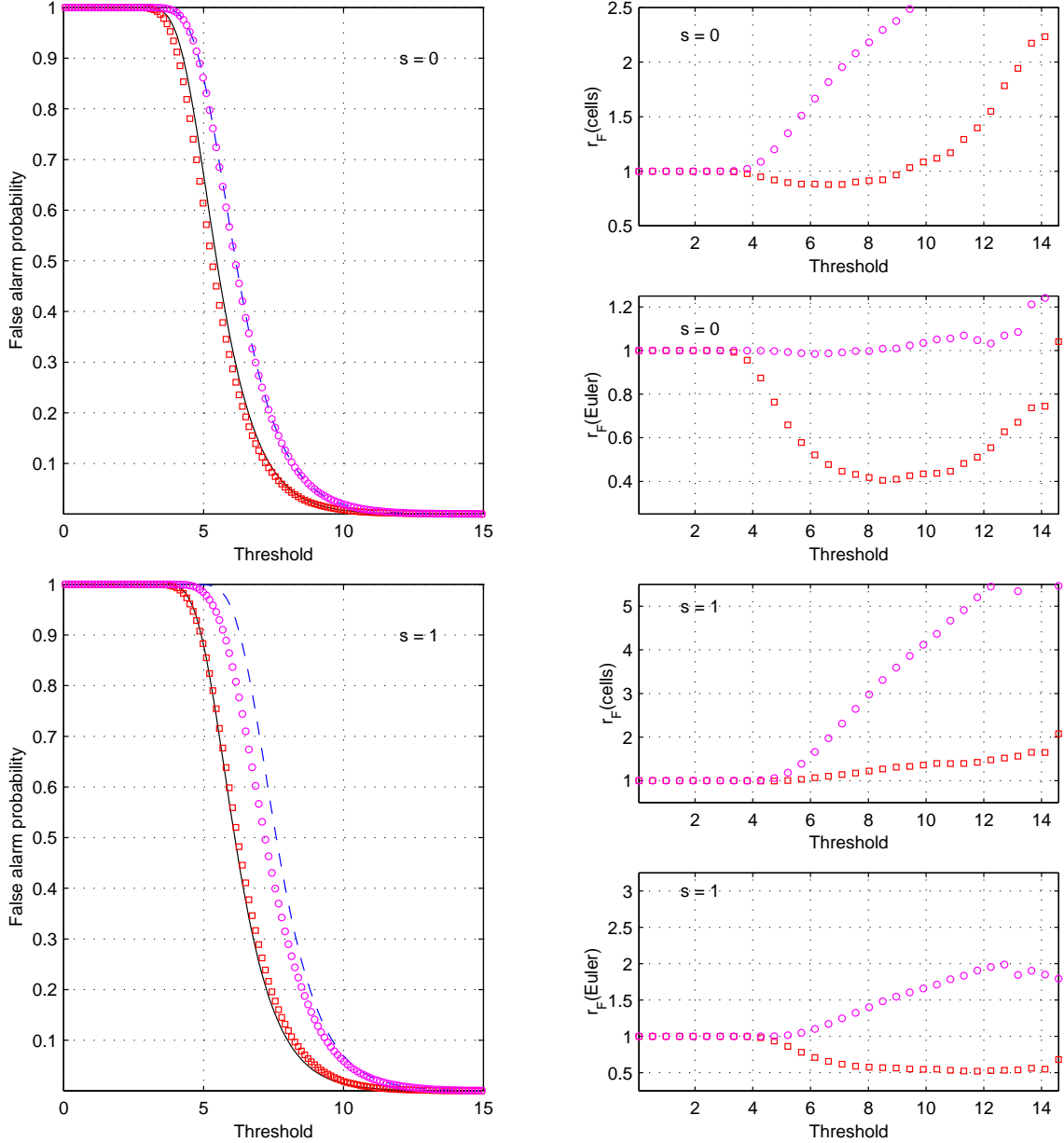}}
\caption{\label{fig03sim}
False alarm probabilities for a monochromatic (three upper plots) and a linearly
frequency modulated signals (three lower plots).  The same random sequences of
length $N=2^8$ were generated as in the simulation in Figure 1 except that
experiment was repeated $10^5$ times. The results of the simulation are marked 
by the squares (no zero padding) and by the circles (for the signal padded with 
$3N$ zeros). The ratio $r_F(\text{cells})$ is the quotient of the false alarm 
probability obtained from the simulations and calculated from Eq.\ (\ref{FP}) 
whereas $r_F(\text{Euler})$ is the quotient of the false alarm probability 
obtained from the simulations and calculated from Eq.\ (\ref{FPf}).}
\end{center}
\end{figure}

\begin{figure}[!ht]
\begin{center}
\scalebox{0.90}{\includegraphics{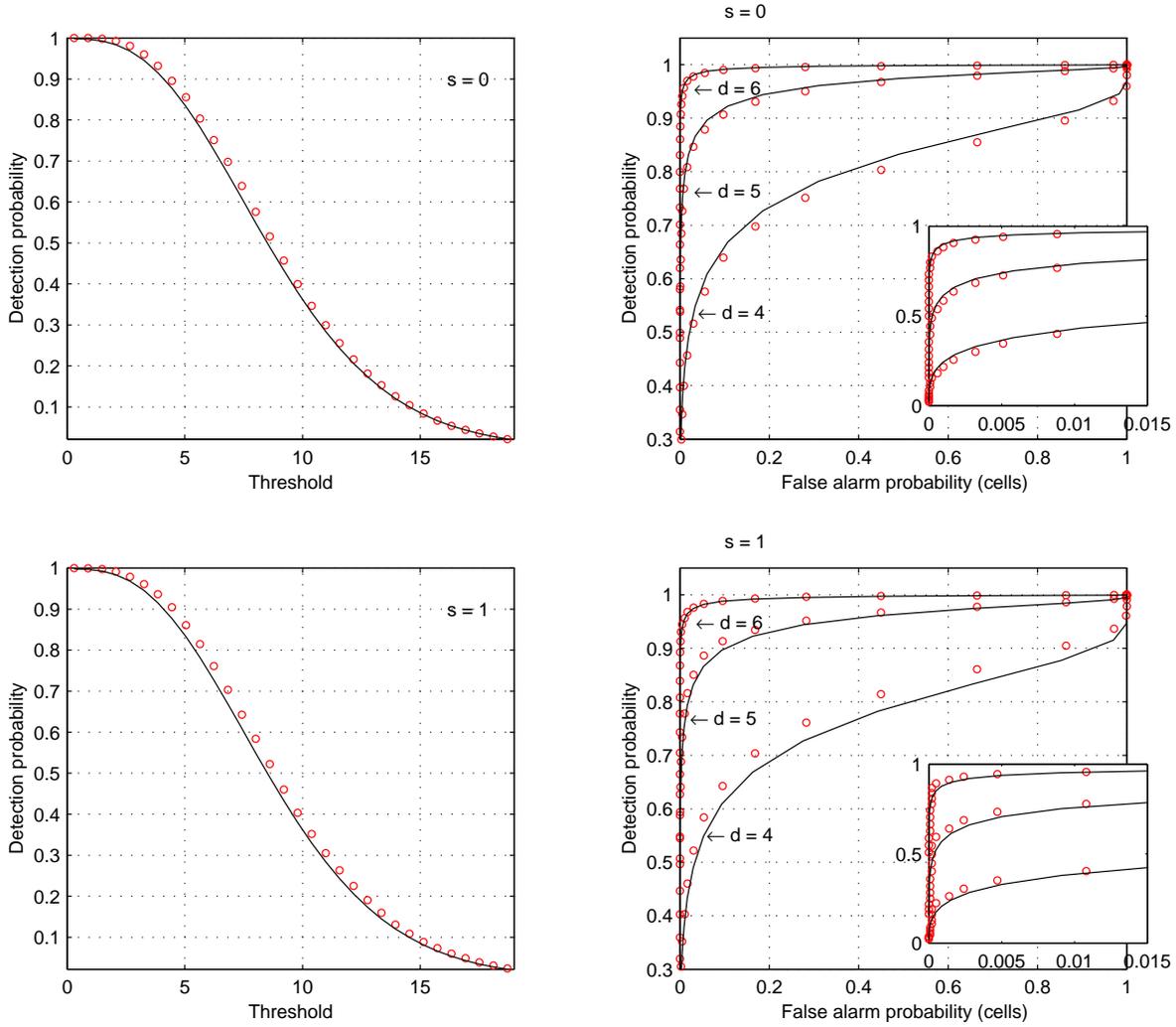}}
\caption{\label{fig04sim}
Probability of detection (plots on the left) and the receiver operating
characteristic (plots on the right) for a monochromatic (upper plots) and a
linearly frequency modulated signal (lower plots).  The same random sequences
of length $N=2^8$ were generated as in the simulation in Figure 1 except that
the experiment was repeated $10^4$ times.  The results of the simulation are
marked by the circles.  Theoretical distributions are given by solid lines.  
Probability of detection is calculated from Eqs.\ (\ref{p1}) and (\ref{PD}) for 
$n=2$ and optimal signal-to-noise ratio $d=4$.  The receiver operating 
characteristics are parametric curves with signal-to-noise ratio $d$
as a parameter, they are calculated from Eqs.\ (\ref{PD}) and (\ref{FP}) for 
$d=4$, 5, and 6.}
\end{center}
\end{figure}

\begin{figure}[!ht]
\begin{center}
\includegraphics{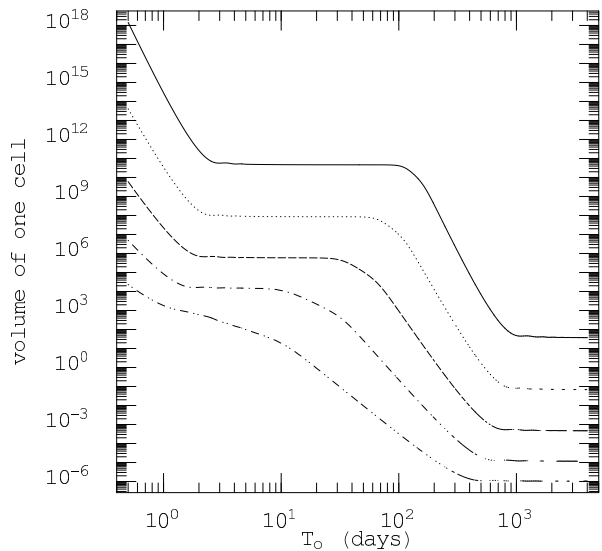}
\caption{\label{fignca0}
Volume of one cell (in units s$^{-2}$) in all-sky searches as a function of the 
observation time $T_o$ for the LIGO Hanford detector (latitude 
$\lambda=46.45^\circ$). We have calculated the volume of one cell from Eq.\ 
(\ref{vca}). The lines shown in the plot correspond to different numbers $s$ of 
spindowns included: $s=4$ (solid), $s=3$ (dotted), $s=2$ (dashed), $s=1$ 
(dotted/dashed), and $s=0$ (double dotted/dashed). We have set $\phi_r=1.456$ 
and $\phi_o=0.123$.}  
\end{center}
\end{figure}

\begin{figure}[!ht]
\begin{center}
\includegraphics{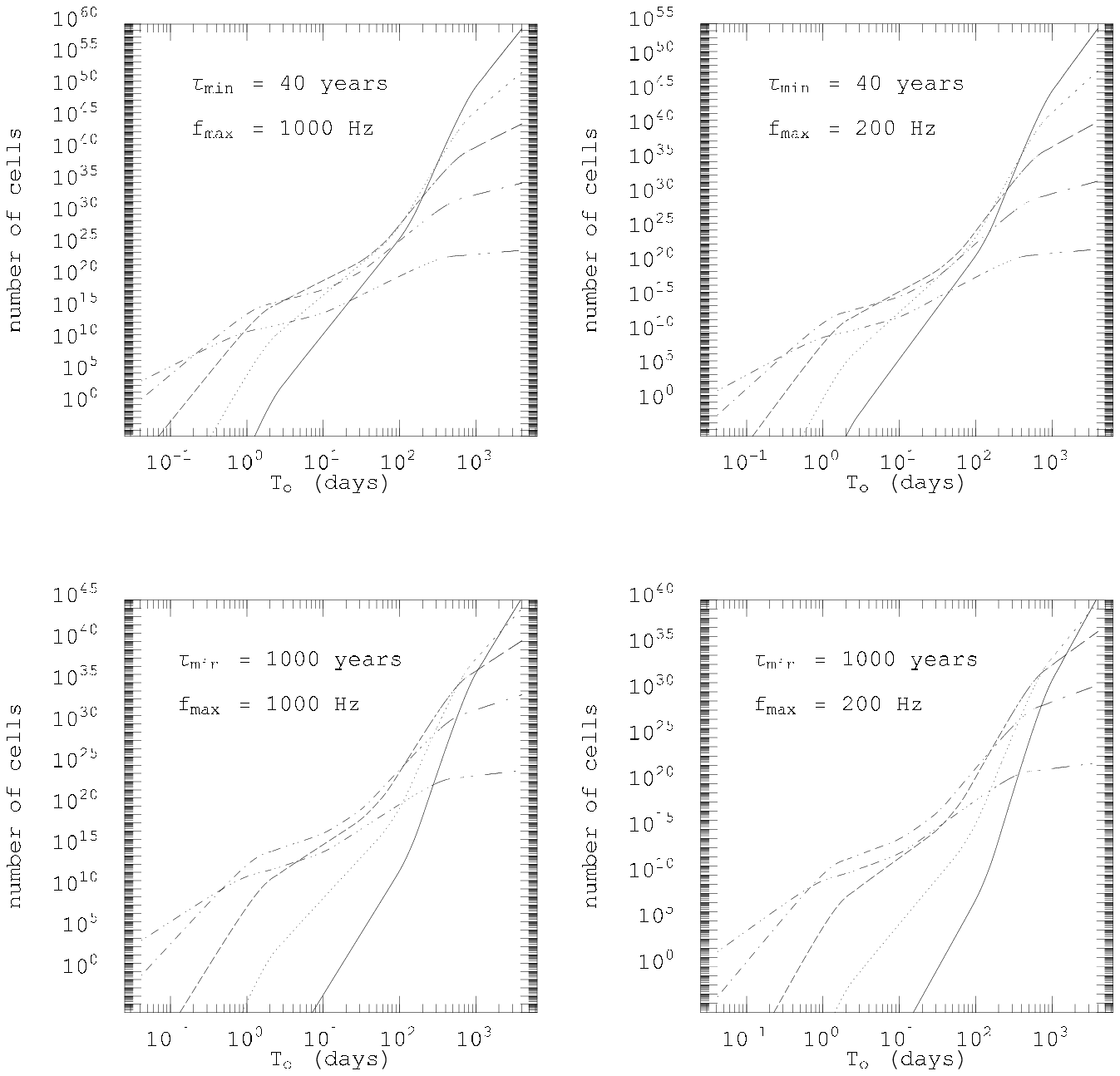}
\caption{\label{fignca1}
Number of cells in all-sky searches as a function of the observation time $T_o$ 
for different values of the minimum spindown age $\tmin$ and the maximum 
gravitational-wave frequency $\fmax$ (the minimum gravitational-wave frequency 
$\fmin=0$). The lines shown in the plots correspond to different numbers $s$ of 
spindowns included: $s=4$ (solid), $s=3$ (dotted), $s=2$ (dashed), $s=1$ 
(dotted/dashed), and $s=0$ (double dotted/dashed). We have assumed the LIGO 
Hanford detector and we have put $\phi_r=1.456$ and $\phi_o=0.123$.}  
\end{center}
\end{figure}

\begin{figure}[!ht]
\begin{center}
\includegraphics{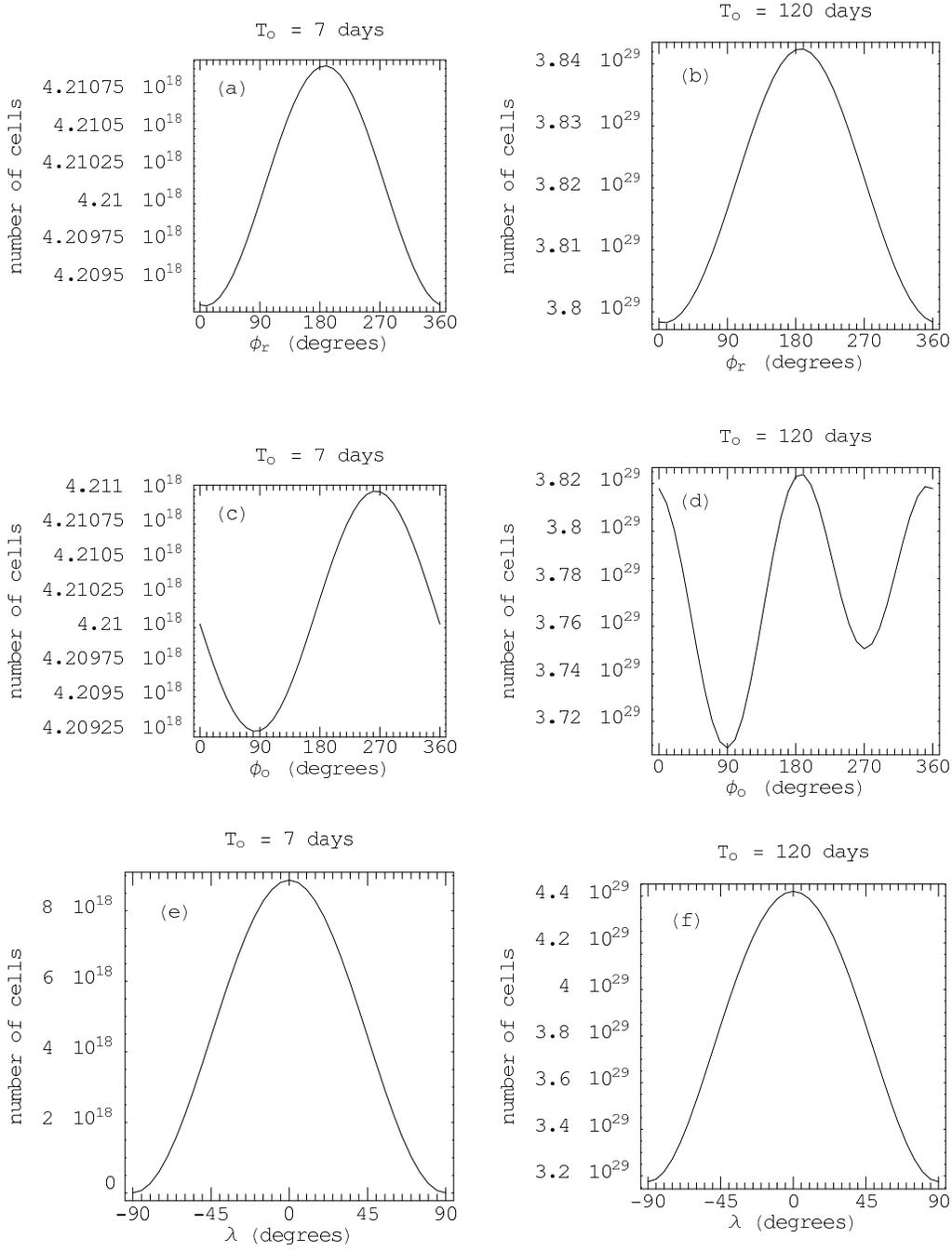}
\caption{\label{fignca2}
Dependence of number of cells in all-sky searches on the angles $\phi_r$, 
$\phi_o$, and the latitude $\lambda$ of the detector's site. We have chosen the 
minimum spindown age $\tmin=40$ years, the maximum gravitational-wave frequency 
$\fmax=1$ kHz, and the minimum gravitational-wave frequency $\fmin=0$. The plots 
(a), (c), and (e) are for the observation time $T_o$ = 7 days (and the number of 
spindowns $s=2$); (b), (d), and (f) are for $T_o$ = 120 days (and the number of 
spindowns $s=3$). In the plots (a), (b), (c), (d) we have used the latitude 
$\lambda=46.45^\circ$ of the LIGO Hanford detector; in (a), (b), (e), (f) we 
have put $\phi_o=0.123$; and in (c), (d), (e), (f) we have used $\phi_r=1.456$.}  
\end{center}
\end{figure}

\begin{figure}[!ht]
\begin{center}
\includegraphics{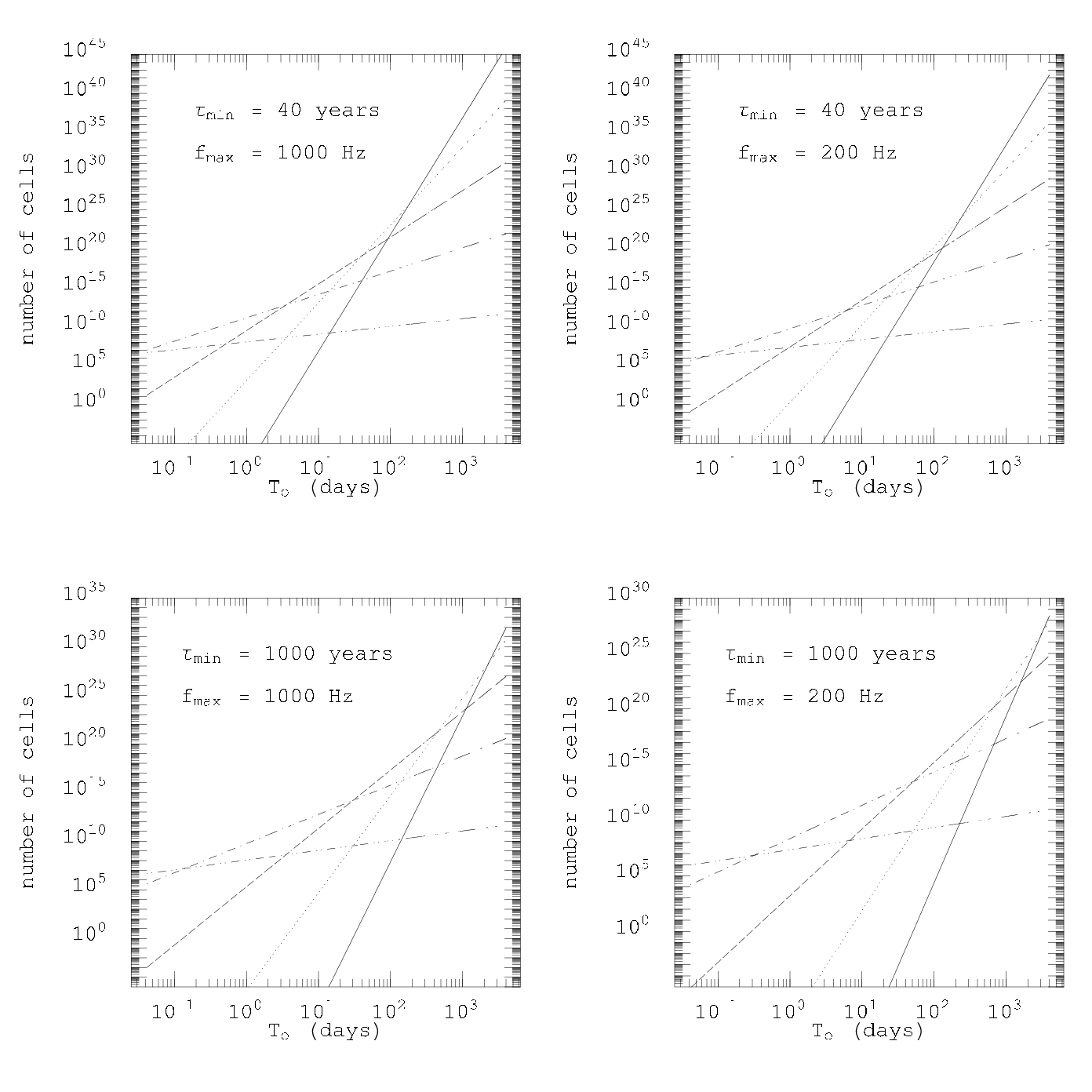}
\caption{\label{figncd}
Number of cells in directed searches as a function of the observation time $T_o$ 
for different values of the minimum spindown age $\tmin$ and the maximum 
gravitational-wave frequency $\fmax$ (the minimum gravitational-wave frequency 
$\fmin=0$). The lines shown in the plots correspond to different numbers $s$ of 
spindowns included: $s=4$ (solid), $s=3$ (dotted), $s=2$ (dashed), $s=1$ 
(dotted/dashed), and $s=0$ (double dotted/dashed).}
\end{center}
\end{figure}

\begin{figure}[!ht]
\begin{center}
\includegraphics{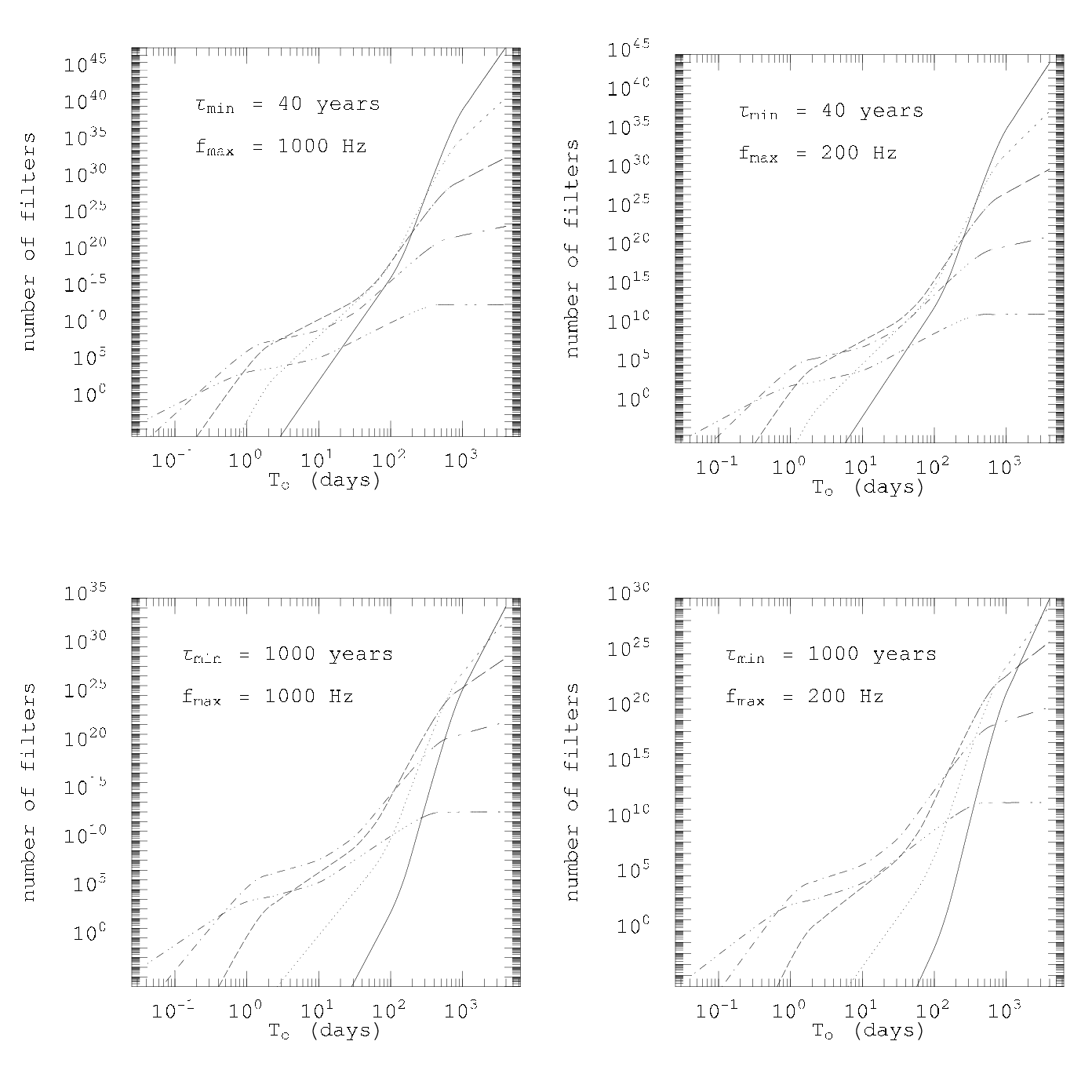}
\caption{\label{fignfa}
Number of filters in all-sky searches as a function of the observation time 
$T_o$ for different values of the minimum spindown age $\tau_{\text{min}}$ and 
the maximum gravitational-wave frequency $f_{\text{max}}$. The lines shown in 
the plots correspond to different numbers $s$ of spindowns included: $s=4$ 
(solid), $s=3$ (dotted), $s=2$ (dashed), $s=1$ (dotted/dashed), and $s=0$ 
(double dotted/dashed). We have assumed the LIGO Hanford detector and we have 
put $\phi_r=1.456$ and $\phi_o=0.123$.}
\end{center}
\end{figure}

\begin{figure}[!ht]
\begin{center}
\includegraphics{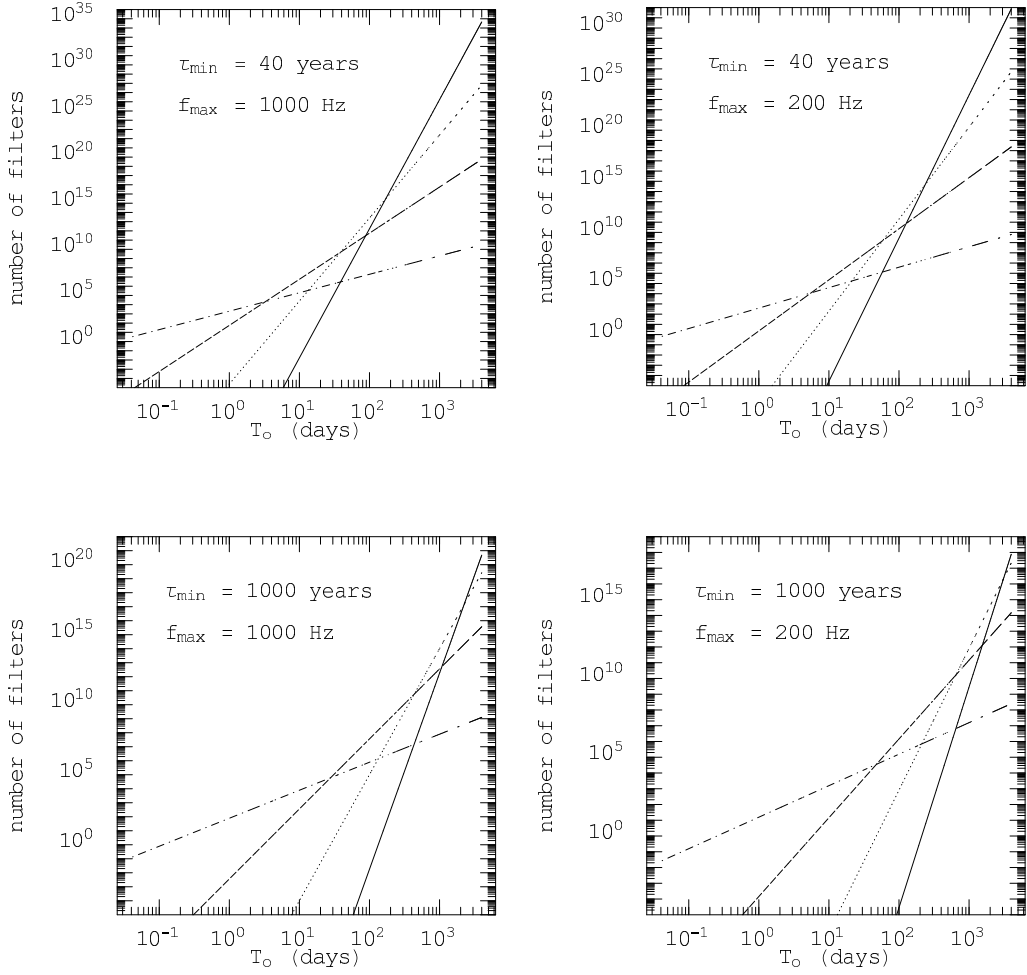}
\caption{\label{fignfd} Number of filters in directed searches as a function of 
the observation time $T_o$ for different values of the minimum spindown age 
$\tmin$ and the maximum gravitational-wave frequency $\fmax$. The lines shown in 
the plots correspond to different numbers $s$ of spindowns included: $s=4$ 
(solid), $s=3$ (dotted), $s=2$ (dashed), $s=1$ (dotted/dashed).}
\end{center}
\end{figure}

\begin{figure}[!ht]
\begin{center}
\scalebox{0.90}{\includegraphics{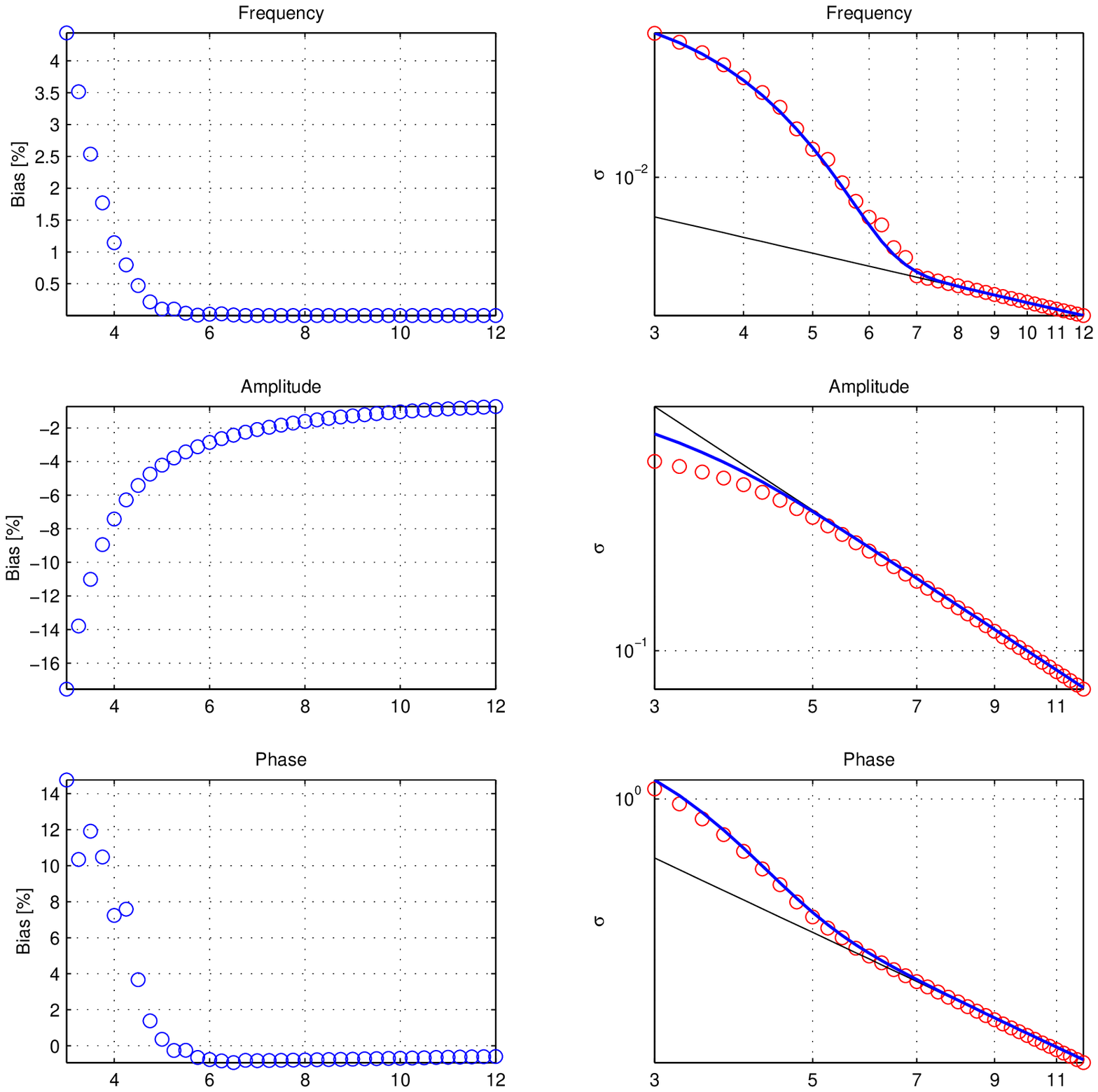}}
\caption{\label{fig1acr} Simulations of the biases (plots on the left) and the 
rms errors (plots on the right) for a monochromatic signal. The results of the 
simulations are marked by the circles. The $x$-axes are labelled by the optimal 
signal-to-noise ratio. The thin solid lines in the plots on the right are 
calculated from the covariance matrix and the thick lines follow from Eqs.\ 
(\ref{err}) and (\ref{pout}).}
\end{center}
\end{figure}

\begin{figure}[!ht]
\begin{center}
\scalebox{0.90}{\includegraphics{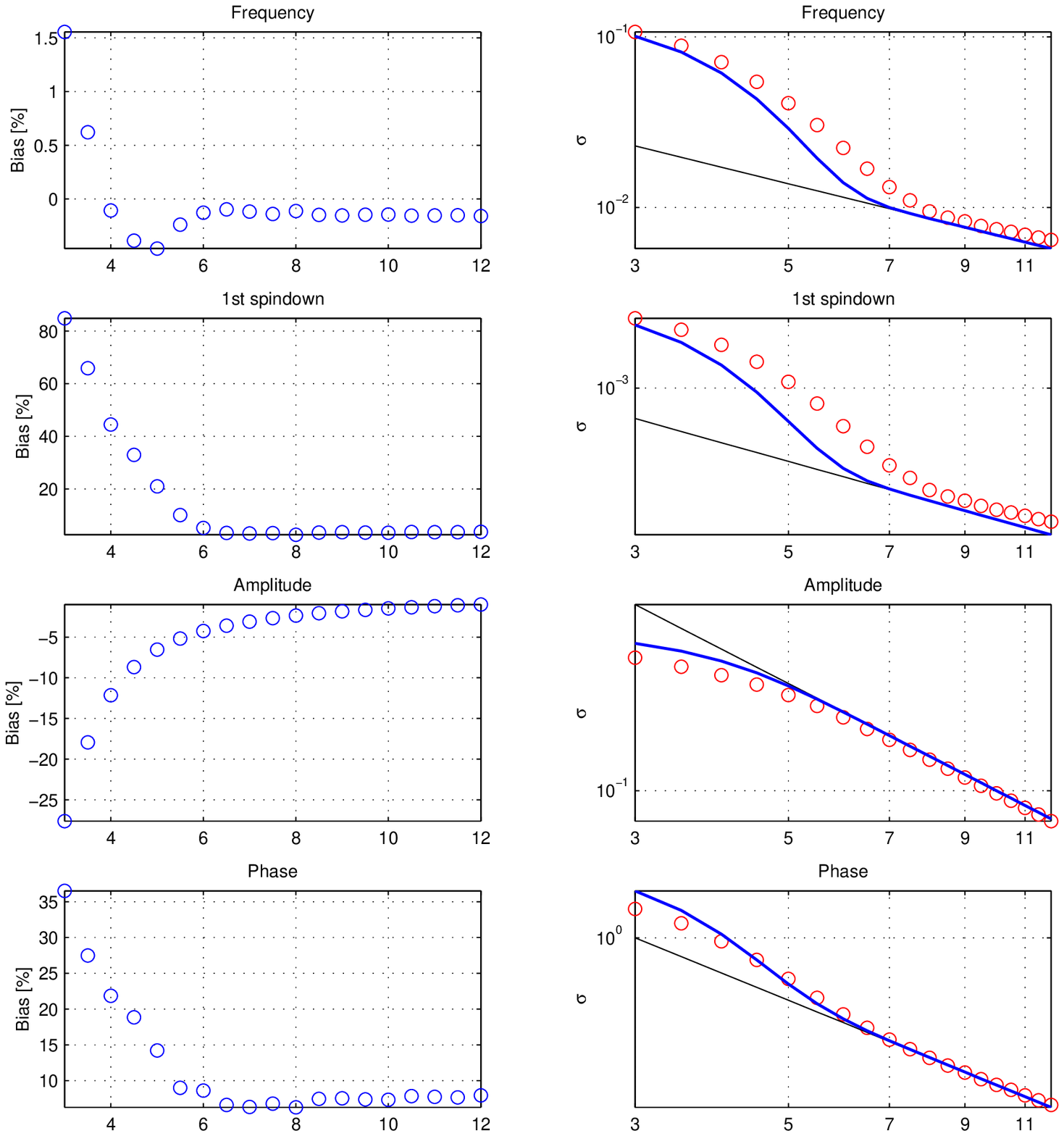}}
\caption{\label{fig1bcr} Simulations of the biases (plots on the left) and the 
rms errors (plots on the right) for a 1-spindown signal. The results of the 
simulations are marked by the circles. The $x$-axes are labelled by the optimal 
signal-to-noise ratio. The thin solid lines in the plots on the right are 
calculated from the covariance matrix and the thick lines follow from Eqs.\ 
(\ref{err}) and (\ref{pout}).}
\end{center}
\end{figure}

\begin{figure}[!ht]
\begin{center}
\scalebox{0.90}{\includegraphics{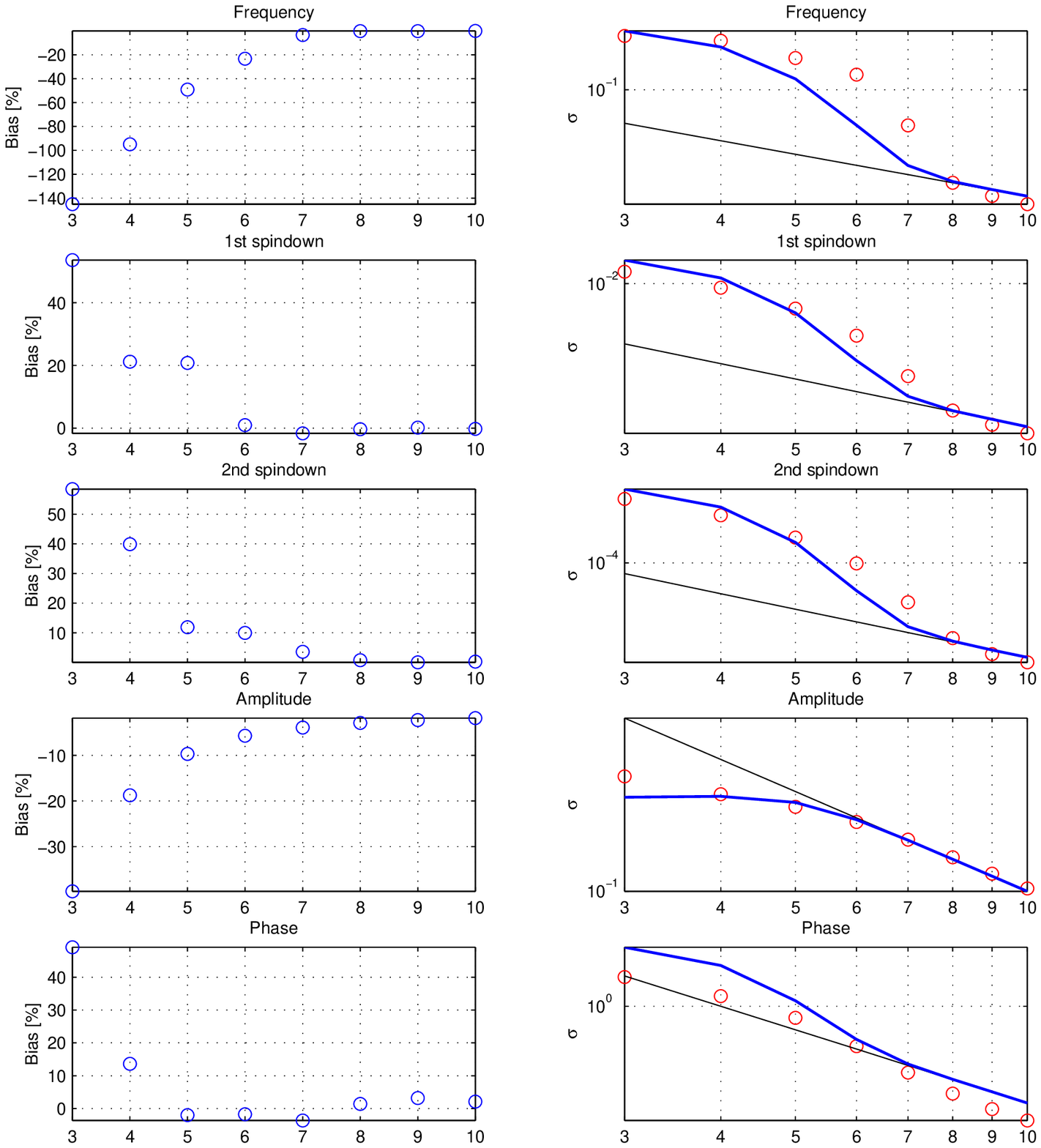}}
\caption{\label{fig2cr} Simulations of the biases (plots on the left) and the 
rms errors (plots on the right) for a 2-spindown signal. The results of the 
simulations are marked by the circles. The $x$-axes are labelled by the optimal 
signal-to-noise ratio. The thin solid lines in the plots on the right are 
calculated from the covariance matrix and the thick lines follow from Eqs.\ 
(\ref{err}) and (\ref{pout}).}
\end{center}
\end{figure}

\begin{figure}[!ht]
\begin{center}
\scalebox{0.90}{\includegraphics{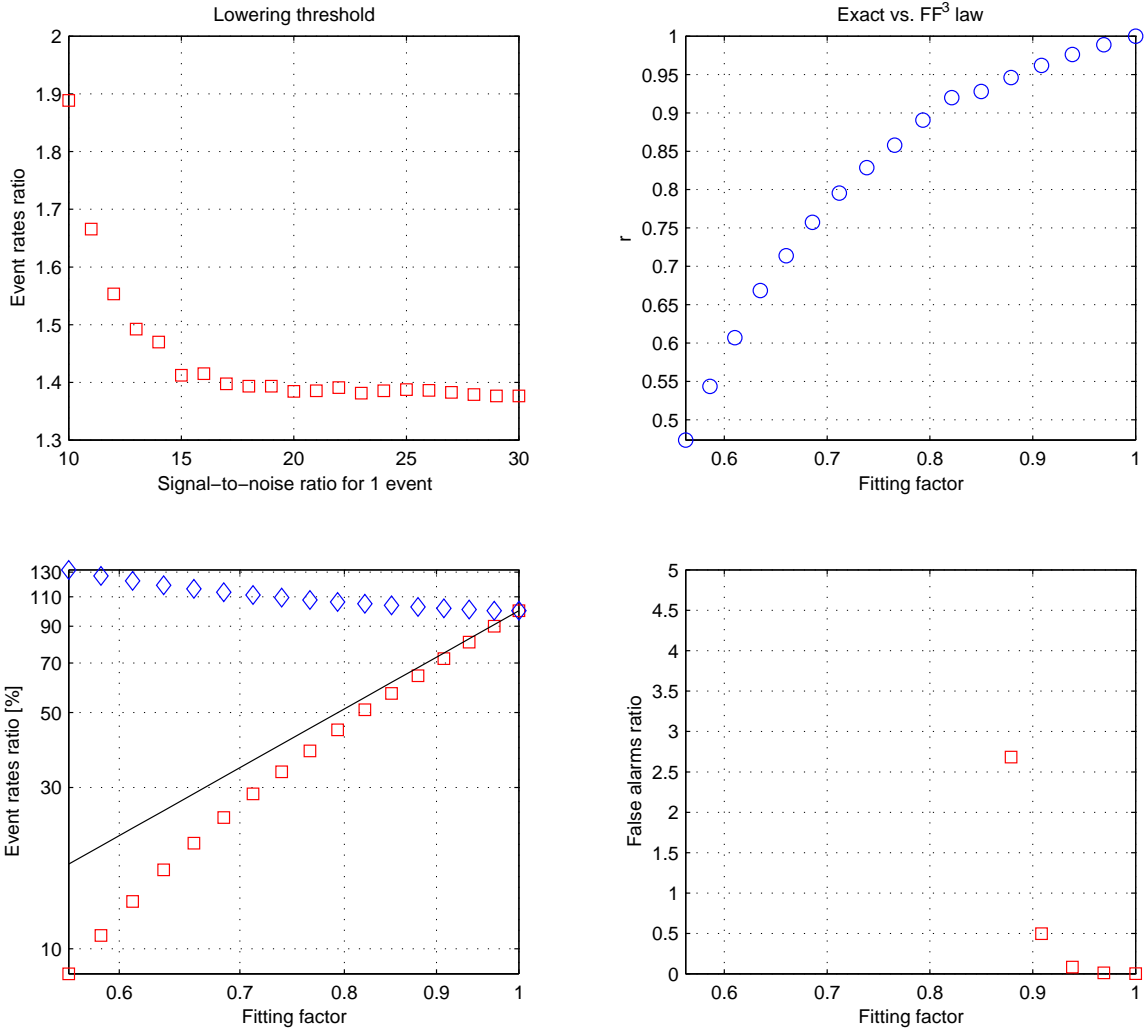}}
\caption{\label{fig05sim} Suboptimal filtering.  In the upper left plot we show
the ratio $N_{sD}(d_{1\text{sub}},\F_{oL})/N_D(d_1,\Fo)$ of the expected number
of the detected events for the suboptimal filtering [calculated from Eq.\
(\ref{NsD})] and the optimal one [calculated from Eq.\ (\ref{ND})] as a function
of the signal-to-noise ratio $d_1$ (the signal-to-noise ratio for which the
number of events is one).  We have assumed that in the suboptimal filter we
lower the threshold according to Eq.\ (\ref{LT}).  We have also put
$\text{FF}=0.91$.  In the right upper plot we give the ratio
$r=[N_{sD}(d_{1\text{sub}},\F_{oL})/N_D(d_1,\Fo)]/\text{FF}^3$ as a function of
the fitting factor (we have used $d_1=16.6$).  In the left lower plot diamonds
mark the ratio $N_{sD}(d_{1\text{sub}},\F_{oL})/N_D(d_1,\Fo)$ of the number of
the detected events for the suboptimal filter with lowered threshold [calculated
from Eqs.\ (\ref{NsD}) and (\ref{LT})] and the number of events detected with
the optimum filter [calculated from Eq.\ (\ref{ND})]; squares denote the ratio
$N_{sD}(d_{1\text{sub}},\Fo)/N_D(d_1,\Fo)$ of the number of events detected by
suboptimal filtering without lowering the threshold and the number of events
detected with the optimum filter; the solid line gives the fraction of the
detected events calculated from FF$^3$ law; all dependencies are shown as
functions of the fitting factor (we have put $d_1=16.6$).  The lower plot on the
right gives the ratio of the expected number of false alarms with the suboptimal
filter and lowered threshold and the expected number of false alarms for the
optimal filter.}
\end{center}
\end{figure}

\end{document}